\documentclass[a4paper,11pt]{article}
\usepackage{comment}
\usepackage{multirow}
\usepackage{adjustbox}
\usepackage{xcolor}
\usepackage[utf8]{inputenc}
\usepackage{booktabs} 
\usepackage{array}    
\usepackage{siunitx}  
\usepackage{longtable} 
\usepackage{caption}  
\usepackage{textgreek} 
\usepackage{ragged2e} 
\pdfoutput=1
\usepackage{jinstpub}
\usepackage{lineno}
\usepackage{tikz}

\title{\boldmath Characterization of UV optical components for photon detector calibration in liquid argon TPCs} 

\author[b]{B. Behera,}
\author[a,c]{M. Bilal Azam}
\author[a,1]{Z. Djurcic,\note{Corresponding author.}}
\author[b]{A. Heindel,}
\author[b]{I. Helgeson,}
\author[a]{T. Hyden,}
\author[b]{D. Leon Silverio,}
\author[a]{S. Magill,}
\author[b,1]{D. A. Martinez Caicedo,}
\author[a]{M. Oberling,}
\author[b]{K. Pickner,}
\author[a]{A. Rafique,}
\author[b]{J. Rodríguez Rondon,}
\author[b]{D. Torres Muñoz,}
\author[b]{C. Winkers,}
\author[a]{L. Xia}

\affiliation[a]{Argonne National Laboratory (ANL), \\ Lemont, IL 60439, USA}
\affiliation[b]{South Dakota School of Mines and Technology,\\ Rapid City, SD 57701, USA}
\affiliation[c]{Illinois Institute of Technology, \\ Chicago, IL 60616, USA}
\emailAdd{zdjurcic@anl.gov, david.martinezcaicedo@sdsmt.edu}

\abstract{
Large liquid argon time projection chambers (LArTPCs) require stable and well-characterized delivery of ultraviolet (UV) light for in situ calibration of photosensors at cryogenic temperatures. This article reports bench-top and cryogenic measurements of the optical components used in a UV light calibration system, including multi-mode fused-silica fibers, SMA-to-SMA connectors, optical fiber feedthroughs, and light-diffuser assemblies. Light loss in several fiber types and SMA connectors was measured across wavelengths from \qtyrange{275}{970}{nm}. In addition, light-loss measurements of the tested fibers after several liquid-nitrogen thermal cycles showed no statistically significant degradation relative to baseline measurements, and high-rate pulsed exposure (30-90 million pulses from a \qty{275}{nm} LED) likewise showed no measurable aging in jacketed fibers. A compact, palm-sized, 3D-printed PEEK diffuser housing with stacked UV-grade fused-silica diffusers yields Lambertian emission and the most uniform angular distribution. Optical components exhibiting improved UV transmission were deployed successfully in multiple DUNE small- and large-scale prototypes, demonstrating reliable operation of UV light calibration systems. These findings inform component selection and calibration procedures for achieving reliable, uniform UV light delivery in large-scale cryogenic detectors such as DUNE.
}


\begin{document}
\begin{tikzpicture}[remember picture,overlay]
    \node[anchor=north east, align=right, font=\small, yshift=-10mm, xshift=-5mm] at (current page.north east) {FERMILAB-PUB-26-0077-LBNF};
\end{tikzpicture}

\maketitle
\flushbottom


\section{Introduction}
A new generation of noble‑element detectors using liquid or gaseous argon and xenon is reshaping searches for rare signals, from dark matter to neutrinoless double beta decay. In accelerator‑based neutrino experiments, large liquid argon time projection chambers (LArTPCs) rely on the detection of ionization charge and scintillation light; the charge which provide essential information for particle identification and reconstruction of spatial information, and the light which provide event timing and aids track reconstruction. Operating photosensors at cryogenic temperatures changes their performance—silicon photomultipliers (SiPMs) and photomultiplier tubes (PMTs) exhibit shifts in breakdown voltage, gain, and photon‑detection efficiency, while dark‑count rates drop—necessitating stable, well‑understood calibration procedures under cold conditions.

Novel LArTPCs implement an in situ light calibration and monitoring system that delivers controlled ultraviolet (UV) pulses synchronized with the photon detection system (PDS) readout as demonstrated in large-scale DUNE prototypes such as ProtoDUNE Single Phase (SP)~\cite{Abi2020} and ProtoDUNE Horizontal Drift (HD)~\cite{Soto-Oton:2024apm}, which have operated at CERN in recent years through dedicated test-beam campaigns~\cite{CERN:ProtoDUNE}.
A light calibration module located outside the cryostat generates programmable UV pulses (amplitude, width, repetition rate, and total count) and distributes triggers from the data‑acquisition (DAQ) system. Light pulses propagate through quartz fibers, traverse an optical feedthrough that bridges the warm–cold boundary; the light enters the cryostat and is emitted from fiber endpoints usually equipped with a light diffuser to illuminate photon detectors on the opposite side of the drift volume record the response~\cite{Abi2020,Soto-Oton:2024apm}. 
Because an efficient and cost-effective \qty{128}{nm} light source matched to liquid-argon vacuum ultraviolet (VUV) scintillation is not available, we use available UV LEDs coupled to commercial fused-silica fibers, and rely on wavelength-shifting at the LArTPC photon-detector surface to convert UV light to detectable wavelengths~\cite{fermilab2018design}.
This paper characterizes the optical components of such a calibration chain across relevant UV and visible wavelengths, with emphasis on cryogenic operation and long‑term stability. 

In this paper, Section \ref{sec:OF} provide a brief introduction how the optical fibers work, and which fibers are tested in this paper. Section \ref{sec:OFQC} summarizes the results of light transmission tests for several multimode fused‑silica fibers (core diameters \qtyrange{400}{600}{\micro\metre}; various coatings) using multiple  LEDs (wavelengths of \numlist{370;465;810;970}\,\unit{nm}). Section \ref{sec:OFTSMA} present the measurement of insertion loss in SMA-to-SMA connectors and multi‑channel optical fiber feedthroughs at \qty{275}{nm} and \qty{367}{nm}. Section \ref{sec:LongFiber} present the estimation of the light attenuation in long fibers and separate SMA-to-SMA connector versus bulk‑fiber loss. Section \ref{sec:CT} present an assessment of  optical fibers mechanical and optical stability via repeated liquid‑nitrogen thermal cycling. Section \ref{sec:DSUV} summarizes the evaluation of potential degradation under high‑rate pulsed UV exposure representative of multi‑year operation using a DUNE-style light calibration module. Section \ref{sec:LDIFF} summarize the characterization of diffuser spatial light profiles, demonstrating a compact 3D‑printed PEEK housing with stacked UV‑grade fused‑silica diffusers that yields near‑Lambertian emission and uniform illumination. Finally, a summary of the research findings is presented in section \ref{sec:Summary}.



\section{Optical Fiber Properties}
\label{sec:OF}

\begin{figure}[h]
    \centering
    \includegraphics[width=0.5\columnwidth]{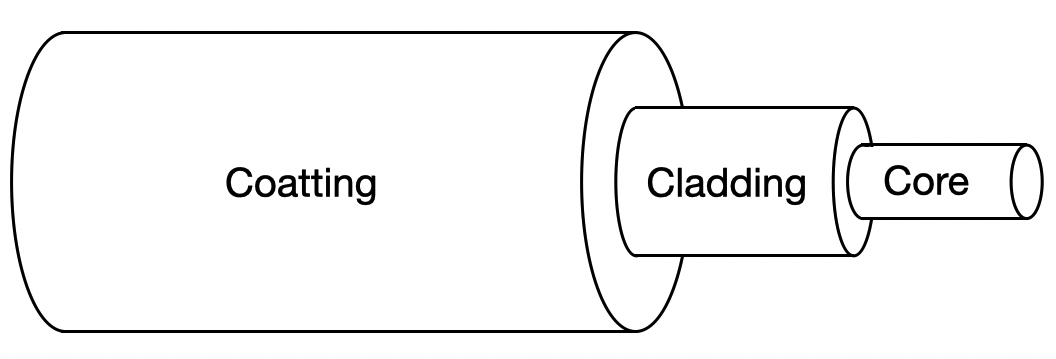}
    \caption{Cross-sectional drawing of optical fiber components, illustrating core, cladding and coating.}
    \label{fig:FiberStructure}
\end{figure}

A typical bare optical fiber consists of three primary components: the core, the cladding, and the coating or buffer (see Figure \ref{fig:FiberStructure}). The core is a cylindrical rod located at the center of the optical fiber and is made of dielectric materials, such as glass or plastic. The core is surrounded by the cladding, which is also made of a dielectric material but has a lower refractive index than the core. The outer layer is the protective coating, typically made of polymer materials, which protects the core and cladding from physical damage, abrasion and moisture. Light travels through the fiber core, where the core-cladding configuration creates an optical waveguide. This configuration keeps the light confined within the core by total internal reflection. Additionally, the coating can reduce reflections at the cladding and improves light transport within the core \cite{fenta2021fibre, okoshi2012optical, keiser2014review}.

\begin{figure}[h]
    \centering
    \includegraphics[width=\columnwidth]{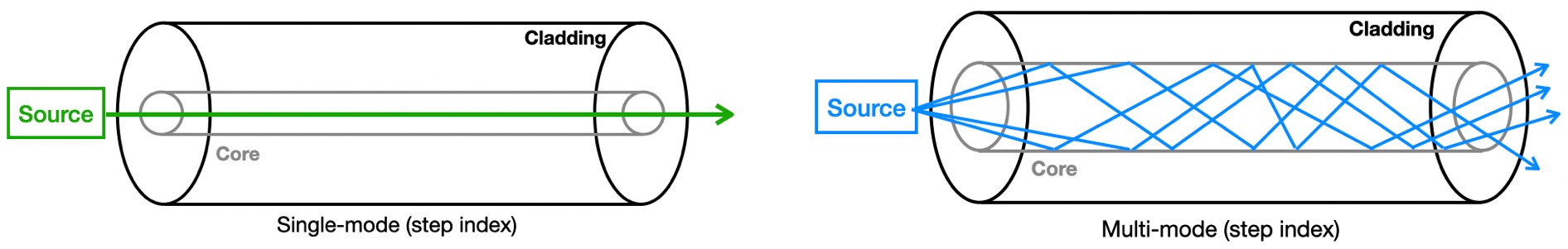}
    \caption{Comparison of light propagation in single-mode (left) and multi-mode (right) optical fibers with step-index profiles.}
    \label{fig:SingleMulti_Mode_StepIndex}
\end{figure}

Optical fibers are classified as single-mode or multi-mode fibers based on their light transmission mode. Single-mode fibers have a narrow core diameter ($\sim$\qtyrange{5}{10}{\micro\metre}) that supports the propagation of only one light mode defined with a single light path, minimizing signal distortion \cite{fenta2021fibre, addanki2018review}. They typically feature a step-index profile and are ideal for long-distance, high-bandwidth applications such as telecommunications (see Figure \ref{fig:SingleMulti_Mode_StepIndex}, left). In contrast, multi-mode fibers have a larger core diameter ($>\qty{50}{\micro\metre}$), which allows multiple light modes to propagate simultaneously and can have either a step-index or gradient-index profile (see Figure \ref{fig:SingleMulti_Mode_StepIndex}, right) \cite{fenta2021fibre, addanki2018review, keiser2014review}. This leads to higher dispersion, reducing signal quality over long distances. However, multi-mode fibers are well-suited for short-distance light transmission (a few meters up to $\sim$\qty{2}{km}) and are compatible with affordable light sources such as LEDs, making them a cost-effective choice. 

This paper presents a characterization of multi-mode optical fibers, studying their light transmission capabilities and suitability for use in cryogenic calibration systems. Measurements of light loss through optical fibers of different types and lengths are presented. Optical fibers and optical feedthroughs needed for fibers cryogenic applications are characterized.
Light transmission performance before and after cryogenic tests (Section \ref{sec:CT}) is compared, and light degradation tests under pulsed LED exposure (Section \ref{sec:DSUV}) are performed.

Multiple optical fibers were selected for the studies presented in this paper with core diameters ranging from \qty{400}{\micro\metre} to \qty{600}{\micro\metre}, while the cladding diameters span from \qty{440}{\micro\metre} to \qty{660}{\micro\metre}. Each fiber is also coated with a protective layer that plays a key role in mechanical integrity and thermal resilience. The coating materials used in the optical fibers included acrylate, polyimide, TECS hard coating, and Tefzel fluoropolymer, each suited for different temperature and environmental requirements. In addition, most of the fibers studied are rated to operate between \qty{-65}{\degreeCelsius} and \qty{+300}{\degreeCelsius}, but this temperature range is still away from the cryogenic conditions necessary to establish their use in particle detectors operating in harsh environments. Then it is important to understand the fibers optical and mechanical performance at cryogenic conditions.

The optical operating wavelength range of the fused-silica (quartz) fibers, typically \qty{180}{nm} to \qty{2400}{nm}, is of particular interest. This broad spectral window covers the UV, visible, and near‑infrared (NIR) regions. Table~\ref{tab:FiberstoTest} summarizes the specifications of each fiber tested, including core/cladding/coating dimensions, coating material, temperature rating, and transmission range.

In the next sections, results on light transmission at UV and visible and NIR wavelengths will be discussed in detail.  Although the emphasis of the measurements presented in this paper was on a UV light calibration systems components deployed in ProtoDUNE SP~\cite{Abi2020} and ProtoDUNE HD~\cite{Soto-Oton:2024apm}, the findings presented here are broadly applicable to future light-calibration systems designed to operate in cryogenic environments.

\begin{table}[htpb]
\centering
\caption{Optical fiber specifications. Information collected from Thorlabs and Molex data sheets. Listed fibers are not rated for cryogenic applications.}
\label{tab:FiberstoTest}
\begin{adjustbox}{max width=\columnwidth,center}
\begin{tabular}{lccccccc}
\toprule
Fiber Tag & Core & Cladding & Coating & Coating & Operating & Numerical & Operating \\
(Supplier) & Diameter [$\mu$m] & Diameter [$\mu$m] & Diameter [$\mu$m] & Material & Temperature [\textdegree C] & Aperture & Wavelength [nm] \\
\midrule
\textbf{FG600AEA (Thorlabs)} 
& 600 $\pm$ 12 & 660 $\pm$ 6 & 750 $\pm$ 20 & Acrylate & -40 to 85 & 0.22 $\pm$ 0.02 & 180 to 1200 \\

\midrule
\multirow{2}{*}{\textbf{FG400UEP (Thorlabs)}} 
& \multirow{2}{*}{400 $\pm$ 8} 
& \multirow{2}{*}{440 $\pm$ 9} 
& \multirow{2}{*}{480 $\pm$ 7} 
& \multirow{2}{*}{Polyimide} 
& \multirow{2}{*}{-65 to 300} 
& \multirow{2}{*}{0.22 $\pm$ 0.02} 
& 400 to 2400 (Low OH) \\
&  &  &  &  &  &  & 250 to 1200 (High OH) \\

\midrule
\multirow{2}{*}{\textbf{FG550UEC (Thorlabs)}} 
& \multirow{2}{*}{550 $\pm$ 19} 
& \multirow{2}{*}{600 $\pm$ 10} 
& \multirow{2}{*}{630 $\pm$ 10} 
& TECS Hard 
& \multirow{2}{*}{-60 to 125} 
& \multirow{2}{*}{0.22 $\pm$ 0.02} 
& 400 to 2200 (Low OH) \\
&  &  &  & Fluoropolymer &  &  & 250 to 1200 (High OH) \\

\midrule
\textbf{UM22-660 (Thorlabs)} 
& 600 $\pm$ 10 & 660 $\pm$ 10 & 710 $\pm$ 10 & Polyimide & -65 to 300 & 0.22 $\pm$ 0.02 & 180 to 850 \\

\midrule
\multirow{2}{*}{\textbf{FT600UMT (Thorlabs)}} 
& \multirow{2}{*}{600 $\pm$ 10} 
& \multirow{2}{*}{630 $\pm$ 10} 
& \multirow{2}{*}{1040 $\pm$ 10} 
& \multirow{2}{*}{Tefzel} 
& \multirow{2}{*}{-65 to 135} 
& \multirow{2}{*}{0.39} 
& 400 to 2100 (Low OH) \\
&  &  &  &  &  &  & 300 to 1200 (High OH) \\

\midrule
\textbf{FVP600660710 (Molex)} 
& 600 $\pm$ 10 & 660 $\pm$ 10 & 710 $\pm$ 10 & Polyimide & -65 to 300 & 0.22 $\pm$ 0.02 & 180 to 1150 (High OH) \\

\midrule
\textbf{FP600URT (Thorlabs)} 
& 600 $\pm$ 10 & 630 $\pm$ 10 & 1040 $\pm$ 30 & Tefzel & -40 to 85 & 0.50 & 300 to 1200 (High OH) \\
\bottomrule
\end{tabular}
\end{adjustbox}
\end{table}

\section{Measurement of Light Transmission through Different Optical Fibers Types }
\label{sec:OFQC}

\begin{figure}[hhh]
    \centering
    \includegraphics[width=0.55\columnwidth]{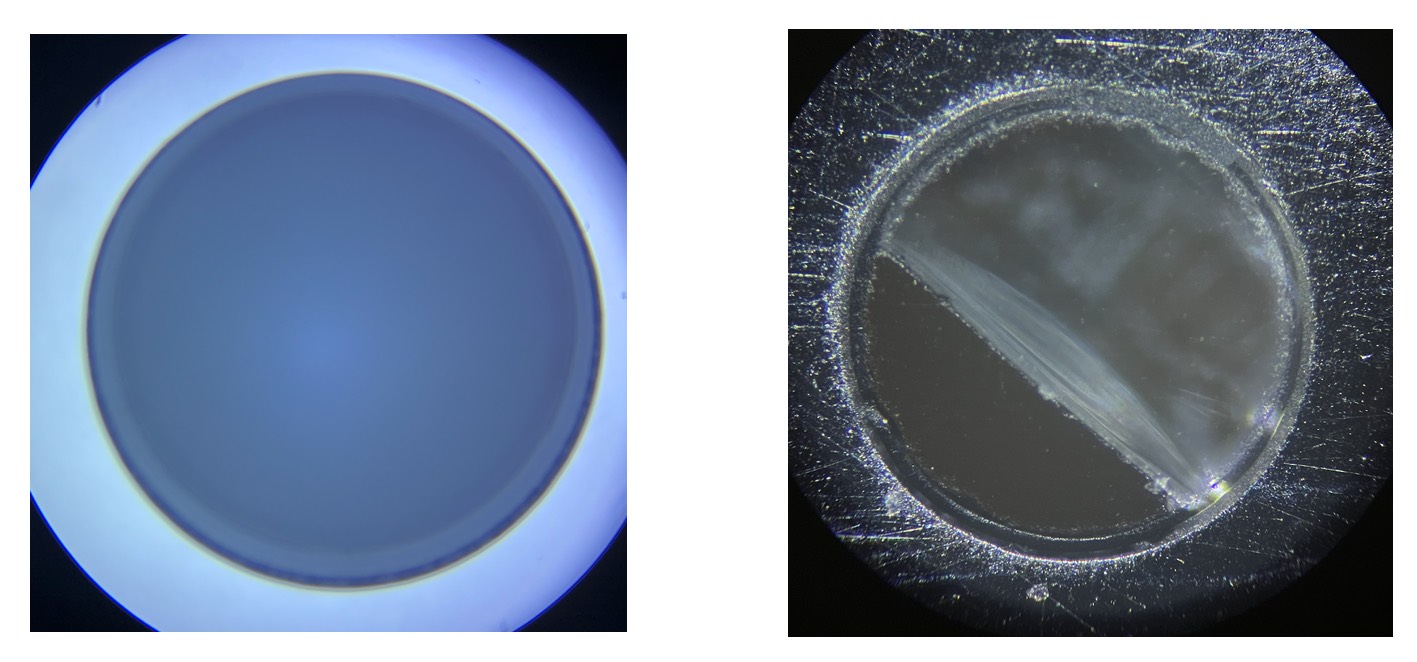}
    \caption{Left: Successful fiber termination with intact core, cladding, and coating. Right: Fiber damaged during cleaving, showing visible damage in core and cladding.}
    \label{fig:FiberTermination}
\end{figure}

The goal of this section is to describe optical fiber transmission properties. A total of 14 fibers were assembled in the laboratory with the procedure described below. The assembly of optical fibers started by cutting the fiber to \qty{1}{m} length. Then, the protective coating was stripped from the fiber’s ends to expose the cladding. The exposed cladding was cleaned with isopropyl alcohol to remove any remaining coating, dirt, or dust. Next, the fiber ends were inserted into SMA 905 connectors (Thorlabs) and secured using cryogenic epoxy (3M Scotch-Weld Epoxy Adhesive 2216). Once the epoxy was cured, each fiber’s end was cleaved close to the SMA connector’s tip and then polished using a commercial polishing machine. Finally, the SMA connector tips were inspected using an optical fiber scope to confirm they were properly assembled and ensure no visible damage at the tip. Figure \ref{fig:FiberTermination} shows a side-by-side comparison of a well-terminated fiber’s end and one with visible damage. 

\begin{figure}[htpb]
    \centering
    \includegraphics[scale=0.22]{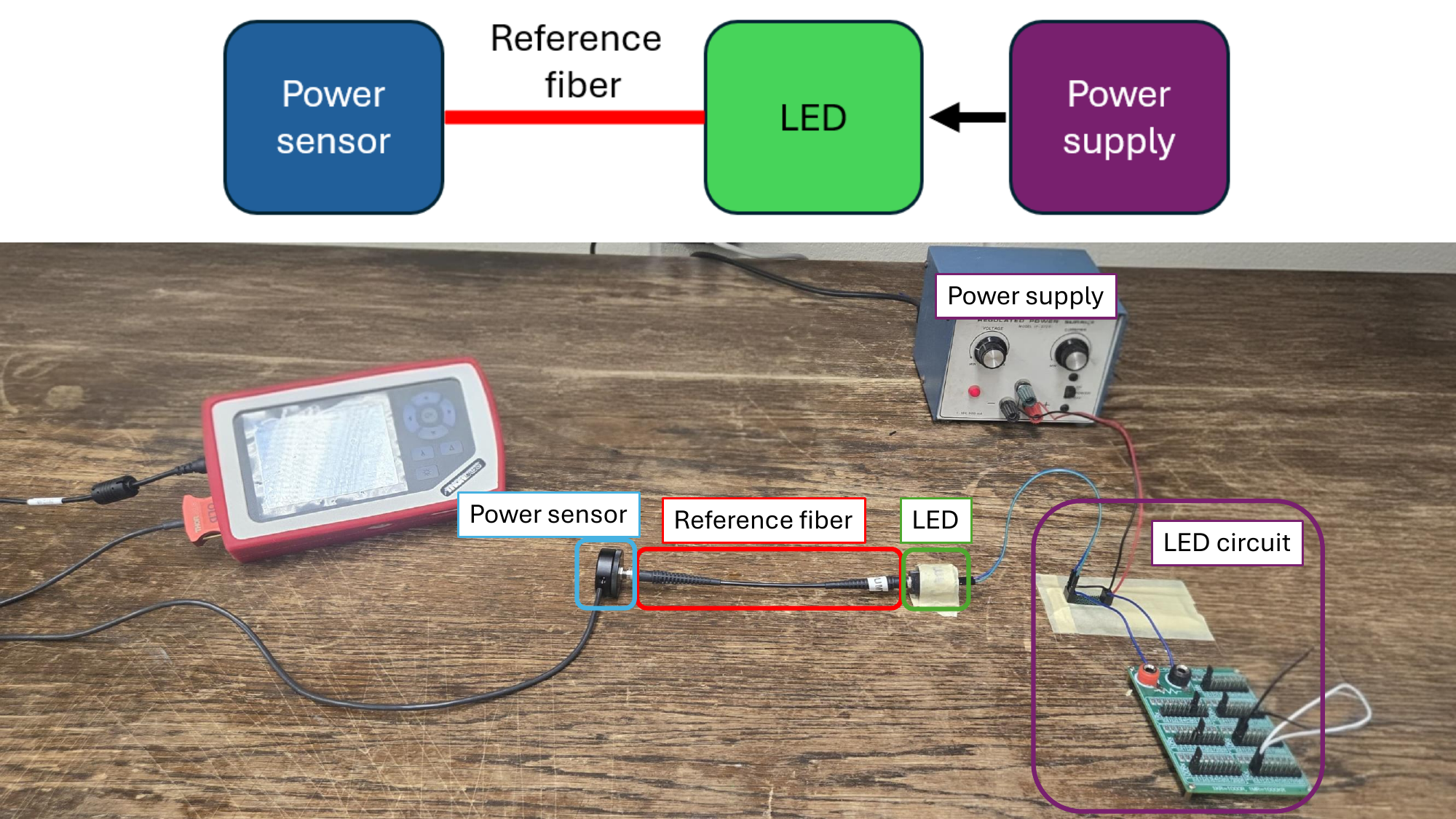}
    \includegraphics[scale=0.22]{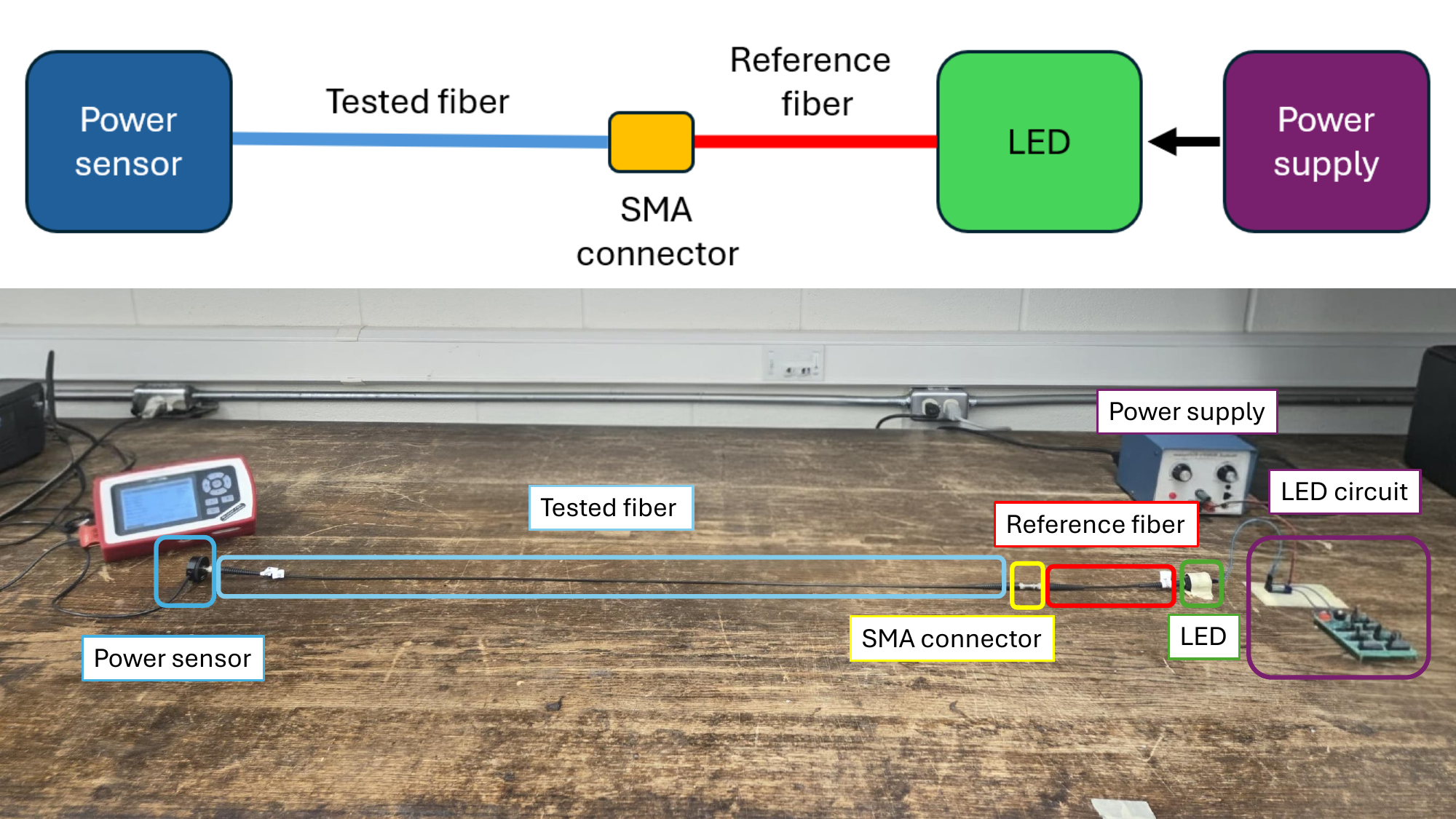}
    \caption{Pictures with its respective diagram of the experimental setup used to measure fiber transparency. Left: Reference configuration, where the LED directly couples into the reference fiber connected to the power sensor. Right: Tested fiber configuration, where one end of the reference fiber is connected to the tested fiber using an SMA-to-SMA connector, and the other end of the tested fiber output at the far end is measured. In both cases, the LED is driven by a regulated 5~V power supply through the resistor-limited circuit visible on the breadboard.}
    \label{fig:bothsetups}
\end{figure}

To characterize the optical performance of different fiber types, their transparency was measured. Transparency is defined as the ratio of the optical power transmitted through a tested fiber to the optical input power provided by the light source, and thus represents the combined effects of light attenuation within the fiber, as well as losses introduced by fiber terminations and SMA-to-SMA connector interfaces. The transparency was determined experimentally by measuring the ratio between the optical power at the input and the output ends of each fiber under controlled conditions. A systematic set of measurements were conducted using four LED sources with wavelengths centered at \qty{370}{nm}, \qty{465}{nm}, \qty{810}{nm}, and \qty{970}{nm} \cite{Thorlabs:LEDsSpecific}. Each LED was used to illuminate fourteen optical fibers in total, seven bare and seven jacketed fibers covering various coating materials and various core, cladding and coating diameters (see Table~\ref{tab:FiberstoTest}). A custom-built setup was developed to ensure mechanical stability and repeatable conditions throughout the entire set of measurements (Figure~\ref{fig:bothsetups}).

The light source consisted of LEDs driven by a simple circuit powered by a regulated \qty{5}{V} DC supply. Each LED was mounted on a compact breadboard alongside a fixed series resistor of approximately \qty{44}{\ohm} for the \qty{370}{nm} LED, \qty{40}{\ohm} for \qty{465}{nm},\qty{54}{\ohm} for \qty{810}{nm}, and \qty{47}{\ohm} for \qty{970}{nm}, which were chosen to match the current requirements for the optimal LED optical output. This electrical configuration provided stable operation of the LED, preventing overheating or overdriving.

The experimental setup to  measure the transparency comprised two configurations: (1) reference setup (Figure~\ref{fig:bothsetups}, left) and (2) testing fiber setup (Figure~\ref{fig:bothsetups}, right). In both configurations, the LED output was coupled into the fibers using a 3D printed holder, and the optical power was measured with a Thorlabs S120VC photodiode power sensor.

In the reference configuration, the LED output was coupled directly to a short reference fiber connected to the power sensor, allowing measurement of the input optical power, denoted as $P_{\text{reference}}$, with minimal loss. In the tested fiber configuration, the reference fiber was connected to the fiber under test via an SMA-to-SMA connector, and the transmitted power at the far end of the tested fiber was measured, yielding $P_{\text{tested fiber}}$. The fiber transparency $T$ was calculated as the ratio:

\begin{equation}\label{eq31}
T = \frac{P_{\text{tested fiber}}}{P_{\text{reference}}}
\end{equation}

Prior to each measurement, the output signal was monitored until stable readings were obtained. The ends of both the tested and reference fibers were visually inspected and cleaned to prevent contamination-related losses. All fibers (tested and reference) were clearly labeled with their corresponding fiber tag and side A or B according to each fiber-end. The same SMA-to-SMA mating connector was used throughout all tests to ensure connection consistency. The mechanical alignment of the setup was fixed for each transparency measurement to reduce systematic errors caused by movement, as even slight misalignments could affect the readings from the power sensor, especially given the low optical power levels measured (in the \unit{\micro\watt} range).

For each of the fourteen fibers tested in configuration (2), the transparency was measured from both sides (A and B). These two measurements were then averaged to obtain the final transparency value reported for each fiber. The measurement sequence involved recording a one-minute reference power reading, followed by a one-minute power measurement of side A for the tested fiber, a second reference measurement, and then a one-minute power measurement of the reversed side B of the tested fiber. This interleaved procedure minimized the effects of LED output fluctuations and improved measurement reproducibility. In configuration (1), used for the reference fibers (with a different reference fiber necessary for each fiber type), side A was always connected to the light source and side B to the SMA-to-SMA connector, establishing a consistent orientation. This same convention was followed in configuration (2), where side B of the reference fiber was always connected to the SMA-to-SMA connector. Maintaining this fixed fiber orientation across all measurements to ensure comparability between tests and to minimize variability arising from potential asymmetries between fiber ends.

\begin{table}[htpb]
\centering
\caption{Transparency measurements for different bare fibers connected with SMA-to-SMA connector at four wavelengths, including their respective errors.}
\label{tab:transparency}
\label{tab:transparency}
\begin{adjustbox}{width=\columnwidth,center}
\begin{tabular}{l|c|c|c|c}
\toprule
\multicolumn{5}{c}{\textbf{Transparency results for bare fibers + SMA-to-SMA connector}} \\
\midrule
\textbf{Fiber} & \textbf{370~nm} $(T \pm \Delta T)$ & \textbf{465~nm} $(T \pm \Delta T)$ & \textbf{810~nm} $(T \pm \Delta T)$ & \textbf{970~nm} $(T \pm \Delta T)$ \\
\midrule
FG600AEA     & $0.827 \pm 0.050$ & $0.841 \pm 0.023$ & $0.798 \pm 0.039$ & $0.737 \pm 0.037$ \\
FG400UEP     & $0.813 \pm 0.041$ & $0.847 \pm 0.022$ & $0.764 \pm 0.055$ & $0.708 \pm 0.034$ \\
FG550UEC     & $0.787 \pm 0.041$ & $0.861 \pm 0.017$ & $0.802 \pm 0.023$ & $0.697 \pm 0.040$ \\
UM22-600     & $0.880 \pm 0.039$ & $0.858 \pm 0.025$ & $0.870 \pm 0.057$ & $0.811 \pm 0.052$ \\
FT600UMT     & $0.734 \pm 0.033$ & $0.883 \pm 0.025$ & $0.923 \pm 0.013$ & $0.858 \pm 0.024$ \\
FVP600660710 & $0.874 \pm 0.042$ & $0.905 \pm 0.030$ & $0.831 \pm 0.020$ & $0.752 \pm 0.038$ \\
FP600URT     & $0.872 \pm 0.036$ & $0.901 \pm 0.014$ & $0.899 \pm 0.009$ & $0.828 \pm 0.031$ \\
\bottomrule
\end{tabular}
\label{table2}
\end{adjustbox}
\end{table}

\begin{table}[htpb]
\centering
\caption{Transparency measurements for different jacketed fibers connected with SMA-to-SMA connector at four wavelengths, including their respective errors.}
\label{tab:transparency_jacket}

\begin{adjustbox}{width=\columnwidth,center}
\begin{tabular}{l|c|c|c|c}
\toprule
\multicolumn{5}{c}{\textbf{Transparency results for jacketed fibers+SMA-to-SMA connector}} \\
\midrule
\textbf{Fiber} & \textbf{370~nm} $(T \pm \Delta T)$ & \textbf{465~nm} $(T \pm \Delta T)$ & \textbf{810~nm} $(T \pm \Delta T)$ & \textbf{970~nm} $(T \pm \Delta T)$ \\
\midrule
FG600AEA     & $0.851 \pm 0.054$ & $0.885 \pm 0.023$ & $0.827 \pm 0.040$ & $0.779 \pm 0.038$ \\
FG400UEP     & $0.848 \pm 0.043$ & $0.859 \pm 0.023$ & $0.814 \pm 0.056$ & $0.726 \pm 0.035$ \\
FG550UEC     & $0.774 \pm 0.040$ & $0.831 \pm 0.017$ & $0.772 \pm 0.023$ & $0.674 \pm 0.039$ \\
UM22-600     & $0.854 \pm 0.039$ & $0.869 \pm 0.025$ & $0.855 \pm 0.058$ & $0.778 \pm 0.051$ \\
FT600UMT     & $0.705 \pm 0.032$ & $0.856 \pm 0.025$ & $0.916 \pm 0.013$ & $0.839 \pm 0.024$ \\
FVP600660710 & $0.851 \pm 0.043$ & $0.883 \pm 0.030$ & $0.851 \pm 0.021$ & $0.730 \pm 0.036$ \\
FP600URT     & $0.848 \pm 0.035$ & $0.887 \pm 0.014$ & $0.902 \pm 0.009$ & $0.866 \pm 0.032$ \\
\bottomrule
\end{tabular}
\end{adjustbox}
\end{table}

\begin{figure}[htpb]
    \centering
    \includegraphics[scale=0.296]{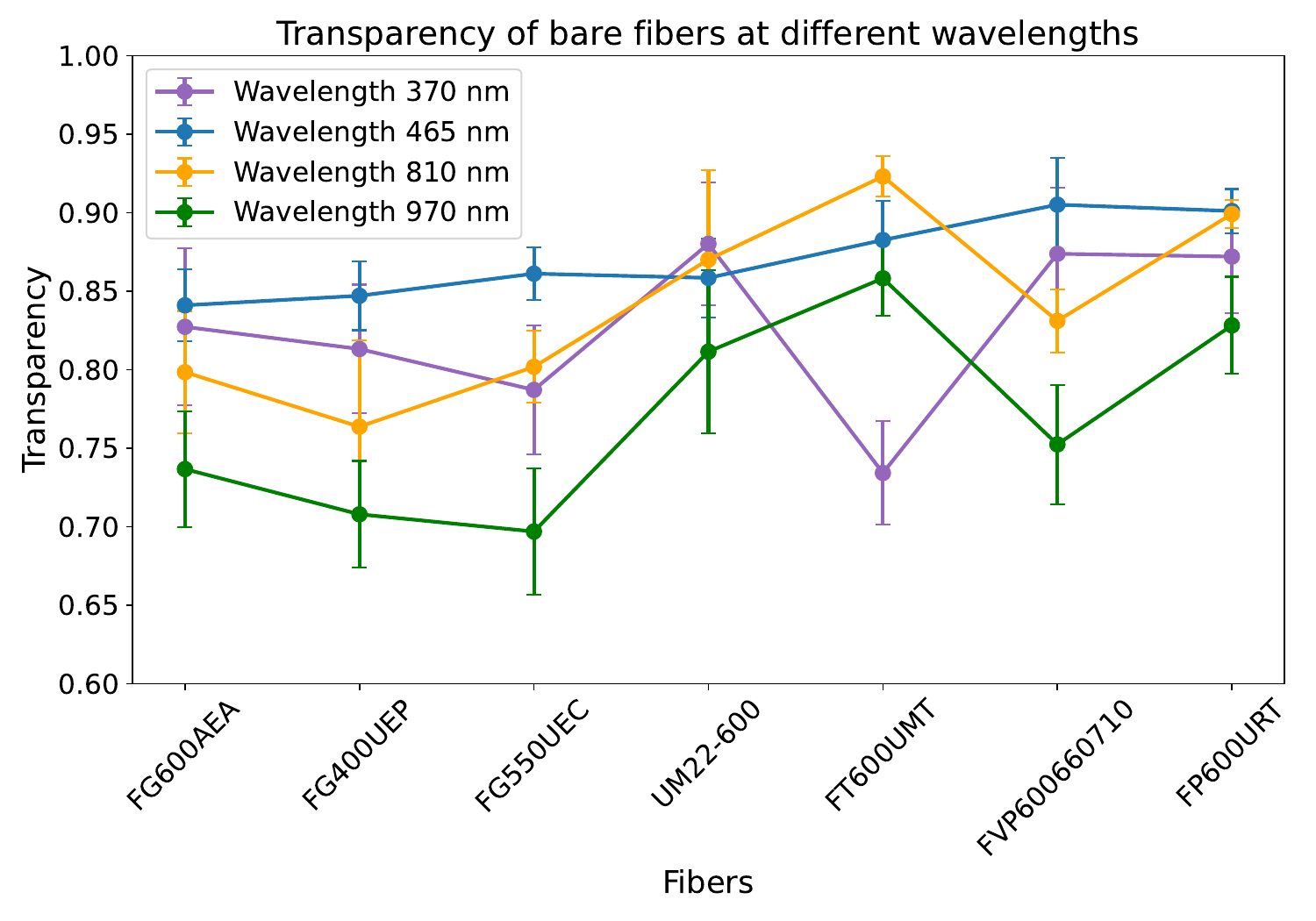}
    \includegraphics[scale=0.296]{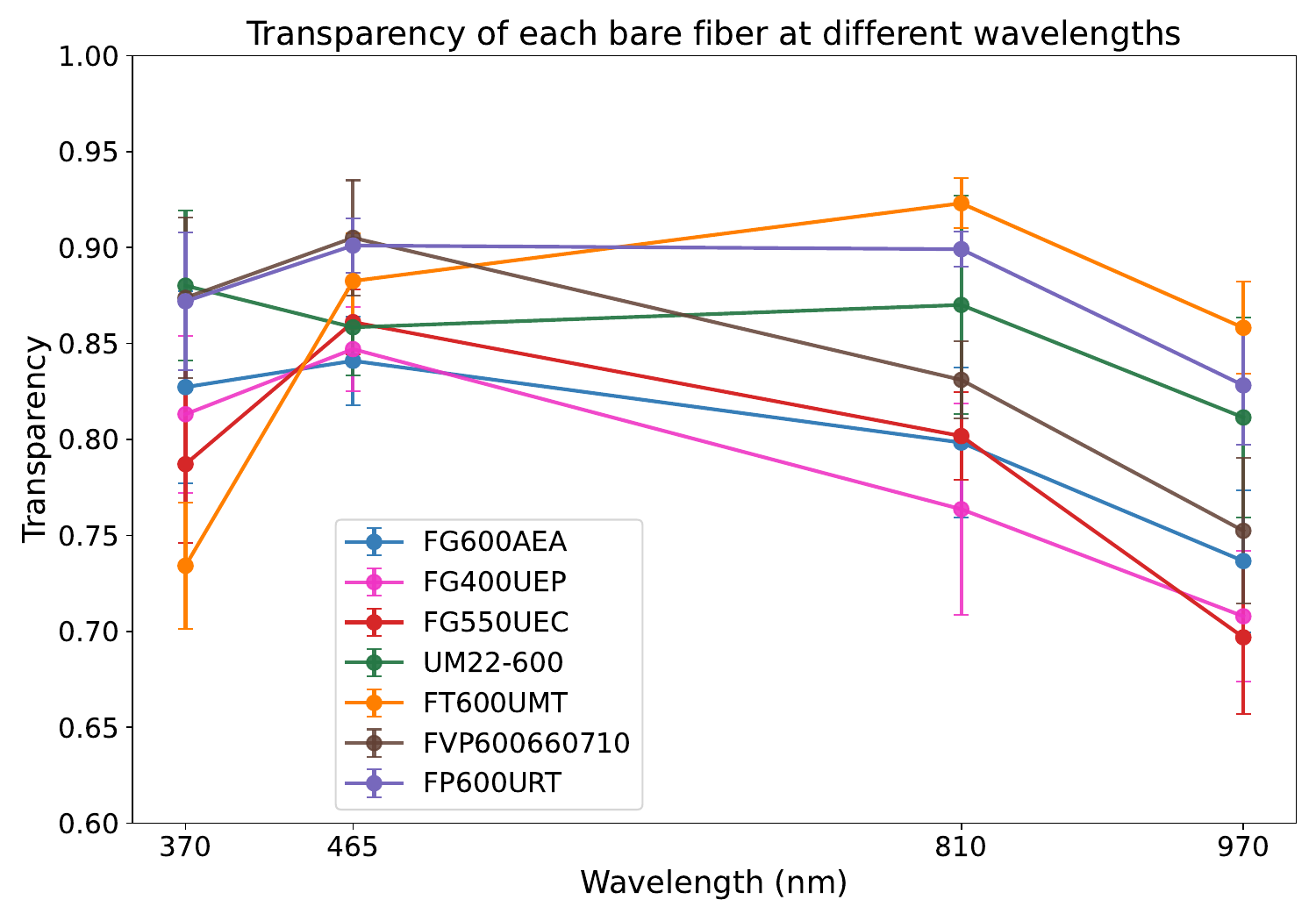}
    \caption{Left: Transparency at four wavelengths (370~nm, 465~nm, 810~nm, and 970~nm) for each bare fiber type. Right: Transparency of each bare fiber type as a function of wavelength. Error bars represent the statistical uncertainty associated with each measurement.}
    \label{fig:baretrans}
\end{figure}

\begin{figure}[htpb]
    \centering
    \includegraphics[scale=0.295]{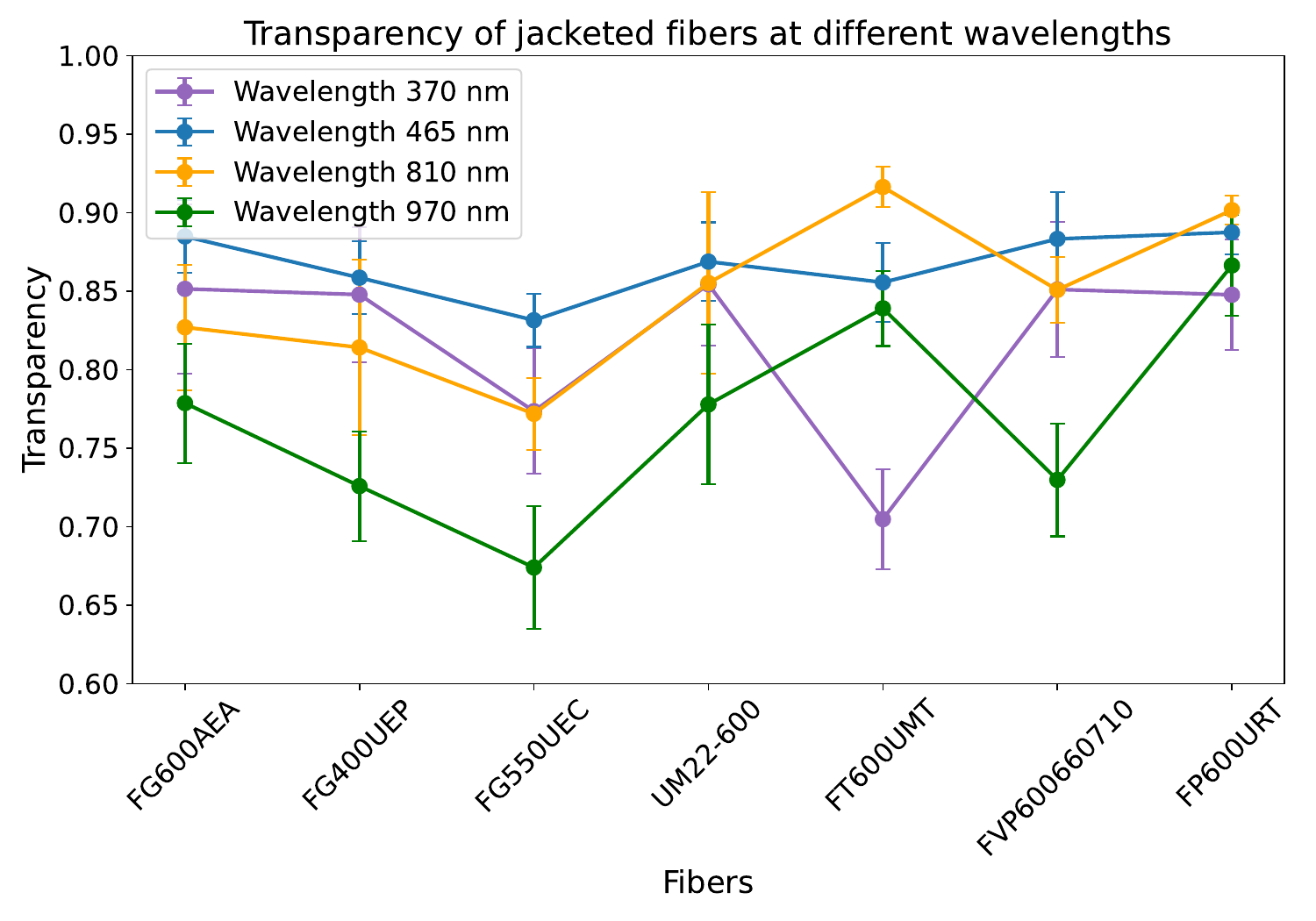}
    \includegraphics[scale=0.295]{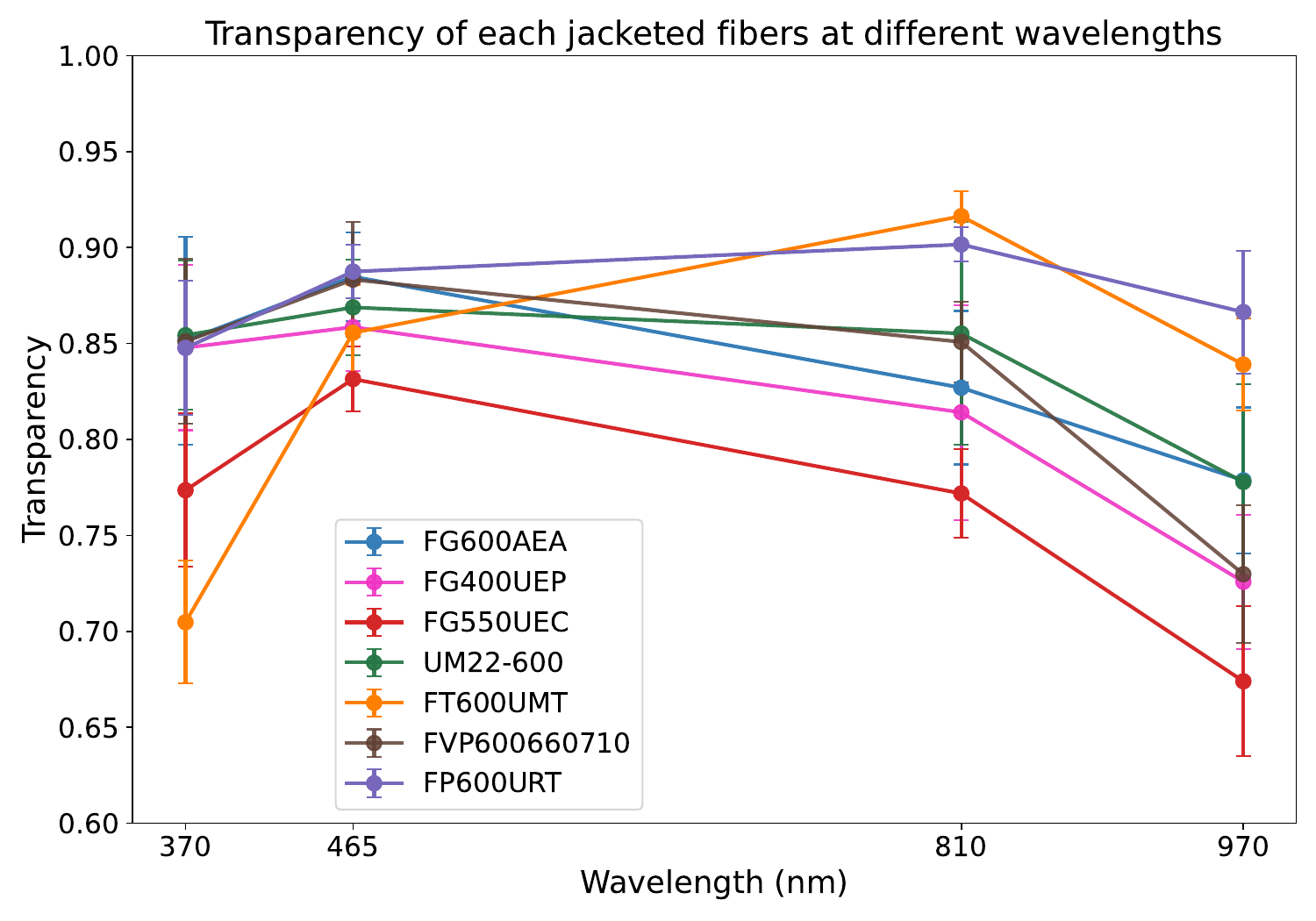}
    \caption{Left: Transparency at four wavelengths (370~nm, 465~nm, 810~nm, and 970~nm) for each jacketed fiber type. Right: Transparency of each jacketed fiber type as a function of wavelength. Error bars represent the statistical uncertainty associated with each measurement.}
    \label{fig:jackettrans}
\end{figure}

Tables~\ref{tab:transparency} and~\ref{tab:transparency_jacket} summarize the transparency measurements and their statistical errors obtained for the different types of bare and jacketed fibers, respectively, at the four tested LED wavelengths. Figures~\ref{fig:baretrans} and~\ref{fig:jackettrans} present the transparency results of each fiber type at 370 nm, 465 nm, 810 nm and 970 nm. The measurements show that the transparency generally decreases at the shortest and longest wavelengths, particularly around \qty{370}{nm}, where absorption in the UV range becomes more significant. This trend is notable for fibers such as FG550UEC and FG400UEP, which exhibit stronger wavelength dependence. In contrast, fibers like FP600URT and FT600UMT show relatively uniform transmission across the measured wavelength spectrum, indicating their potential suitability for broadband applications.

The comparison between bare and jacketed fibers shows no significant difference in transparency across the tested wavelengths, with most variations falling within the statistical uncertainties. This suggests that the presence of a protective jacket does not substantially affect the optical transmission under the controlled measurement conditions used here. The effect of coating material, however, is more evident: polyimide-coated fibers such as FG400UEP and UM22-600 consistently demonstrate high transparency in the UV and visible regions, reinforcing their suitability for cryogenic environments where both optical performance and material resilience are essential. Additionally, all tested fibers exhibit good transmission in the IR (\qty{810}{nm} and \qty{970}{nm}), indicating their general effectiveness across a broad spectral range.

Of particular relevance, FP600URT and FVP600660710 exhibit strong UV transmission (above $87\%$ at \qty{370}{nm}), while the FVP600660710 transmits efficiently down to \qty{270}{nm}. FVP600660710 was deployed in ProtoDUNE-SP~\cite{Abi2020} while both FP600URT and FVP600660710 were successfully tested in a recent run of the ProtoDUNE-HD UV-light calibration system \cite{NP02:2025AnnualReport}. Their reliable performance under cryogenic conditions and across the tested wavelengths makes them promising candidates for deployment in particle detectors operating in cryogenic environments.


\section{Optical Feedthrough and SMA-to-SMA Connector Interface Loss}
\label{sec:OFTSMA}

\begin{figure}[!htbp]
\centering
\includegraphics[height=4.0cm,width=0.7\textwidth]{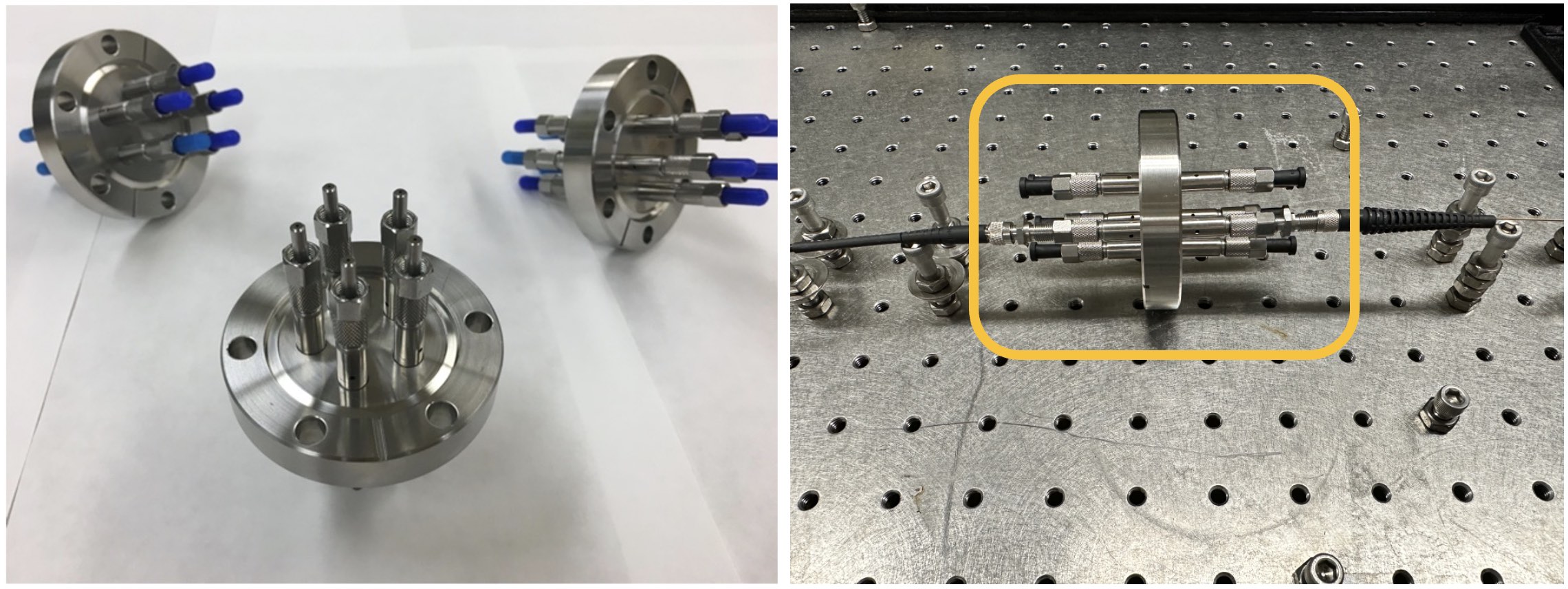}
\caption{Left: Photograph of three optical feedthroughs. Right: Photograph of a feedthrough installed on the optical table during testing.}
\label{fig:test_stand_2}
\end{figure}

Optical fiber feedthroughs are used to separate the warm and cold sides of the LArTPC, while providing light transport through the cryostat boundary. Such an optical feedthrough, used in~\cite{Abi2020, Soto-Oton:2024apm}, is presented in Figure~\ref{fig:test_stand_2}. 

The optical feedthrough used in this study to interface a reference and tested fiber has a short internal fused-silica fiber segment of approximately \qty{3.5}{inches}. The fiber segment is terminated with SMA-to-SMA connectors on both ends. Therefore, each feedthrough channel has two optical connectors.
Across the wavelength range studied (\qtyrange{275}{970}{nm}), the attenuation of this short segment is below $0.5\%$, and therefore negligible compared with losses arising from the SMA-to-SMA connector interfaces. The measurements presented in the following subsections focus on quantifying the losses introduced by the SMA-to-SMA connection on each side of the feedthrough.

Sections~\ref{subsec:name2} and ~\ref{subsec:name1} describe two complementary methods used to extract the loss associated with these interfaces. The first method is based on optical-power measurements using a calibrated power meter, while the second method relies on photodiode-response measurements performed on a dedicated test stand. In both cases, the feedthrough is inserted into a reference optical chain such that the chain contains two SMA-to-SMA connectors (one on each side of the feedthrough). 

In particular, the light propagation losses through fibers, SMA-to-SMA connectors, and through the optical cryostat feedthrough at wavelengths of \qty{275}{nm} and \qty{367}{nm} are of special interest due to deployment in recent LArTPC detectors~\cite{Abi2020, Soto-Oton:2024apm}.

\subsection{SMA-to-SMA Connector Attenuation Using Optical Power Measurements}~\label{subsec:name2}
\begin{figure}[ht]
    \centering
\includegraphics[width=0.7\linewidth]{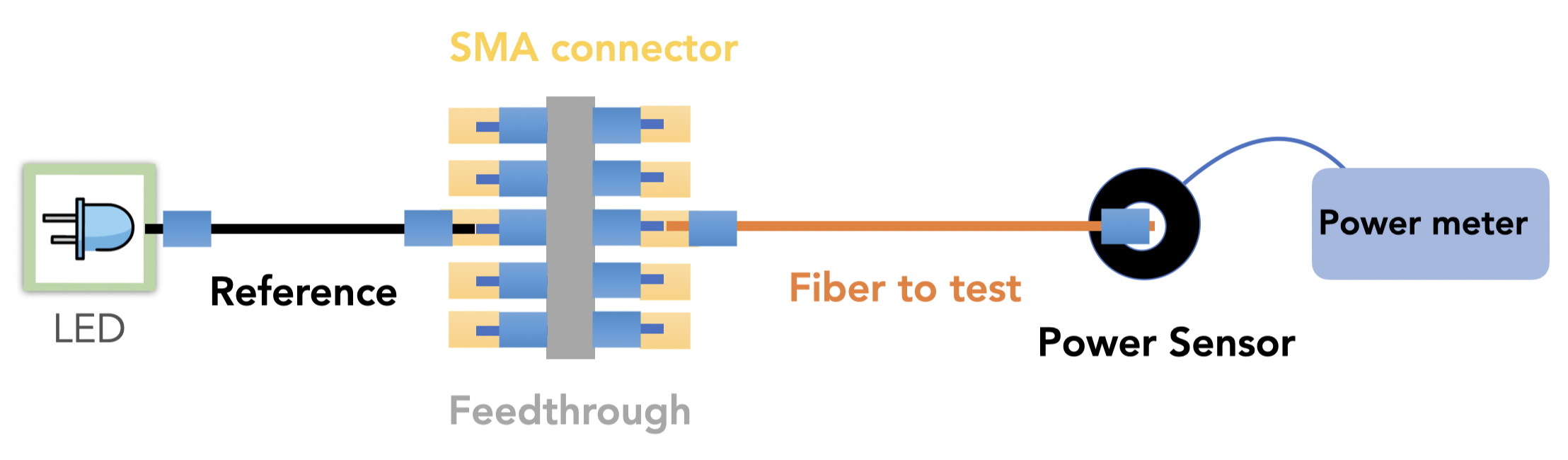}
    \caption{Schematic of the optical power measurement setup through a feedthrough. Light emitted from an LED is coupled into a reference fiber, routed through an SMA-to-SMA connector coupled with feedthrough, and transmitted via the test fiber to a calibrated power sensor. The detected optical power is recorded using a power meter to evaluate transmission efficiency and loss characteristics of the test fiber.}
    \label{fig:feedthrough_setup}
\end{figure}

To characterize multimode fiber performance, systematic optical measurements were conducted at \qty{370}{nm}, \qty{465}{nm}, \qty{810}{nm}, and \qty{970}{nm} wavelengths. The experimental setup (see Figure ~\ref{fig:feedthrough_setup}) employed LED sources coupled to fibers, with transmitted power recorded by a calibrated power meter. For comprehensive loss characterization, measurements included both direct test fiber outputs and complete optical chains incorporating SMA-to-SMA connectors and a feedthrough assembly.

\begin{figure}[htpb]
    \centering
    \includegraphics[scale=0.296]{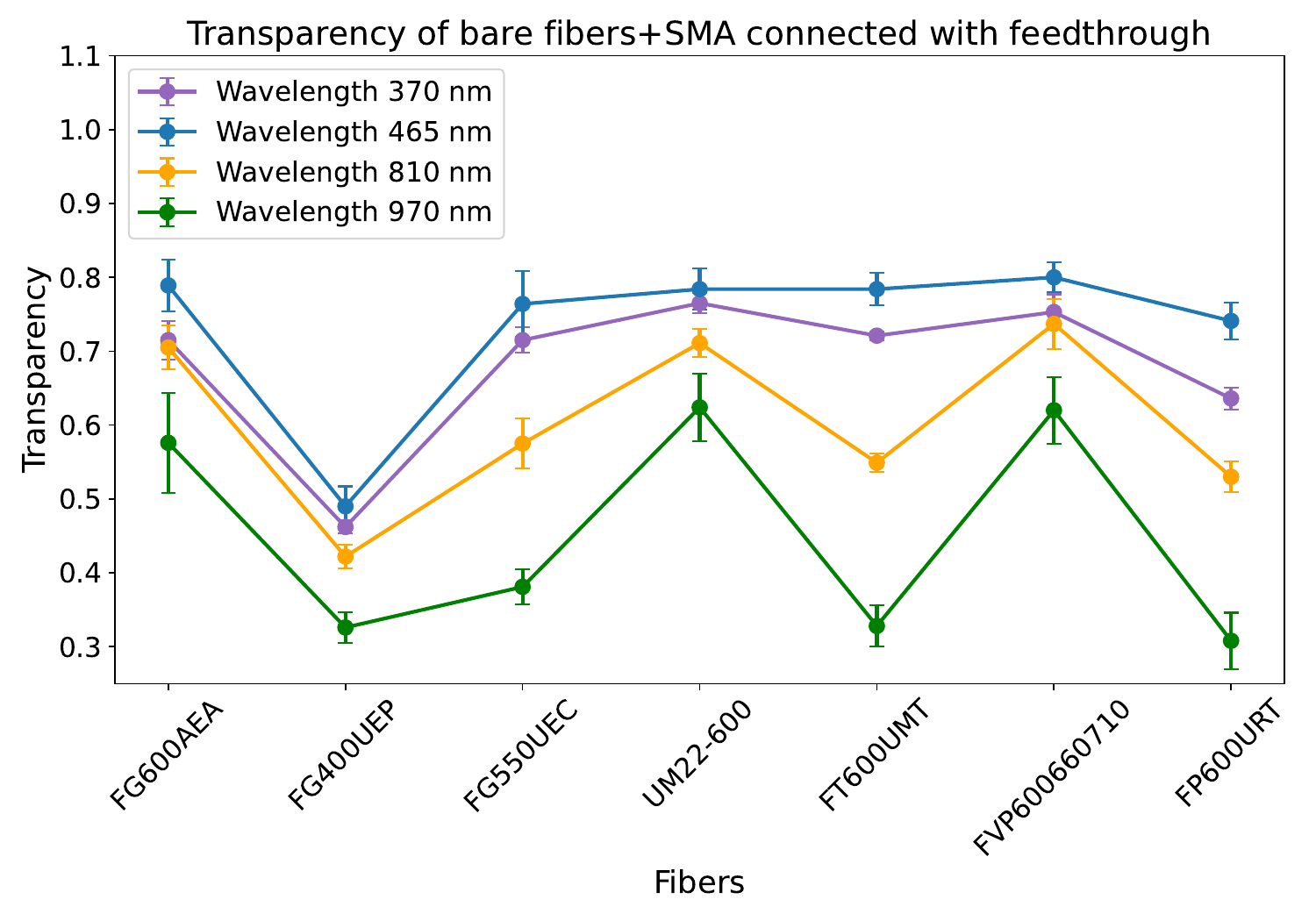}
    \includegraphics[scale=0.296]{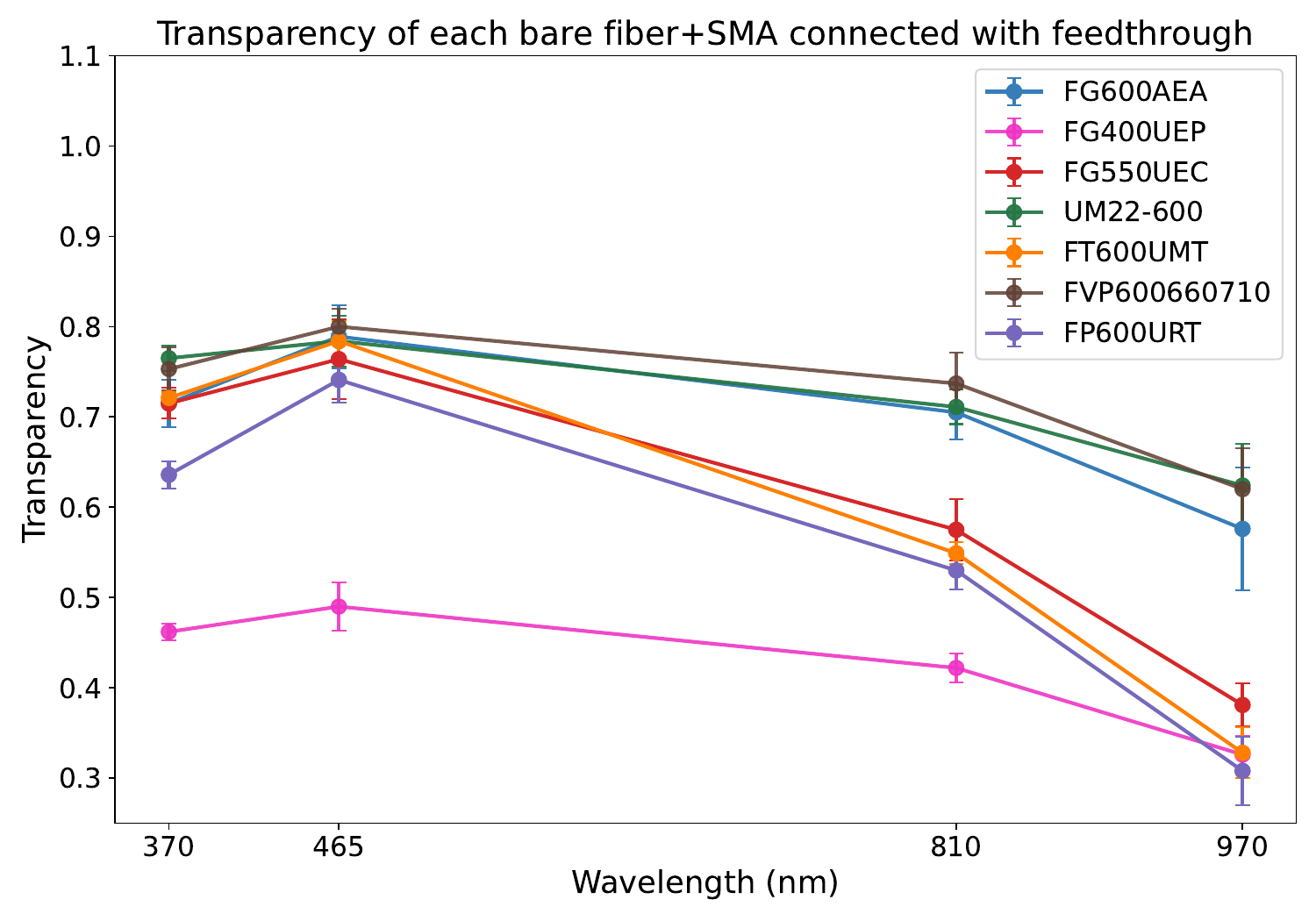}
    \caption{Left: Transparency at four wavelengths (370~nm, 465~nm, 810~nm, and 970~nm) for each bare fiber type connected with feedthrough. Right: Transparency of each bare fiber connected with feedthrough as a function of wavelength. Error bars represent the statistical uncertainty associated with each measurement.}
    \label{fig:barefeed}
\end{figure}

The transparency results listed in Table~\ref{tab:powerloss} and represented in Figure~\ref{fig:barefeed}, were calculated as follows:
\begin{equation}\label{eq41}
T_{\text{full}} = \frac{P_{\text{full}}}{P_{\text{reference}}},
\end{equation}

where $P_{\text{full}}$ represent the power loss through the reference fiber, two SMA-to-SMA connectors (each in one end of the optical feedthrough), optical fiber feedthrough, and the tested fiber (see Figure \ref{fig:feedthrough_setup}).

To isolate the contribution from the two SMA-to-SMA connectors connected to the optical feedthrough ends, the transparency of the full optical chain (Table \ref{tab:powerloss}) is compared with the reference fiber plus SMA-to-SMA connector plus tested fiber transparency (Table \ref{table2}). The SMA-to-SMA connector loss is therefore defined as the difference between Eq. \ref{eq31} and Eq. \ref{eq41} as:

\begin{equation}
\begin{aligned}
    \text{SMA-to-SMA (Loss)} &= T_{\text{}} - T_{\text{full}}\\  
\end{aligned}
\label{eq:SMA_loss}
\end{equation}

This definition makes it possible to estimate the transparency loss in the SMA-to-SMA connector for each wavelength using the transparency results reported in Section ~\ref{sec:OFQC}.

This subtraction cancels the loss contributions from the reference fiber, the tested fiber, and the SMA-to-SMA connector, isolating the loss of a single SMA-to-SMA connector, under the assumption that the power losses related to the feedthrough are negligible for all tests conducted. Table~\ref{tab:powerloss_dB} shows the transparency results obtained for different fiber types and wavelengths for a single SMA-to-SMA connector.

\begin{table}[htpb]
\centering
\caption{Measured transparencies of the full optical chain for all tested fibers and wavelengths. These values include the contribution from the reference fiber, two SMA-to-SMA connectors (both ends of feedthrough), the optical feedthrough, and the tested fiber, including their respective errors.}
\label{tab:powerloss}

\begin{adjustbox}{width=\columnwidth,center}
\begin{tabular}{l|c|c|c|c}
\toprule
\multicolumn{5}{c}{\textbf{Transparency results for bare fibers+SMA-to-SMA connectors with Feedthrough}} \\
\midrule
\textbf{Fiber} & \textbf{370~nm} $(T \pm \Delta T)$ & \textbf{465~nm} $(T \pm \Delta T)$ & \textbf{810~nm} $(T \pm \Delta T)$ & \textbf{970~nm} $(T \pm \Delta T)$ \\
\midrule
FG600AEA     & $0.715 \pm 0.026$ & $0.789 \pm 0.035$ & $0.705 \pm 0.030$ & $0.576 \pm 0.068$ \\
FG400UEP     & $0.462 \pm 0.009$ & $0.490 \pm 0.027$ & $0.422 \pm 0.016$ & $0.326 \pm 0.021$ \\
FG550UEC     & $0.715 \pm 0.017$ & $0.764 \pm 0.044$ & $0.575 \pm 0.034$ & $0.381 \pm 0.024$ \\
UM22-600     & $0.765 \pm 0.014$ & $0.784 \pm 0.028$ & $0.711 \pm 0.019$ & $0.624 \pm 0.046$ \\
FT600UMT     & $0.721 \pm 0.006$ & $0.784 \pm 0.022$ & $0.549 \pm 0.012$ & $0.328 \pm 0.028$ \\
FVP600660710 & $0.753 \pm 0.024$ & $0.800 \pm 0.020$ & $0.737 \pm 0.034$ & $0.620 \pm 0.045$ \\
FP600URT     & $0.636 \pm 0.015$ & $0.741 \pm 0.025$ & $0.530 \pm 0.021$ & $0.308 \pm 0.038$ \\
\bottomrule
\end{tabular}
\end{adjustbox}
\end{table}

\begin{table}[htpb]
\centering
\caption{Single SMA-to-SMA connector power losses derived from the difference between the full optical-chain transparency and the bare-fiber transparency. The results quantify the attenuation introduced by a single SMA-to-SMA connector at each wavelength, and each value is reported with its corresponding error.}
\label{tab:powerloss_dB}
\begin{adjustbox}{max width=\columnwidth,center}
\begin{tabular}{l|c|c|c|c}
\toprule
\multicolumn{5}{c}{\textbf{Fractional power loss for a single SMA-to-SMA connector at all tested wavelengths}} \\
\midrule
\textbf{Fiber} & \textbf{370~nm} & \textbf{465~nm} & \textbf{810~nm} & \textbf{970~nm} \\
\midrule
FG600AEA     & $0.112 \pm 0.056$ & $0.052 \pm 0.042$ & $0.093 \pm 0.049$ & $0.161 \pm 0.077$ \\
FG400UEP     & $0.352 \pm 0.042$ & $0.357 \pm 0.035$ & $0.342 \pm 0.057$ & $0.382 \pm 0.040$ \\
FG550UEC     & $0.072 \pm 0.044$ & $0.097 \pm 0.047$ & $0.227 \pm 0.041$ & $0.316 \pm 0.047$ \\
UM22-600     & $0.115 \pm 0.041$ & $0.074 \pm 0.038$ & $0.159 \pm 0.060$ & $0.187 \pm 0.069$ \\
FT600UMT     & $0.013 \pm 0.034$ & $0.099 \pm 0.033$ & $0.374 \pm 0.018$ & $0.530 \pm 0.037$ \\
FVP600660710 & $0.121 \pm 0.048$ & $0.105 \pm 0.036$ & $0.094 \pm 0.039$ & $0.132 \pm 0.059$ \\
FP600URT     & $0.237 \pm 0.039$ & $0.160 \pm 0.029$ & $0.369 \pm 0.023$ & $0.520 \pm 0.049$ \\
\bottomrule
\end{tabular}
\end{adjustbox}
\end{table}

\begin{figure}[htpb]
    \centering
    \includegraphics[scale=0.296]{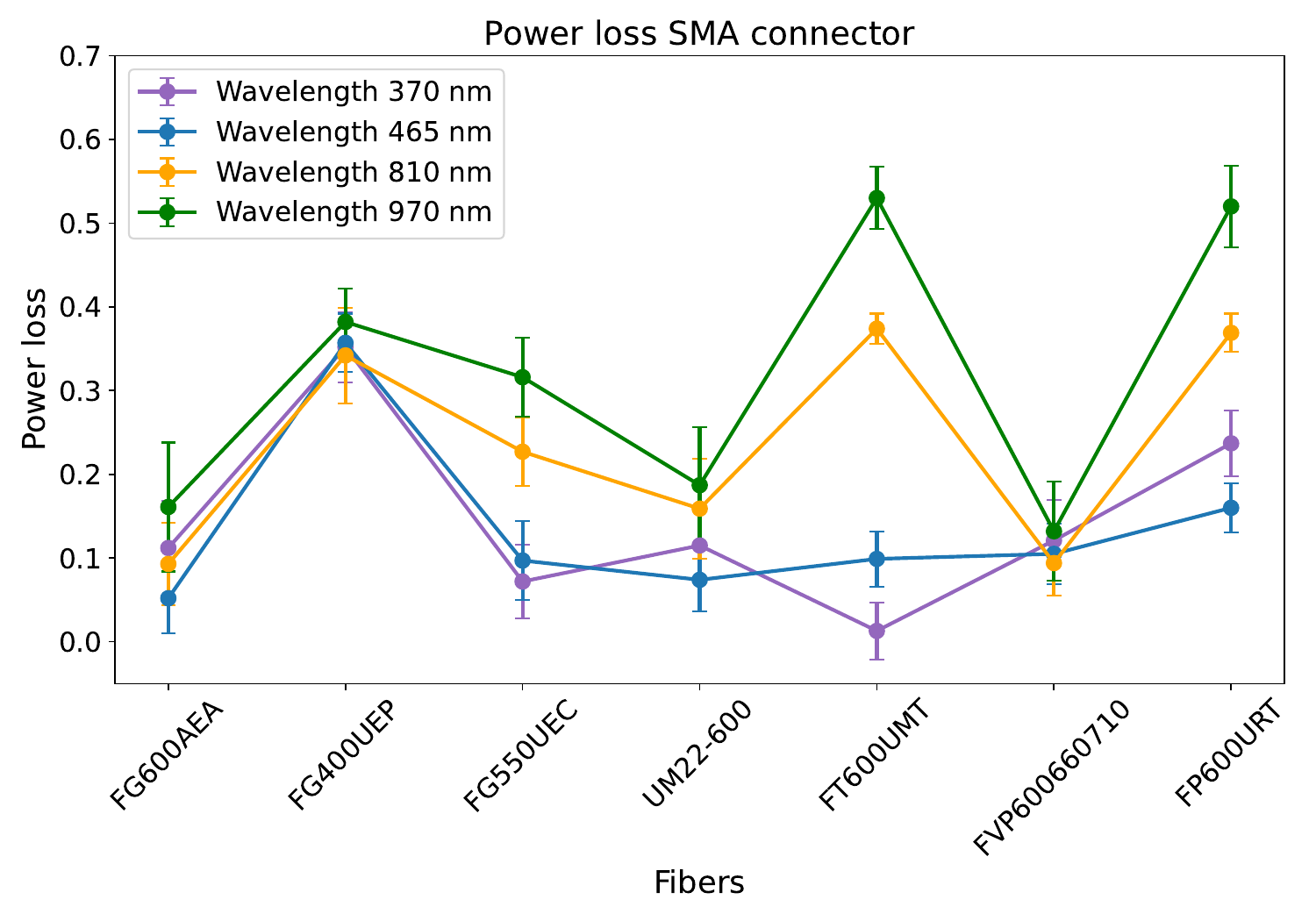}
    \includegraphics[scale=0.296]{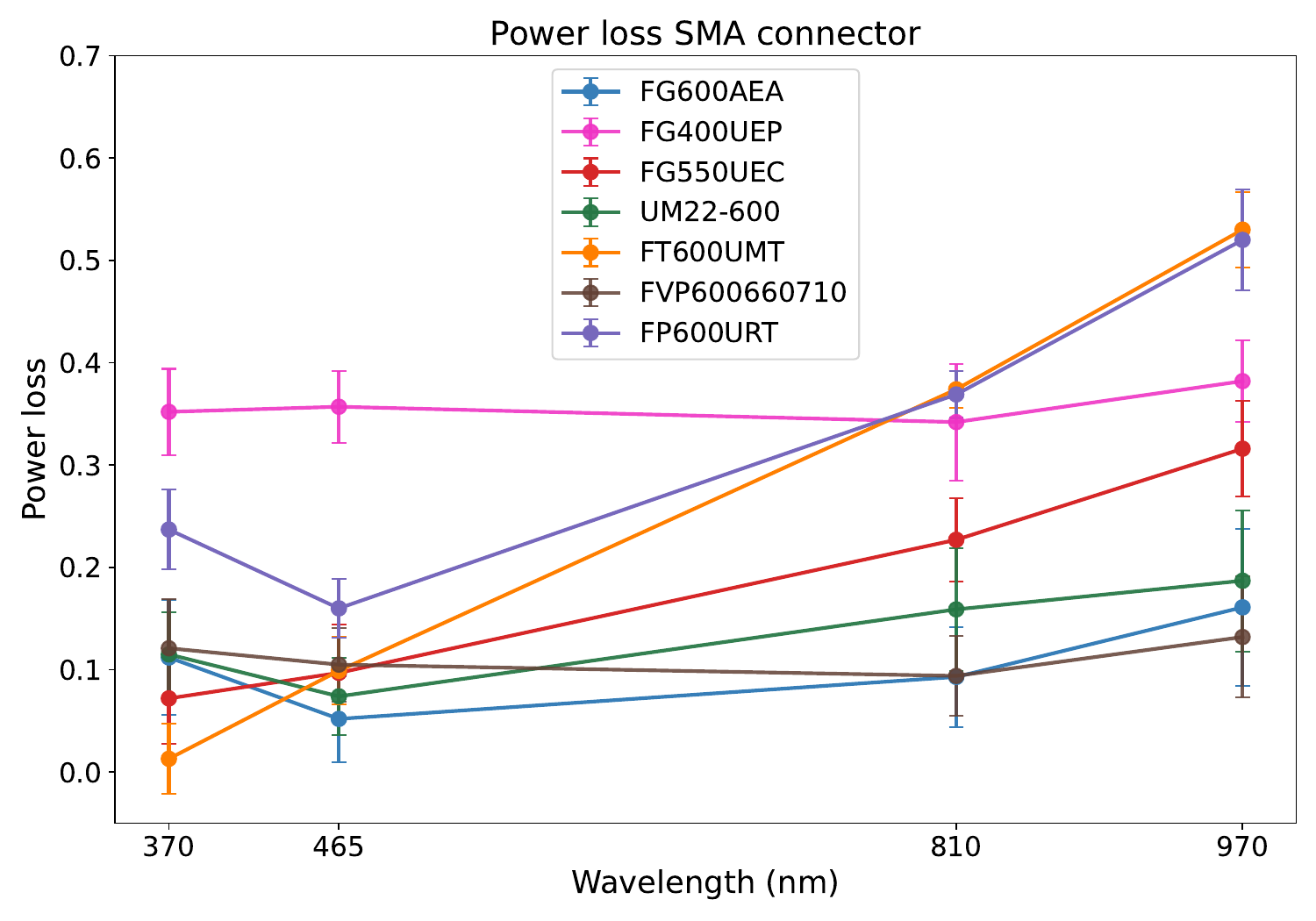}
    \caption{Left: Power loss of a single SMA-to-SMA connector at four wavelengths (370~nm, 465~nm, 810~nm, and 970~nm) for each bare fiber type. Right: Power loss of a single SMA-to-SMA connector connected to each bare fiber as a function of wavelength. Error bars represent the statistical uncertainty associated with each value.}
    \label{fig:SMAloss}
\end{figure}

The Figure~\ref{fig:SMAloss} summarizes the wavelength-dependent transmission characteristics of seven multi-mode fibers. The left panel illustrates power loss variations across fiber types at each wavelength, revealing that FG400UEP consistently exhibits the highest attenuation, while FVP600660710 demonstrates superior transmission stability. The right panel traces across the spectral range for individual fibers, clearly showing the progressive increase in attenuation toward longer wavelengths, particularly pronounced in FT600UMT and FP600URT fibers. Results indicate that \qty{970}{nm} experiences the most significant power dissipation across all fiber types, with losses approximately 2–3 times higher than FVP600660710, for example, when compared at \qty{370}{nm}. Such an attenuation effect could be explained with the higher numerical aperture value compared to other fiber types. 

\subsection{SMA-to-SMA Connector Attenuation Using Photodiode (PD) Response Measurements}~\label{subsec:name1}

\begin{figure}[!htbp]
\centering
\includegraphics[height=6.0cm,width=0.65\textwidth]{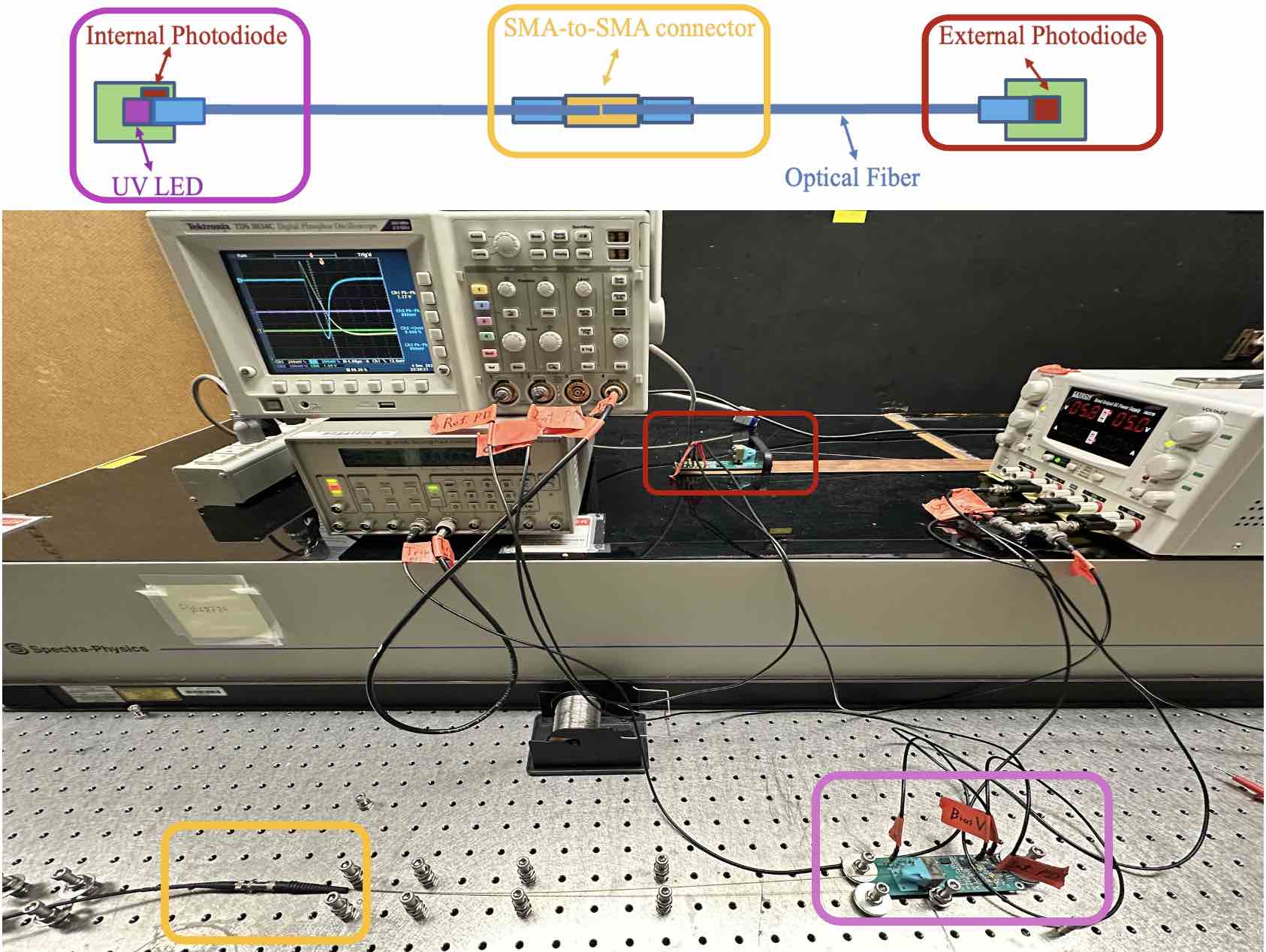}
\caption{Top: Schematic of the experimental setup used to test light sources (LEDs, photodiodes), optical fibers and SMA-to-SMA connectors, and optical feedthroughs. Bottom: Photograph of the test stand showing the light sources, photodiodes (PD), optical fibers, connectors, and feedthroughs arranged on the optical table. }
\label{fig:test_stand_1}
\end{figure}

The measurements described in this section are performed on the test stands presented in Figure~\ref{fig:test_stand_1} and Figure~\ref{fig:test_stand_2}.
The test stand is used to test light sources, optical fibers and SMA-to-SMA connectors, and optical fiber feedthroughs.

\begin{figure}[!htbp]
\centering
\includegraphics[height=4.0cm,width=0.8\textwidth]{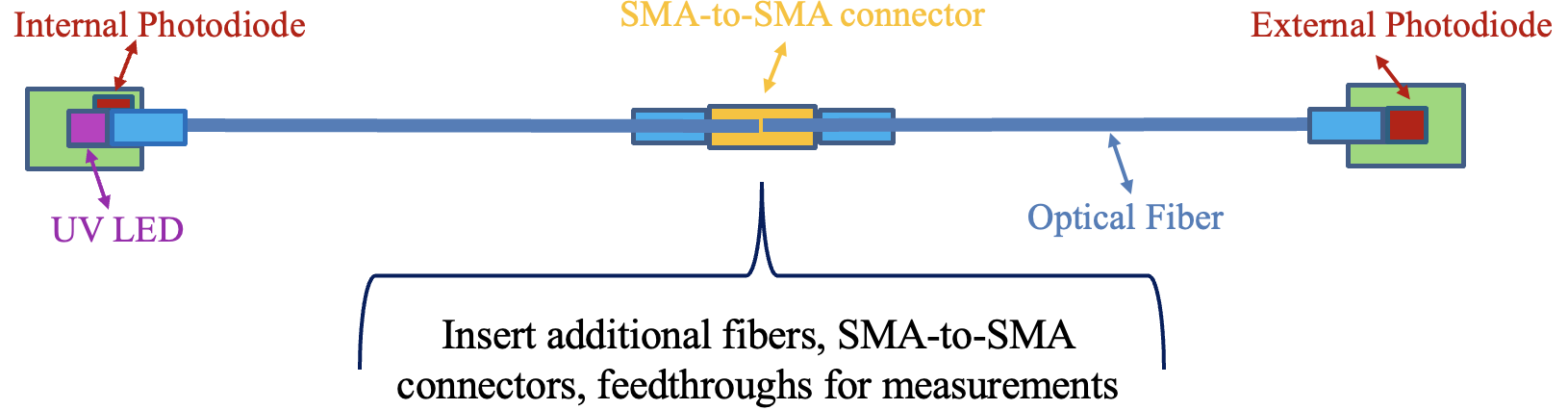} 
\caption{Sketch of the experimental design for the tests of light sources (LEDs, photodiodes), optical fibers and connectors, and optical feedthroughs.}
\label{fig:ref_sketch}
\end{figure}

This section describes the measurement of the light loss through an optical fiber feedthrough. The feedthrough shown in Figure~\ref{fig:test_stand_2} (right) is inserted into the reference set-up sketched in Figure~\ref{fig:ref_sketch}, so it essentially adds another optical connection with an SMA-to-SMA connector.  The optical fiber length through the feedthrough is \qty{3.5}{inches}. This measurement allows the determination of the optical loss driven by the single SMA-to-SMA added connector, with respect to the baseline sketched in Figure~\ref{fig:ref_sketch}.

Light loss measurements were performed using LEDs of \qty{275}{nm} and \qty{367}{nm} respectively; in both cases the light throughput is measured with respect to the reference measurement. Figure~\ref{fig:test_stand_2} (right) shows one of the optical feedthroughs used in the measurements. Measurements were performed using two 5-channel optical fiber feedthroughs, called ``A'' and ``B''. The results are summarized in Table \ref{tab:OFTgeneraltable}.

\begin{table}[!htp]
\centering
\caption{Optical feedthrough (OFT) light attenuation measurements with 275 nm and 367 nm light.}
\label{tab:OFTgeneraltable}
\begin{tabular}{|c|c|c|c|c|c|c|} 
\hline
 & \multicolumn{3}{|c|}{\textbf{275 nm}} & \multicolumn{3}{|c|}{\textbf{367 nm}} \\ \hline \hline
 &  &  Internal & External &  &  Internal & External   \\
 & \# &  PD [mV] & PD [mV] & \# &  PD [V] & PD [mV]  \\ \hline \hline
Baseline & 1 & 776--792 & 254--260 & \multirow{2}{*}{1} & \multirow{2}{*}{ 1.48--1.49} & \multirow{2}{*}{712--724} \\ \cline{2-4}
measurements & 2 & 776--792 & 240--246 & & & \\ \hline \hline
& 1 & 780--792 &  206--215 & 1  & 1.49 & 684--692 \\ \cline{2-7}
Feedthrough A & 2  & 780--788 & 202--208 & 2  & 1.49 & 564--572 \\ \cline{2-7}
(channel 1-5) & 3  & 776--796 & 228--236 & 3  & 1.49 & 632--644 \\ \cline{2-7}
& 4  & 780--792 & 214--218 & 4  & 1.49 & 640--652 \\ \cline{2-7}
& 5  & 776--788 & 214--222 & 5  & 1.49 & 584--592 \\ \hline \hline
 & 1 & 776--792 & 206--214 & 1 & 1.49 & 616--624 \\ \cline{2-7}
Feedthrough B & 2 & 780--788 & 206--216 & 2 & 1.49 & 656--664 \\ \cline{2-7}
(channel 1-5) & 3 & 776--788 & 198--204 & 3 & 1.49 & 620--628 \\ \cline{2-7}
& 4 & 780--792 & 216--226 & 4 & 1.49 & 640--652 \\ \cline{2-7}
& 5 & 780--792 & 202--210 & 5 & 1.49 & 600--608 \\ \hline
\end{tabular}
\end{table}
Table~\ref{tab:OFTsummarytable} presents the summary of the measurements listed in Table~\ref{tab:OFTgeneraltable}. The Light loss through the single SMA-to-SMA connector is determined to be $(15.2 \pm 5.1)\%$ and $(12.4 \pm 5.0)\%$, for \qty{275}{nm} and \qty{367}{nm} light, respectively. In addition, measurements for the light loss in a single SMA-to-SMA connector were presented in section \ref{subsec:name2} and Table \ref{tab:powerloss_dB}. It can be noted that for the fiber FVPFVP600660710 at 370 nm the light loss for a single SMA-to-SMA connector was found to be $(12.1 \pm 4.8)\%$.

\begin{table}[!htp]
\centering
\caption{Calculation of the total light loss for a single SMA-to-SMA connector using two different wavelengths.}
\label{tab:OFTsummarytable}
\begin{tabular}{|c|c|c|} 
\hline
Wavelength & Measurements &  Total value     \\
 \hline
\multirow{4}{*}{275 nm} & Baseline ($I_\text{o}$) & 250 $\pm$ 10.3 mV \\ 
 & Feedthrough A \& B ($I_\text{OFT}$) &  212 $\pm$ 9.4 mV \\  
 & Ratio ($I_\text{OFT}/I_\text{o}$) & 0.848 $\pm$ 0.051 \\ 
 & Total light loss ($1 - I_\text{OFT}/I_\text{o}$) & 0.152 $\pm$ 0.051 \\ \hline \hline
\multirow{4}{*}{367 nm} & Baseline ($I_\text{o}$) & 717 $\pm$ 5.0 mV\\ 
 & Feedthrough A \& B ($I_\text{OFT}$) & 628.3 $\pm$ 35.6 mV \\ 
 & Ratio ($I_\text{OFT}/I_\text{o}$) & 0.876 $\pm$ 0.05 \\ 
 & Total light loss ($1 - I_\text{OFT}/I_\text{o}$) & 0.124 $\pm$ 0.050 \\ \hline
\end{tabular}
\end{table}

\begin{figure}[!htbp]
\centering
\includegraphics[height=5.0cm,width=0.9\textwidth] 
{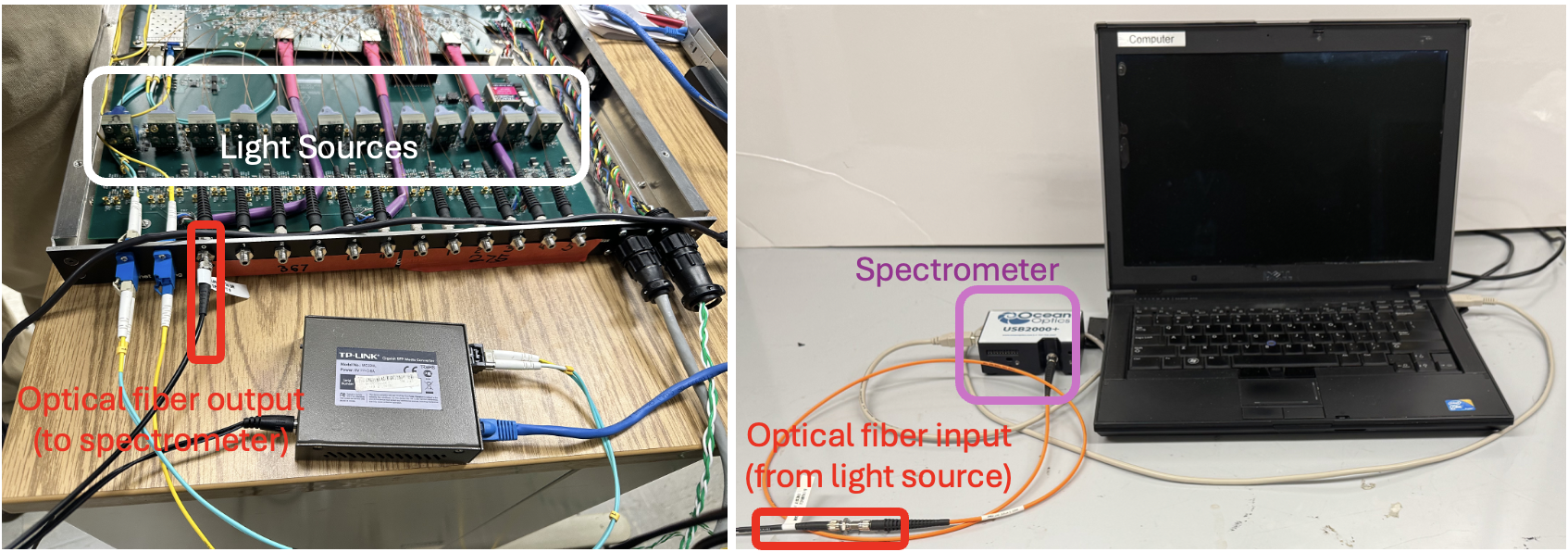}\\
\caption{The spectrometer setup at ANL used to measure and confirm the LED wavelengths. 
}
\label{fig:spectrometer_photo}
\end{figure}

\begin{figure}[!htbp]
\centering
\includegraphics[width=0.44\textwidth]{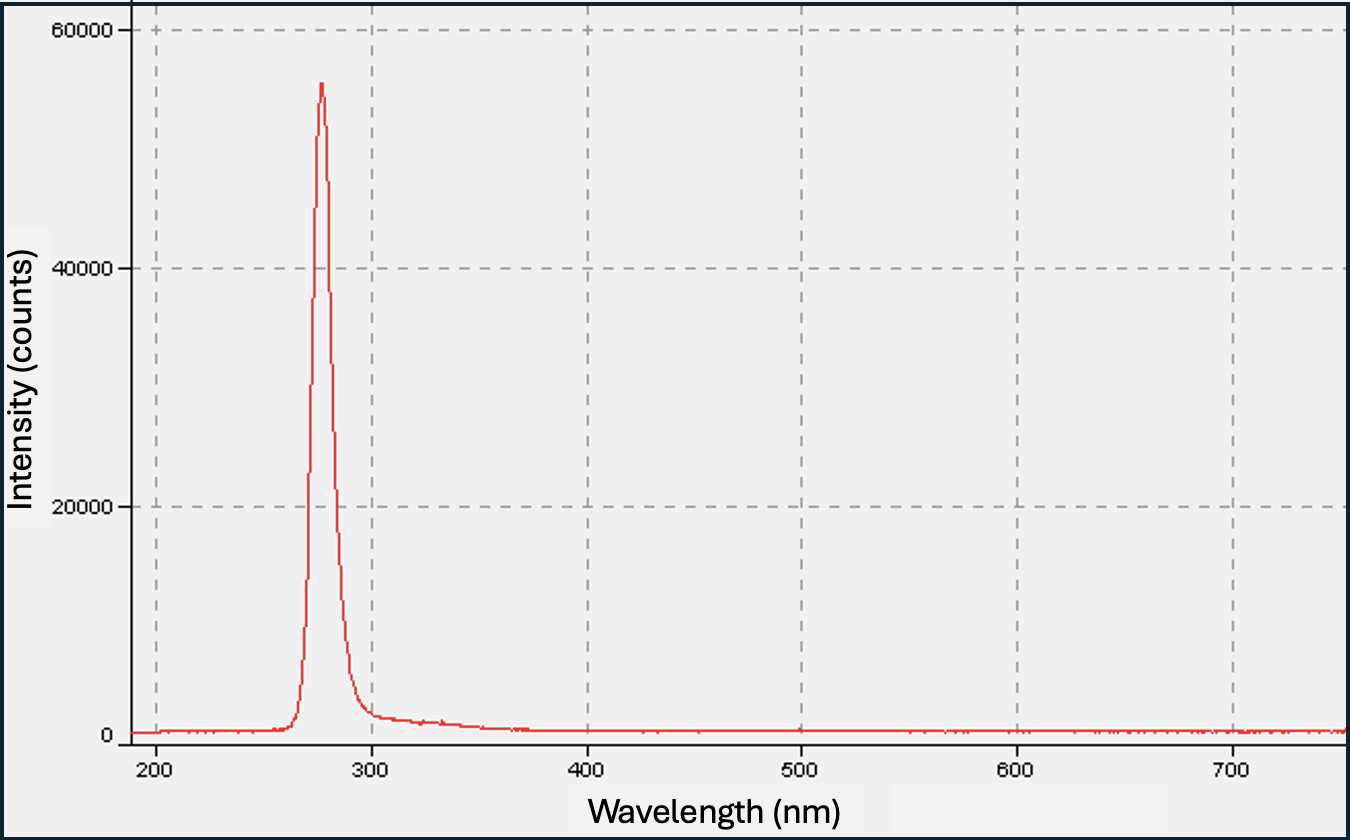} \hspace{3mm}
\includegraphics[width=0.44\textwidth]{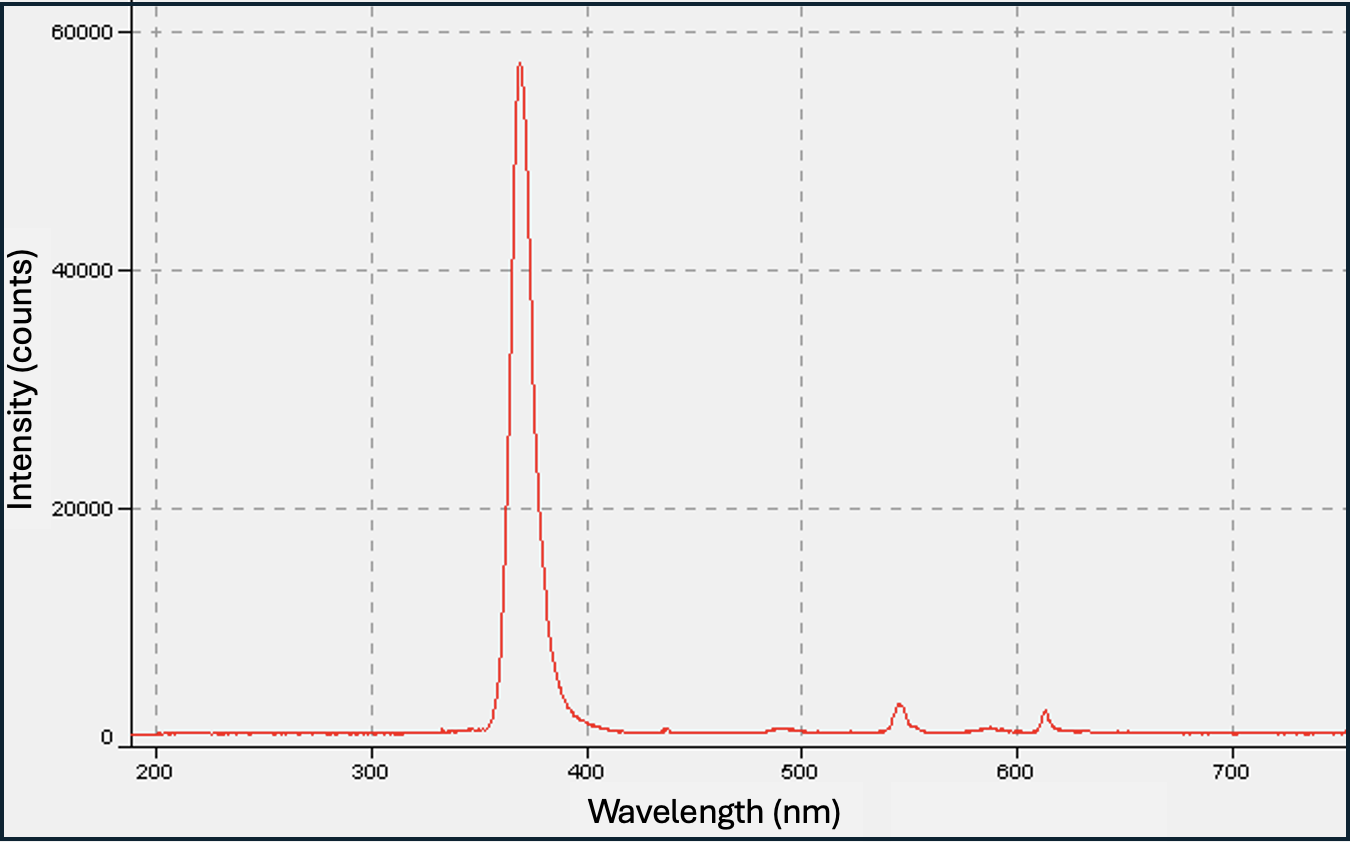}
\caption{
Measured wavelengths of tested 275 nm and 367 nm LEDs.}
\label{fig:spectrometer_photo_2}
\end{figure}

The wavelengths of the two LED light sources used in the test stand, shown in Figure \ref{fig:test_stand_1}, were independently tested using an Ocean Optics USB2000+ spectrometer. The setup is shown in Figure~\ref{fig:spectrometer_photo}  in which the spectrometer software is displayed on the laptop screen and the orange and blue cable is the output of the spectrometer unit.  A fiber from the LED is the input to the spectrometer unit coming from the light calibration module. Figure~\ref{fig:spectrometer_photo_2} shows the output of the spectrometer, verifying the LED wavelengths of \qty{275}{nm} and \qty{367}{nm}.

\section{Measurement of the Light Loss through the Long Fibers}
\label{sec:LongFiber}

\begin{figure}[!htbp]
    \centering
    \includegraphics[height=2.4cm,width=0.57\textwidth]{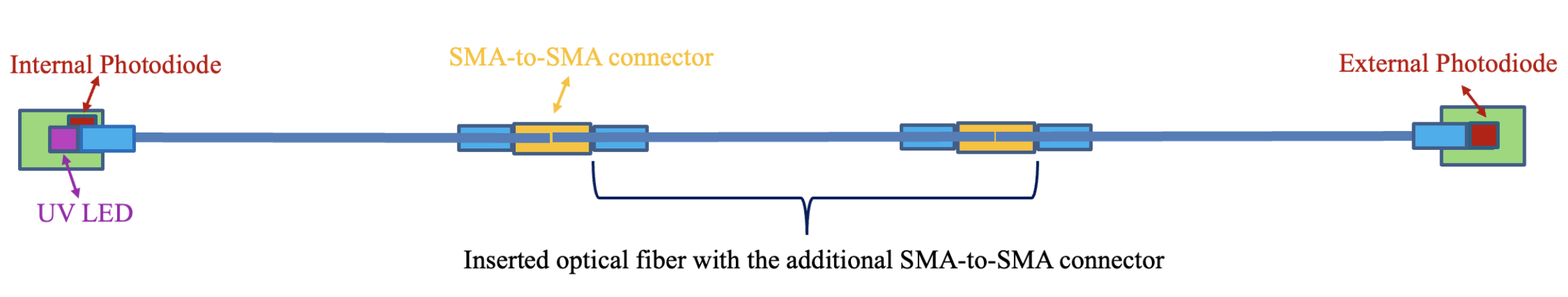} \hspace{1cm} \includegraphics[height=2.0cm,width=0.35\textwidth]{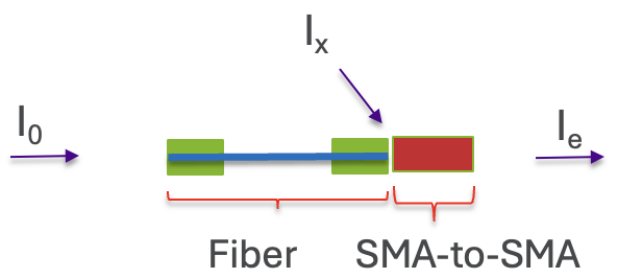} 
    \caption{Left: Sketch of the experimental design for the single fiber attenuation tests. The tested fiber is inserted with one additional SMA-to-SMA connector. Note that the fiber has been added together with the additional SMA-to-SMA connector. Right: Sketch with the definition of light propagation losses.}
    \label{fig:1fiber_1sma2sma}
\end{figure}

Longer fibers will experience larger attenuation effects above the measurement uncertainties, when compared to the short 1 m fibers used above, and will help provide more accurate measurements of the fiber attenuation length.
In this section, we focus on the measurements of two fiber types, FVP600660710 and FP600URT (presented in Table~\ref{tab:FiberstoTest}). The FVP600660710 was tested in earlier ProtoDUNE SP~\cite{Abi2020} while both FP600URT and FVP600660\hspace{0pt}710 were deployed in recent ProtoDUNE HD~\cite{Soto-Oton:2024apm}. Fiber FVP600660710 is expected to transmit both \qty{275}{nm} and \qty{367}{nm} light efficiently, while the FP600URT is rated for wavelength above \qty{300}{nm}. 
The situation discussed in this section is presented in Figure~\ref{fig:1fiber_1sma2sma}, while the baseline measurement was performed according to the sketch in Figure~\ref{fig:ref_sketch}. Therefore, this measurement is performed with tested fiber with one additional SMA-to-SMA connector inserted to the baseline configuration. The SMA-to-SMA loss has been measured in Section~\ref{sec:OFTSMA}. Therefore, the measurements presented here will provide a clean determination of fiber attenuation at \qty{275}{nm} and \qty{367}{nm}. Figure~\ref{fig:test_stand_2} shows the test stand detail with inserted fiber ready for measurements.

\subsection{Light Attenuation of Single 5.7 m Long FVP600660710 Fiber}~\label{sec:single_fiber_and_connector}

The default fiber type used in the test stand is the FVP600660710 fiber (see Table \ref{tab:FiberstoTest}). The data presented in Table~\ref{tab:table_combined_fiber} provides the raw experimental readings necessary to determine the attenuation properties of the single \qty{5.7}{m} long FVP600660710 optical fiber at \qty{275}{nm} and \qty{367}{nm} wavelengths.
\begin{table}[!htp]
\centering
\caption{Light attenuation measurements for 5.7 m FVP600660710 fiber with 275 nm and 367 nm light. All values are in millivolts (mV).}
\label{tab:table_combined_fiber}
\begin{tabular}{ |p{1.0cm}|p{2.2cm}|p{2.2cm}|p{2.2cm}|p{2.2cm}| } 
\hline
\multirow{2}{*}{Type} & \multicolumn{2}{c|}{275 nm} & \multicolumn{2}{c|}{367 nm} \\ \cline{2-5}
 & Internal PD (mV) & External PD (mV) & Internal PD (mV) & External PD (mV) \\ \hline
\multicolumn{5}{|c|}{Baseline Measurements (No Fiber)} \\ \hline
1  & 744--760 & 466--474 & 1540 & 968--976 \\  
2  & 720--740 & 492--502 & 1540 & 968--980 \\
3  & 748--762 & 490--496 & -- & -- \\
4  & 764--772 & 488--494 & -- & -- \\ \hline
\multicolumn{5}{|c|}{Fiber Measurements (5.7 m)} \\ \hline
1  & 696--728 & 314--320 & 1530 & 712--728 \\ 
2  & 696--708 & 322--328 & 1540 & 720--728 \\ 
3  & 700--712 & 310--318 & 1540 & 704--720 \\ 
4  & 696--716 & 326--334 & 1540 & 752--765 \\ 
5  & 704--716 & 288--292 & 1540 & 754--780 \\
6  & 696--724 & 292--300 & 1540 & 704--720 \\ 
7  & 724--744 & 294--300 & 1540 & 728--740 \\ 
8  & 704--732 & 306--312 & 1540 & 704--718 \\
9  & 776--784 & 314--322 & 1540 & 704--718 \\ \hline
\end{tabular}
\end{table}
\textbf{Baseline Characterization.} Four separate baseline measurements were first recorded for \qty{275}{nm} wavelength without the 5.7 m long fiber in the path. Similarly, two separate baseline measurements were first recorded for \qty{367}{nm} wavelength. This establishes the reference light intensity ($I_0$) detected directly from the source. The observed external photodiode average value (e.g., $487.8 \pm 12.7$ mV for \qty{275}{nm}, and $973.0 \pm 6.0$ mV for \qty{367}{nm}) represent the inherent stability of the light source and the measurement system. Averaging these multiple measurements provides a robust and statistically significant value for $I_0$, mitigating the effect of any short-term fluctuations.\\
\textbf{Fiber and SMA-to-SMA Connector Transmission Measurement.} Nine separate measurements were then taken with the \qty{5.7}{m} fiber (inserted with SMA-to-SMA connectors as shown in Figure~\ref{fig:1fiber_1sma2sma}) for both \qty{275}{nm} and \qty{367}{nm} light. The purpose of taking multiple measurements over different fibers is to account for light transport variations and ensure that the result is reproducible. In general, this variability may arise from slight differences in connector mating or small changes in alignment, as well as from the light source and photo-diode readings. The average transmitted values for the external photodiode are $310.7 \pm 14.2$ mV for \qty{275}{nm}, and $727.7 \pm 22.7$ mV for \qty{367}{nm} light.\\
\textbf{Separation of Fiber Attenuation from the SMA-to-SMA Connector Light Loss.}
Figure~\ref{fig:1fiber_1sma2sma} (right) shows the total light transmission through the system composed of a fiber of length $x = \qty{5.7}{m}$ and a single SMA-to-SMA connector is given by the ratio $R = I / I_\text{o}$, where $I_\text{o}$ is the baseline intensity and $I$ is the intensity measured after the combined system. The total loss is simply $1 - R$. This total effect can be separated into the loss from the fiber itself and the loss from the SMA-to-SMA connector. The SMA-to-SMA connector's performance is characterized by its light survival fraction $S_\text{c}$, previously measured in a separate test described in detail in Section \ref{subsec:name1}. The light survival fraction of the fiber alone, $S_\text{f}$, is defined by the exponential decay law $S_\text{f} = e^{-\alpha x}$, where $\alpha$ is the attenuation coefficient we wish to determine. The combined system's ratio $R$ is the product of the two survival fractions, $R = S_\text{f} \cdot S_\text{c}$. Therefore, the fiber's survival fraction is derived as $S_\text{f} = R / S_\text{c}$, and its attenuation coefficient is calculated by $\alpha = -\frac{1}{x} \ln S_\text{f}$. The Table~\ref{tab:fiber_attenuation_summary} summarizes the results of applying these formulas to the measurements at \qty{275}{nm} and \qty{367}{nm}.\\
\textbf{Fiber Attenuation Results.}
The measured total light loss, which includes the combined effect of the \qty{5.7}{m} fiber and one SMA-to-SMA connector, is significantly higher for \qty{275}{nm} light ($36.3\%$) than for \qty{367}{nm} light ($25.2\%$) as shown in 
Table~\ref{tab:fiber_attenuation_summary}.
After correcting for the transmission loss of the SMA-to-SMA connector itself ($S_\text{c}$), the intrinsic attenuation of the fiber alone was isolated. The resulting fiber attenuation coefficient $\alpha$ confirms the strong wavelength dependence, with the value for \qty{275}{nm} (\qty{0.050}{\per\meter}) being approximately twice that for \qty{367}{nm} (\qty{0.028}{\per\meter}). This shows that shorter wavelengths are attenuated more effectively within the fiber material.
\begin{table}[hhh!]
\centering
\caption{Summary of total light loss and derived fiber attenuation coefficients for 5.7 m FVP600660710 fiber.}
\label{tab:fiber_attenuation_summary}

\begin{tabular}{|l|c|c|}
\hline
\textbf{Parameter} & \textbf{275 nm} & \textbf{367 nm} \\ \hline
Baseline Intensity, $I_\text{o}$ (mV) & $487.8 \pm 12.7$ & $973.0 \pm 6.0$ \\ \hline
Transmitted Intensity, $I$ (mV) & $310.7 \pm 14.2$ & $727.7 \pm 22.7$ \\ \hline
Transmission Ratio, $R = I / I_\text{o}$ & $0.6369 \pm 0.0334$ & $0.748 \pm 0.024$ \\ \hline
\textbf{Total Loss (Fiber + SMA-to-SMA connector)} & $\mathbf{(36.3 \pm 3.4)\%}$ & $\mathbf{(25.2 \pm 2.4)\%}$ \\ \hline \hline
SMA-to-SMA connector Survival, $S_\text{c}$ & $0.848 \pm 0.051$ & $0.876 \pm 0.050$ \\ \hline
Fiber Survival, $S_\text{f} = R / S_\text{c}$ & $0.751 \pm 0.060$ & $0.854 \pm 0.056$ \\ \hline
\textbf{Fiber Attenuation Coefficient, $\alpha$ (m$\mathbf{^{-1}}$)} & $\mathbf{0.050 \pm 0.014}$ & $\mathbf{0.028 \pm 0.011}$ \\ \hline
\end{tabular}
\end{table}

\subsection{Light Attenuation of Single Long FP600URT Fiber}

\begin{figure}[!htbp]
    \centering
    \includegraphics[height=5.0cm,width=0.6\textwidth]{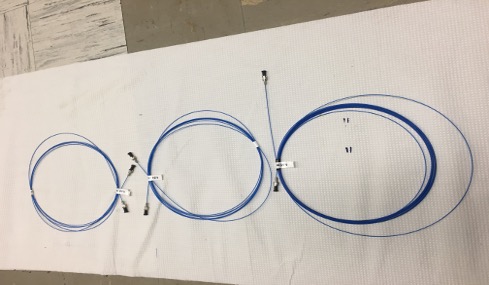} 
    \caption{FP600URT fiber samples used in the light attenuation measurements, arranged from left to right with fiber lengths of 3.50 m, 4.83 m, and 9.55 m.}
    \label{fig:tefzel_1sma2sma}
\end{figure}

Three FP600URT fibers (see Table \ref{tab:FiberstoTest}) shown in Figure~\ref{fig:tefzel_1sma2sma}, with lengths of \qty{3.50}{m}, \qty{4.83}{m}, and \qty{9.55}{m} are prepared for tests with \qty{275}{nm} and \qty{367}{nm} light.

The experimental situation is again similar to the sketch presented in Figure~\ref{fig:1fiber_1sma2sma}. However, in this case, the fiber $\#3$ (\qty{9.55}{m}) is installed between fibers $\#1$ (\qty{3.50}{m}) and $\#2$ (\qty{4.83}{m}). In this way, we provide the baseline measurement with the same core, cladding, and coating diameters for baseline fibers $\#1$ and $\#2$, and for the tested fiber $\#3$.\\
\textbf{Fiber Attenuation Results.} Table~\ref{tab:tefzel_fiber_results} shows the results of the measurement. We use the SMA-to-SMA connector loss value measured with the FP600URT fiber at 370 nm presented in Table \ref{tab:powerloss_dB}.
When going to \qty{275}{nm} wavelength, the following is observed:
\begin{itemize}
    \item FP600URT fibers $\#1$ and $\#2$ absorb much of \qty{275}{nm} UV light so, cannot be used for the reference measurement as there is not enough light budget left for testing the Tefzel fiber $\#3$. Results for such a case are listed in parenthesis of Table~\ref{tab:tefzel_fiber_results} with transmitted intensity measured at $<\qty{5}{mV}$.
    
    \item As the above test resulted in a limit, the approach here was modified to estimate the Tefzel $\#3$ fiber attenuation length for \qty{275}{nm} light. The FVP600660710 were used for baseline measurements, and the FP600URT fiber $\#3$ (\qty{9.55}{m}) was installed in between. These results are presented in \qty{275}{nm} column of Table~\ref{tab:tefzel_fiber_results}.
\end{itemize}
The measured fiber attenuation coefficient $\alpha = \qty{0.214}{\per\meter}$ at 275 nm demonstrate a strong light attenuation and highlight that the FP600URT fiber is not designed to be used for UV light transmission of less than 300 nm (see Table \ref{tab:FiberstoTest}).

\begin{table}[hhh!]
\centering
\caption{Summary of total light loss and derived fiber attenuation coefficients for 9.55 m FP600URT fiber with a single SMA-to-SMA connector.}
\label{tab:tefzel_fiber_results}
\begin{tabular}{|l|c|c|}
\hline
\textbf{Parameter} & \textbf{275 nm} & \textbf{367 nm} \\ \hline
Baseline Intensity, $I_\text{o}$ (mV) & $91.6 \pm 0.8$ ($9.4 \pm 0.8$) & $486.7 \pm 9.1$ \\ \hline
Transmitted Intensity, $I$ (mV) & $10.1 \pm 0.7$ ($<$ 5) & $327.0 \pm 4.5$ \\ \hline
Transmission Ratio, $R = I / I_\text{o}$ & $0.110 \pm 0.008$ & $0.672 \pm 0.016$ \\ \hline
\textbf{Total Loss (Fiber + SMA-to-SMA Connector)} & $\mathbf{(89.0 \pm 0.8)\%}$ & $\mathbf{(32.8 \pm 1.6)\%}$ \\ \hline \hline
SMA-to-SMA Connector Survival, $S_\text{c}$ & $0.848 \pm 0.051$ & $0.763 \pm 0.039$ \\ \hline
Fiber Survival, $S_\text{f} = R / S_\text{c}$ & $0.130 \pm 0.012$ & $0.881 \pm 0.049$ \\ \hline
\textbf{Fiber Attenuation Coefficient, $\alpha$ (m$\mathbf{^{-1}}$)} & $\mathbf{0.214 \pm 0.016}$ & $\mathbf{0.013 \pm 0.010}$ \\ \hline
\end{tabular}
\end{table}

\subsection{Light Loss Measured through Combinations of Multiple FVP600660710 Fibers: DUNE-style Mockup}~\label{sec:dune_style_test} 
The goal is to characterize optical losses through the multiple optical components that include optical fiber, optical feedthrough and SMA-to-SMA connector, to form an expectation on light losses in DUNE far detector, assuming the application of FVP600660710 fibers. To that end we mounted the test that mimics the DUNE far detector in terms of anticipated number of warm fiber (one warm fiber from the electronics calibration module to the optical feedthrough), three cold fibers from optical feedthrough to diffusers in cryostat, and all fibers connected together with SMA-to-SMA connectors.
Therefore, this subsection considers light losses of \qty{275}{nm} and \qty{367}{nm} light through the combination of four \qty{4.7}{m} fibers and an optical feedthrough, with SMA-to-SMA connectors. 
Note that we use four \qty{4.7}{m} fibers in this test, but the actual fiber length in DUNE far detector will likely differ in lengths. This results will form an expectation on the order of magnitude of optical loss in the DUNE-line UV-light calibration systems.
\begin{table}[hhh!]
\centering
\caption{Summary of total light loss and derived fiber attenuation coefficients for four 4.7 m FVP600660710 fibers with the optical feedthrough.}
\label{tab:mockup_test_results}
\begin{tabular}{|l|c|c|}
\hline
\textbf{Parameter} & \textbf{275 nm} & \textbf{367 nm} \\ \hline
Baseline Intensity, $I_\text{o}$ (mV) & $477.0 \pm 3.0$ & $717.0 \pm 5.0$ \\ \hline
Transmitted Intensity, $I_\text{OFT+4fibers+4sma-to-sma}$ (mV) & $99.6 \pm 6.0$ & $321.4 \pm 14.5$ \\ \hline
Transmission Ratio, $R_\text{OFT+4fibers+4sma-to-sma}$ & $0.209 \pm 0.026$ & $0.448 \pm 0.020$ \\ \hline
\textbf{Total Loss (Fiber + SMA-to-SMA connector)} & $\mathbf{(79.1 \pm 2.6)\%}$ & $\mathbf{(55.2 \pm 2.0)\%}$ \\ \hline \hline
SMA-to-SMA connector Survival, $S_\text{c}$ & $0.848 \pm 0.051$ & $0.876 \pm 0.050$ \\ \hline
Single Fiber Survival, $S_\text{f} = R^{1/4} / S_\text{c}^{5/4}$ & $0.831 \pm 0.053$ & $0.965 \pm 0.060$ \\ \hline
\end{tabular}

\end{table}
In this case we insert four \qty{4.7}{m} FVP600660710 fibers with the optical feedthrough, which is effectively four fibers and five connector points in the line, added to the baseline configuration. The results are presented in Table~\ref{tab:mockup_test_results}.\\
\textbf{275 nm Propagation.} One finds that about $20.9\%$ of \qty{275}{nm} light will survive the optical path tested in this section. In addition, $I_\text{OFT + 4 fiber + 4 sma-to-sma} (\qty{275}{nm})/I_\text{o} (\qty{275}{nm}) = S^4_\text{f} (\qty{275}{nm}) \cdot S^5_\text{c} (\qty{275}{nm}) $. 
From here, we can determine the single fiber-only light attenuation loss to be
\begin{align*}
    S_\text{f} (\qty{275}{nm}) &= (R_\text{OFT + 4  fiber + 4 sma-to-sma} (\qty{275}{nm}))^{(1/4)}/S_\text{c} (\qty{275}{nm})^{(5/4)} \\ &= (0.209 \pm 0.026)^{(1/4)} / (0.848 \pm 0.051)^{(5/4)} \\ &= (0.676 \pm 0.005)  / (0.814 \pm 0.053) = 0.831 \pm 0.060.  
\end{align*}\\
The single \qty{4.7}{m} fiber light attenuation length measured in this section ($x \pm dx$) is consistent with the value measured in section~\ref{sec:single_fiber_and_connector} ($\alpha = 0.050 \pm 0.012$), within the measurement uncertainties. \\
\textbf{367 nm Propagation.} One finds that about $44.8\%$ of \qty{367}{nm} light will survive the optical path. In addition, $I_\text{OFT + 4 fiber + 4 sma-to-sma} (\qty{367}{nm})/I_\text{o} (\qty{367}{nm}) = S^4_\text{f} (\qty{367}{nm}) \cdot S^5_\text{c} (\qty{367}{nm}) $. 
From here, we can determine the single fiber-only light attenuation loss to be
\begin{align*}
    S_\text{f} (\qty{367}{nm}) &= (R_\text{OFT + 4  fiber + 4 sma-to-sma} (\qty{367}{nm}))^{(1/4)}/S_\text{c} (\qty{367}{nm})^{(5/4)} \\ &= (0.448 \pm 0.020)^{(1/4)} / (0.876 \pm 0.050)^{(5/4)} \\ &= (0.818 \pm 0.004)  / (0.8475 \pm 0.0530) = 0.965 \pm 0.060.  
\end{align*}\\
The single \qty{4.7}{m} fiber light attenuation length measured in this section ($x \pm dx$) is consistent with the value measured in Section~\ref{sec:single_fiber_and_connector} ($\alpha = 0.028 \pm 0.011$), within the measurement uncertainties. 

The longer optical path with more optical connectors in-line will lead to larger attenuation and connector light losses, as illustrated in this section. For both \qty{275}{nm} and \qty{367}{nm} light, these losses ($79.1\%$ and $55.2\%$ in this case, respectively) need to be accounted for in the initial light budget to ensure optimal light levels for the calibration in large cryostats.


\section{Optical Fiber Thermal Cycling in Cryogenic Conditions}
\label{sec:CT}

\begin{figure}[htpb]
    \centering
    \includegraphics[width=0.3\columnwidth]{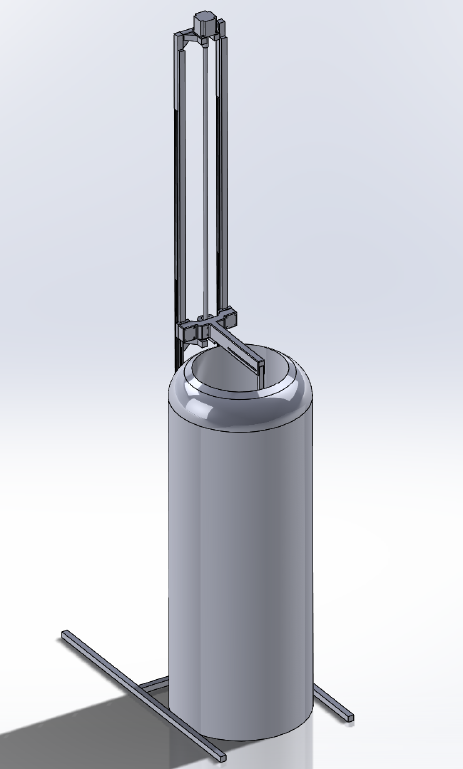}
    \includegraphics[width=0.22\columnwidth]{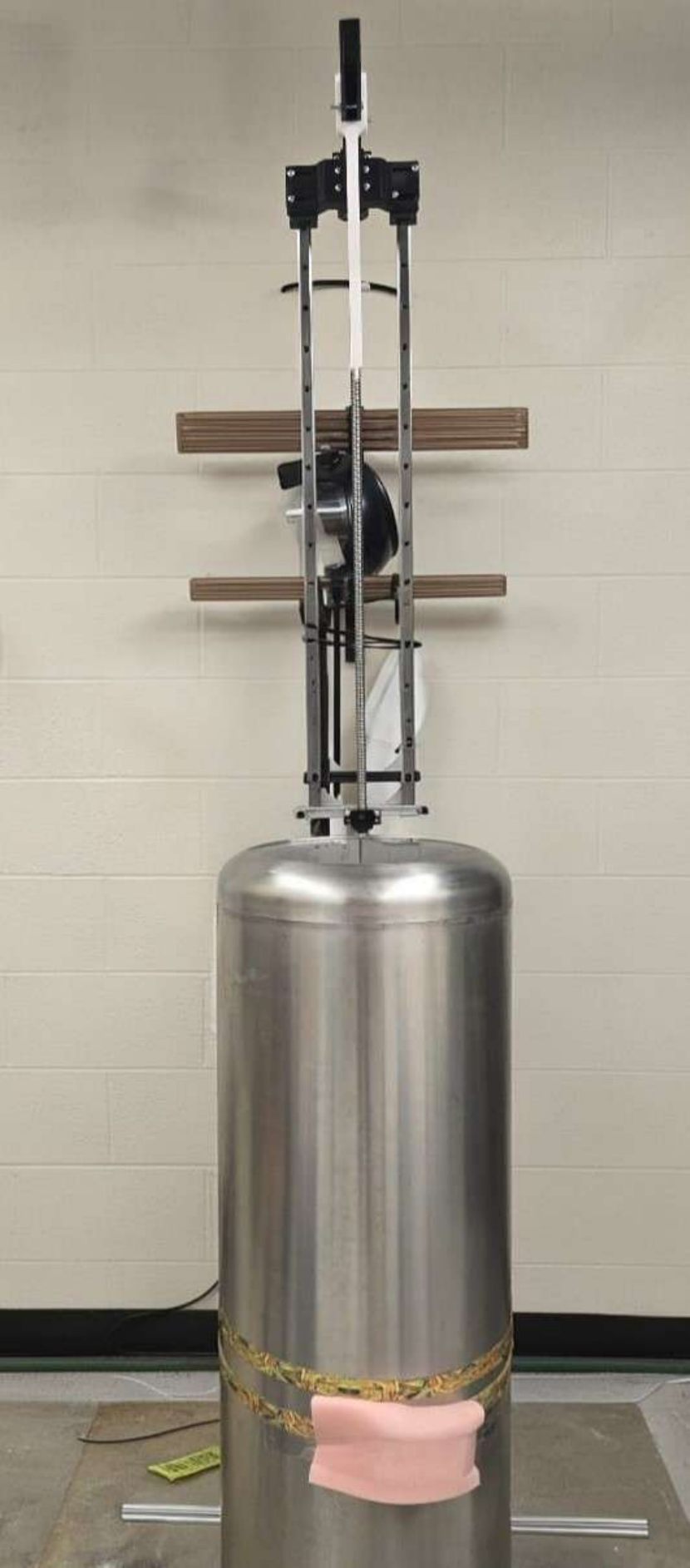}
    \caption{Cryogenic Thermal Cycle Apparatus (CTCA) used to test the optical fibers components under thermal stress. The setup includes the structural aluminum frame, 3D-printed support components, and the cryostat filled with liquid nitrogen, which allows for repeated thermal cycling of optical fiber assemblies.}
    \label{fig:Dewar}
    \end{figure}

The mechanical structural integrity and optical performance of the fibers is a critical characteristic that was evaluated through the design, assembly, and operation of a Cryogenic Thermal Cycle Apparatus (CTCA). This system was designed to thermally stress the fibers and their optical components through repeated immersion in liquid nitrogen, emulating conditions that could be expected in high energy physics detectors, and other technology applications such as space exploration or instrumentation intended for cryogenic environments. These tests are essential for ensuring the reliability and durability of the optical fiber assemblies during long‑term detector operation. The CTCA consists of an aluminum frame built with structural extruded bars and various 3D printed parts, with vertical motion controlled by an Arduino microcontroller. This allowed for fully automated immersion and removal sequences with programmable timing. Figure~\ref{fig:Dewar} shows the complete CTCA setup, including the cryostat used to place the optical components that will be exposed to thermal cycles.

\begin{figure}[htpb]
    \centering
    \includegraphics[width=0.416\columnwidth]{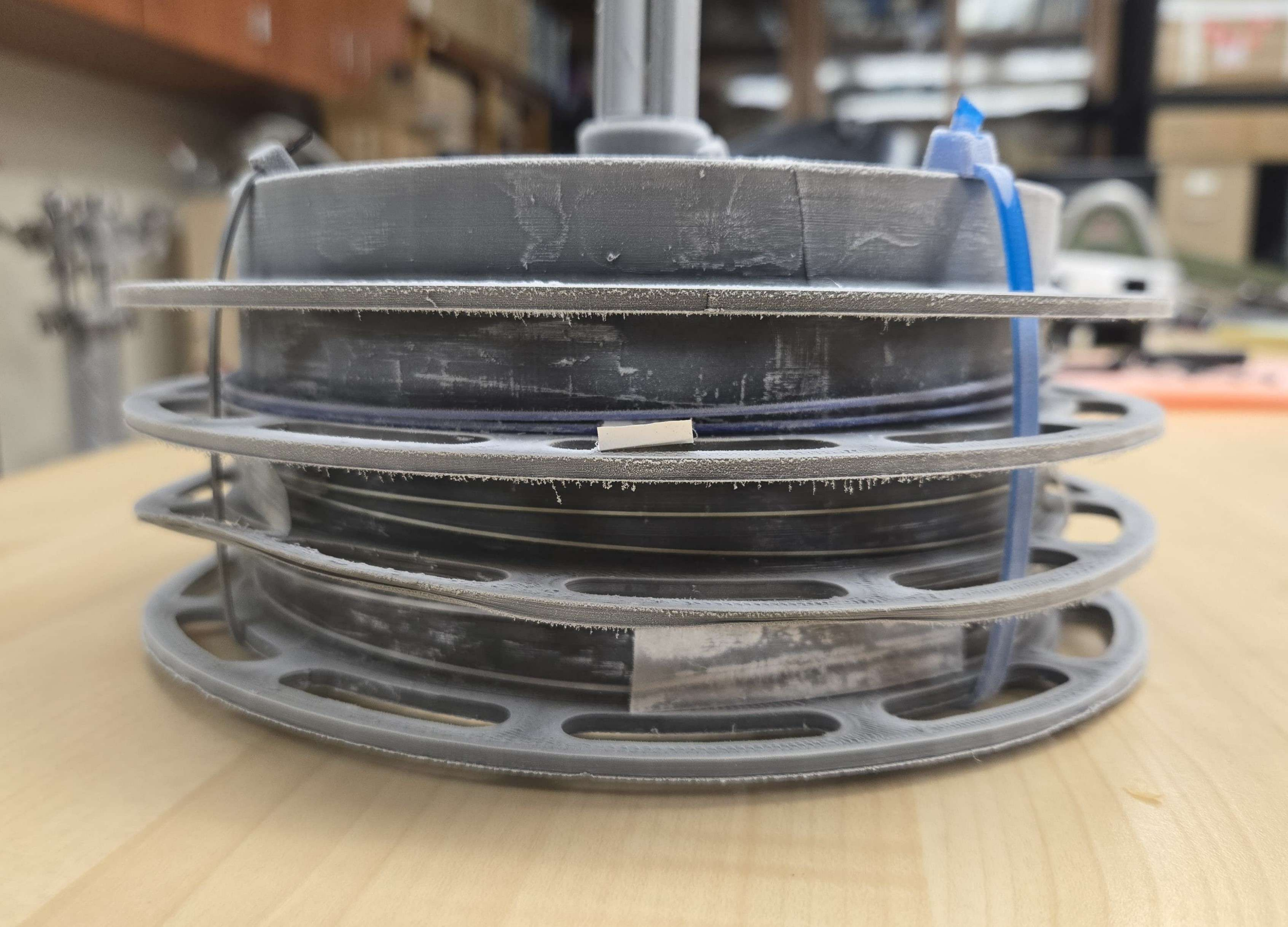}
    \caption{Bare optical fibers looped around custom 3D-printed wheels after completion of cryogenic cycling. Each wheel has a diameter of 16 cm and a height of 2 cm, allowing the fiber to complete two loops with a controlled bend radius. The wheels were mounted on the movable arm of the CTCA and repeatedly immersed in liquid nitrogen as part of the thermal cycling procedure.}
    \label{fig:3dprintedthermal}
    \end{figure}

For this test, the seven bare optical fibers characterized in Section~\ref{sec:OFQC} were used. Each fiber was looped twice around a 3D-printed wheel with a diameter of \qty{16}{cm} and a width of \qty{2}{cm}, resulting in a total length of approximately \qty{1}{m} per fiber (see Figure~\ref{fig:3dprintedthermal}). Four identical wheels were used in total, each designed to hold a single fiber while maintaining a consistent bend radius. These wheels were mounted on the movable arm at the bottom of the CTCA system, which was automated to lower and raise the fibers for submersion in and extraction from the liquid nitrogen. This mechanism enabled precise and repeatable thermal cycling of the fiber samples, as illustrated in Figure~\ref{fig:Dewar}.

The thermal cycling protocol was executed using the Arduino to automate a sequence consisting of $0.5$ minutes of immersion transition, $2$ minutes fully submerged in liquid nitrogen, $0.5$ minutes of removal transition, and $3$ minutes out of the liquid nitrogen. This defined a complete cycle with duration of $6$ minutes. Each fiber tested was subjected to $30$ such cycles, amounting to a total exposure time of $3$ hours per fiber.

\begin{figure}[htpb]
    \centering
    \includegraphics[scale=0.295]{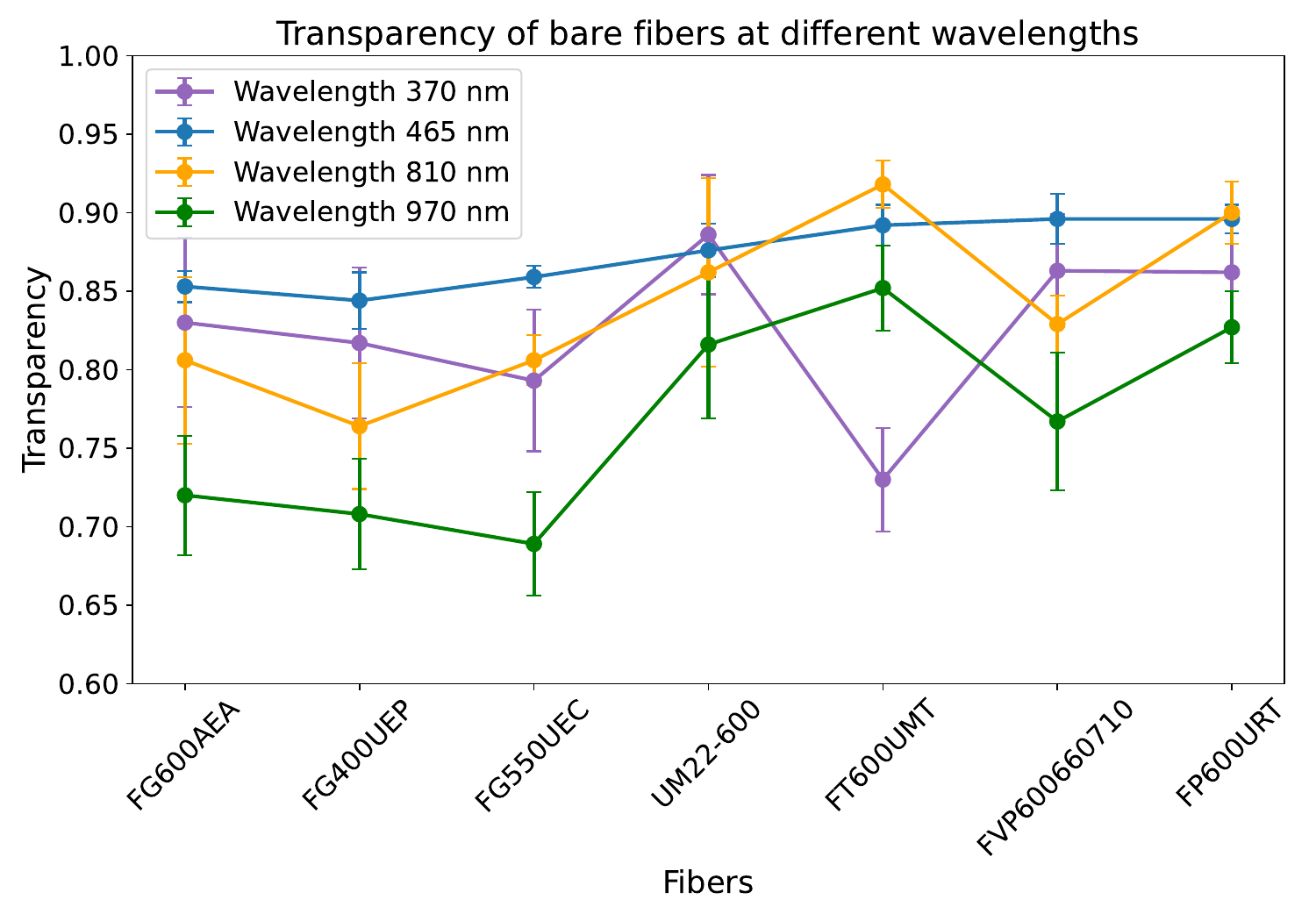}
    \includegraphics[scale=0.295]{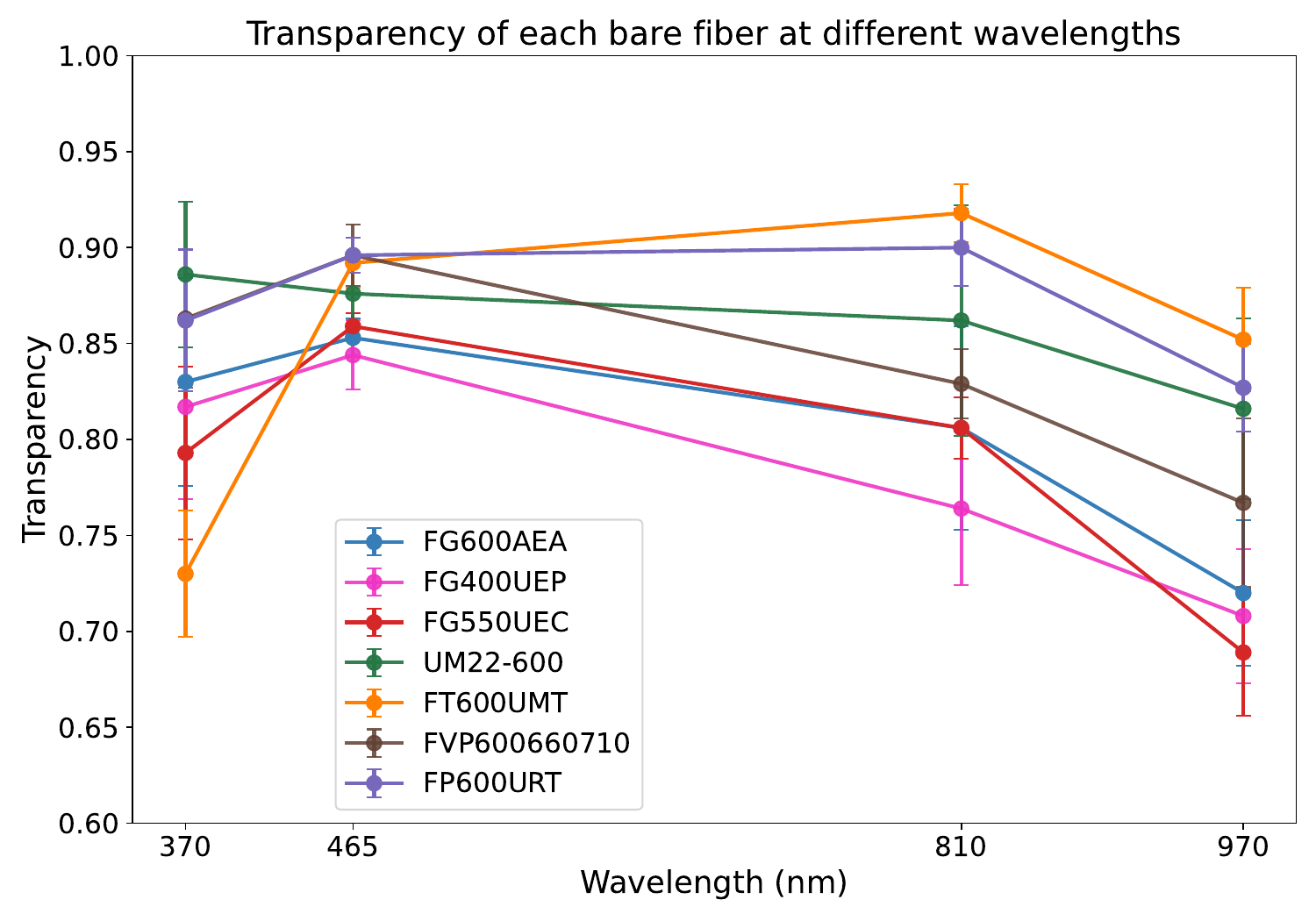}
    \caption{Left: Transparency at four wavelengths (370~nm, 465~nm, 810~nm, and 970~nm) for each bare fiber type after 30 cryogenic thermal cycles. Right: Transparency of each bare fiber as a function of wavelength after LN2. Error bars represent the statistical uncertainty associated with each measurement.}
    \label{fig:bareafterLN2}
\end{figure}

After completing the thermal cycling, the bare fibers were visually inspected for any signs of mechanical degradation such as cracking, crushing, or deformation.    No evidence of such degradation was observed, indicating good mechanical resilience of the bare fibers under cryogenic testing. With the goal to quantify any changes in optical performance, the transparency measurements described in the Section~\ref{sec:OFQC} were repeated after the thermal cycling process and are summarized in Table~\ref{tab:afterLN2}. Corresponding plots are also shown in Figure~\ref{fig:bareafterLN2}. The results show that the transparency values remain consistent within the statistical uncertainties, indicating that thermal cycling did not degrade the optical transmission properties of the tested fibers.

\begin{table}[htpb]
\centering
\caption{Transparency measurements for different bare fibers after undergoing $30$ cryogenic thermal cycles. Measurements were taken at four wavelengths, and the corresponding errors represent the standard deviation from repeated trials.}
\label{tab:afterLN2}

\begin{adjustbox}{width=\columnwidth,center}
\begin{tabular}{l|c|c|c|c}
\toprule
\multicolumn{5}{c}{\textbf{Transparency results for bare fibers  SMA-to-SMA connector after cryogenic cycling}} \\
\midrule
\textbf{Fiber} & \textbf{370~nm} $(T \pm \Delta T)$ & \textbf{465~nm} $(T \pm \Delta T)$ & \textbf{810~nm} $(T \pm \Delta T)$ & \textbf{970~nm} $(T \pm \Delta T)$ \\
\midrule
FG600AEA     & $0.830 \pm 0.054$ & $0.853 \pm 0.010$ & $0.806 \pm 0.053$ & $0.720 \pm 0.038$ \\
FG400UEP     & $0.817 \pm 0.048$ & $0.844 \pm 0.018$ & $0.764 \pm 0.040$ & $0.708 \pm 0.035$ \\
FG550UEC     & $0.793 \pm 0.045$ & $0.859 \pm 0.007$ & $0.806 \pm 0.016$ & $0.689 \pm 0.033$ \\
UM22-600     & $0.886 \pm 0.038$ & $0.876 \pm 0.017$ & $0.862 \pm 0.060$ & $0.816 \pm 0.047$ \\
FT600UMT     & $0.730 \pm 0.033$ & $0.892 \pm 0.013$ & $0.918 \pm 0.015$ & $0.852 \pm 0.027$ \\
FVP600660710 & $0.863 \pm 0.036$ & $0.896 \pm 0.016$ & $0.829 \pm 0.018$ & $0.767 \pm 0.044$ \\
FP600URT     & $0.862 \pm 0.037$ & $0.896 \pm 0.009$ & $0.900 \pm 0.020$ & $0.827 \pm 0.023$ \\
\bottomrule
\end{tabular}
\end{adjustbox}
\end{table}

\begin{table}[htpb]
\centering
\caption{Ratio of transparency values measured before and after cryogenic cycling for each bare fiber. The errors represent the propagated standard deviation from repeated measurements at each LED wavelength.}
\label{tab:ratioLN2}

\begin{adjustbox}{max width=\columnwidth,center}
\begin{tabular}{l|c|c|c|c}
\toprule
\multicolumn{5}{c}{\textbf{\shortstack{Transparency ratio (after / before cryogenic cycling)\\
for bare fibers + SMA-to-SMA connector}}} \\
\midrule
\textbf{Fiber} & \textbf{370~nm} & \textbf{465~nm} & \textbf{810~nm} & \textbf{970~nm}\\
\midrule
FG600AEA     & $1.003 \pm 0.089$ & $1.014 \pm 0.030$ & $1.009 \pm 0.083$ & $0.978 \pm 0.071$ \\
FG400UEP     & $1.004 \pm 0.078$ & $0.996 \pm 0.033$ & $1.001 \pm 0.089$ & $1.000 \pm 0.069$ \\
FG550UEC     & $1.007 \pm 0.078$ & $0.997 \pm 0.021$ & $1.005 \pm 0.035$ & $0.988 \pm 0.074$ \\
UM22-600     & $1.006 \pm 0.062$ & $1.021 \pm 0.036$ & $0.991 \pm 0.095$ & $1.005 \pm 0.087$ \\
FT600UMT     & $0.995 \pm 0.063$ & $1.010 \pm 0.032$ & $0.995 \pm 0.021$ & $0.993 \pm 0.042$ \\
FVP600660710 & $0.987 \pm 0.063$ & $0.990 \pm 0.037$ & $0.998 \pm 0.032$ & $1.020 \pm 0.078$ \\
FP600URT     & $0.988 \pm 0.059$ & $0.994 \pm 0.018$ & $1.001 \pm 0.024$ & $0.998 \pm 0.047$ \\
\bottomrule
\end{tabular}
\end{adjustbox}
\end{table}

\begin{figure}[htpb]
    \centering
    \includegraphics[scale=0.295]{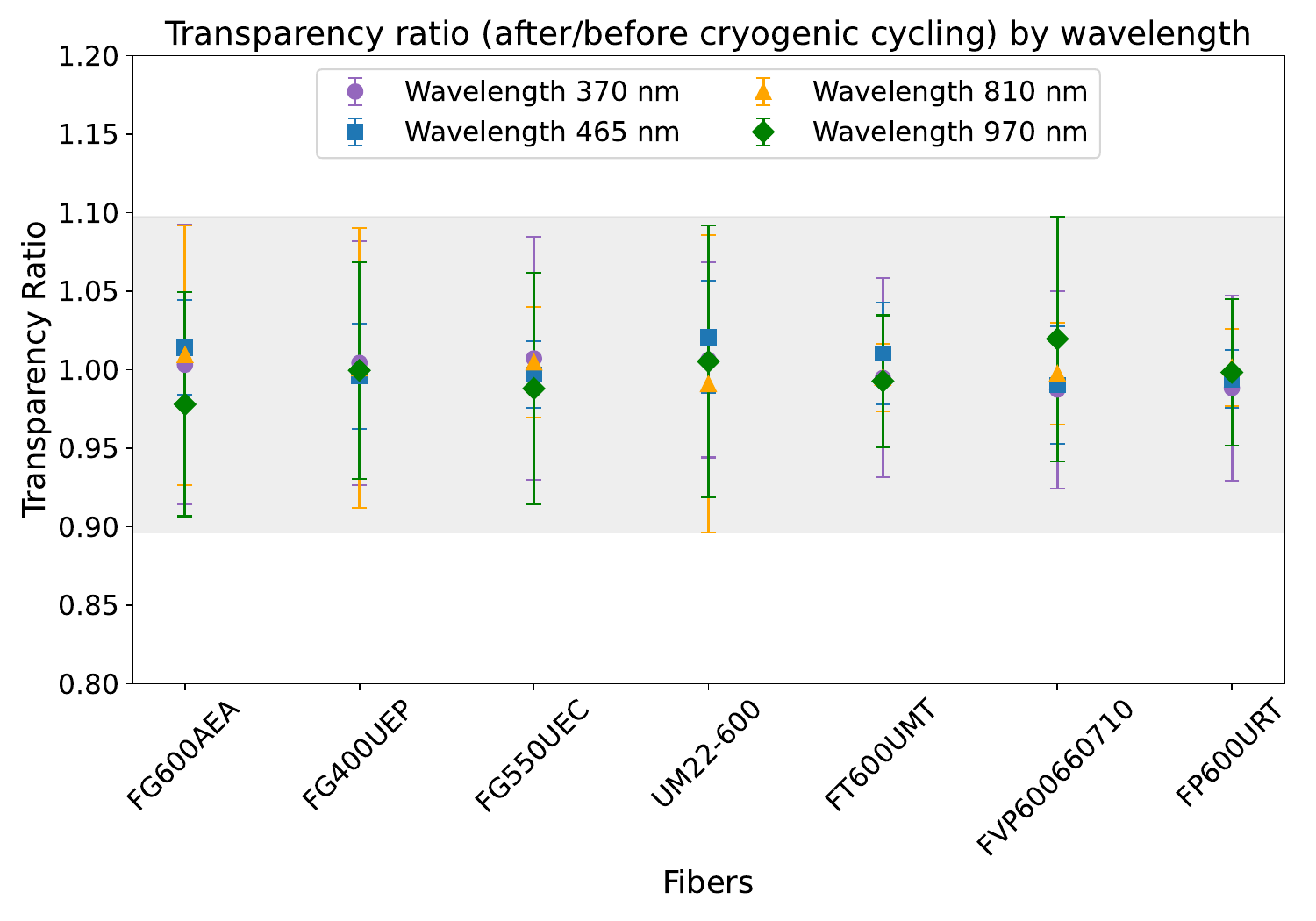}
    \includegraphics[scale=0.295]{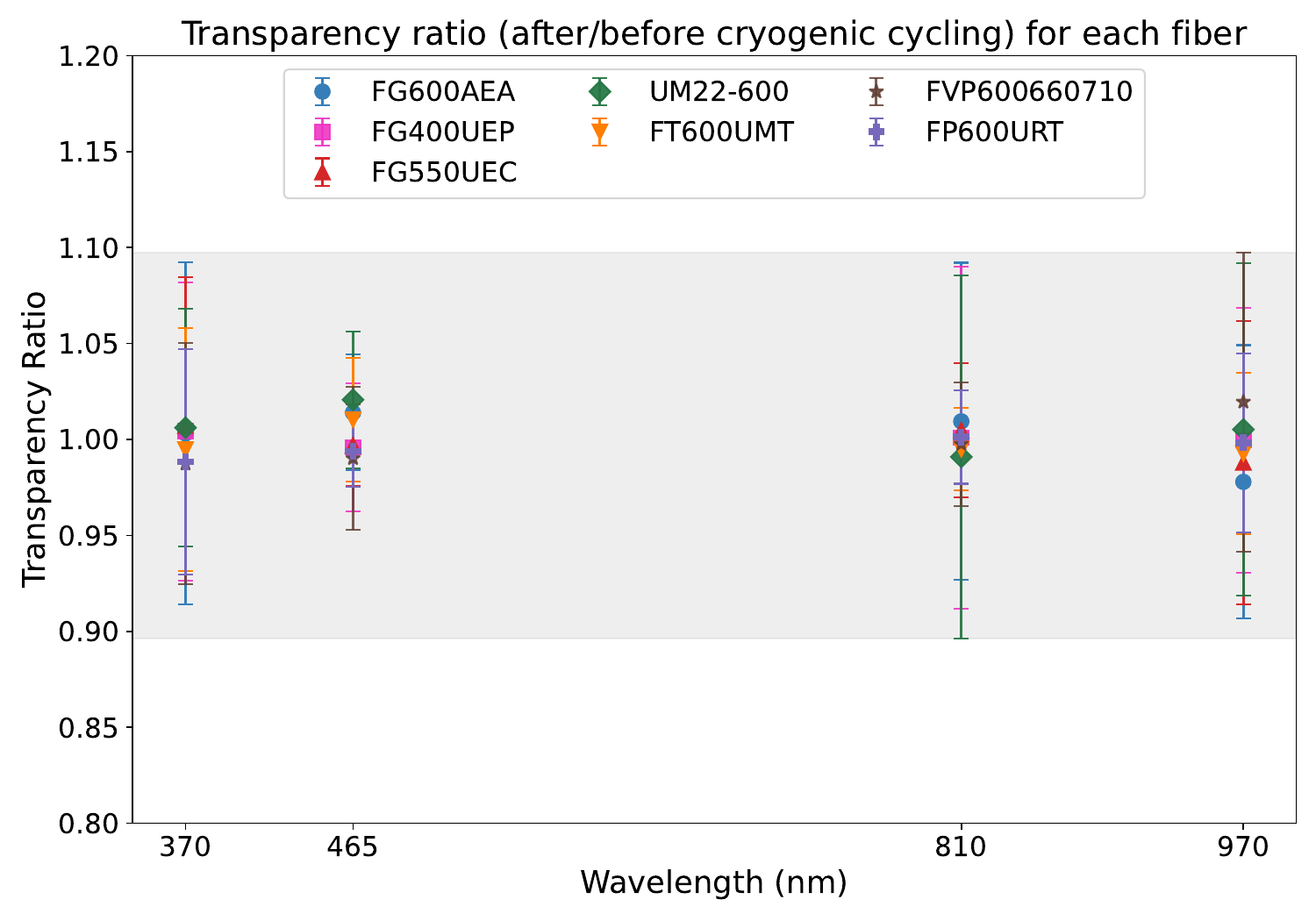}
    \caption{Ratio of transparency values measured after and before cryogenic cycling for each bare fiber. The shaded region represents an approximate ±10\% range around unity, within which all measurements fall when statistical uncertainties are included, indicating no significant degradation in optical performance due to thermal cycling. Left: Transparency ratios grouped by LED wavelength, showing the consistency of each fiber’s performance covering the tested spectral range. Right:  Ratios grouped by fiber type, showing the stability of transparency at each wavelength.}
    \label{fig:Ratiobare}
\end{figure}

To further  evaluate the impact of cryogenic thermal  cycling on the optical performance of the bare  fibers, the ratio  between the measured transparency values before  and after the  thermal cycles was calculated for each LED wavelength. As shown in Table~\ref{tab:ratioLN2}, all central values of the transparency ratios are consistent with unity, with deviations ranging from $\pm2.2\%$. When the associated statistical  measurements uncertainties are included, the deviations remain within $\pm10\%$, as illustrated in Figure~\ref{fig:Ratiobare}. This indicates a good transparency stability across the multiple tested wavelengths and fiber types. The observed variations are within statistical uncertainties, indicating no statistically significant change in transparency due to thermal cycling. These results shown the mechanical and optical resilience of the tested fibers under repeated cryogenic cycling, a key requirement for these fibers when planned to be used in particle physics detectors and other technologies that operate in cryogenic environments. In particular, the FP600URT and FVP600660710 fibers demonstrated reliable performance and were successfully deployed in the ultraviolet light calibration system of the ProtoDUNE-HD detector located at CERN \cite{NP02:2025AnnualReport}.


\section{Test of potential fiber degradation caused by a long UV-light exposure}
\label{sec:DSUV}

\begin{figure}[htpb]
    \centering
    \includegraphics[scale=0.38]{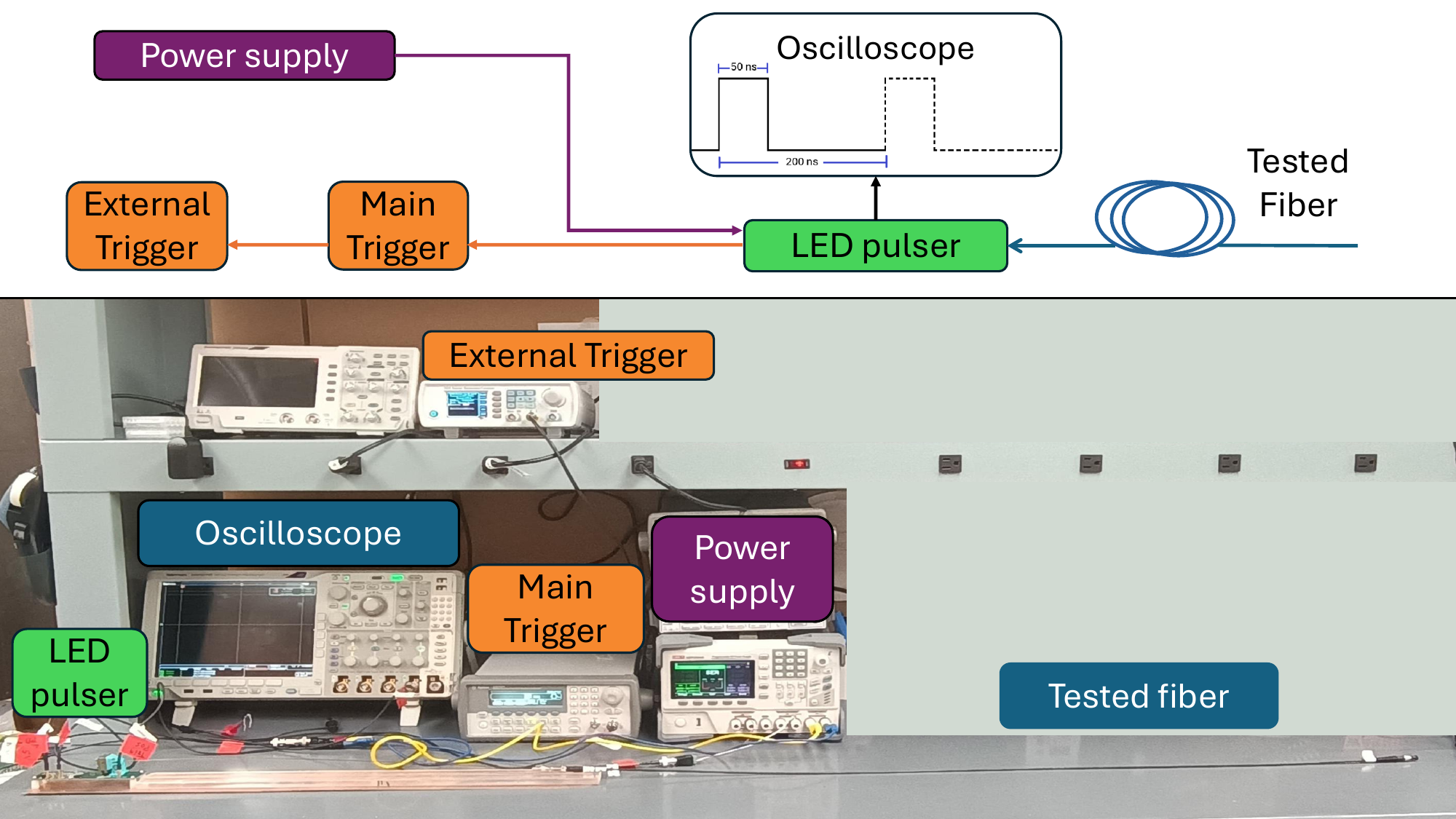}
    \caption{Degradation test setup. The LED pulser includes a permanently installed bare optical fiber that delivers UV light pulses, with the output wavelength of 285 nm to match the expected conditions of the DUNE light calibration system. The LED pulser was powered by an external power supply, and the pulse rate was controlled via a function generator used as an external trigger, which enabled the timing of the LED pulses sent by the main trigger. An oscilloscope was used to monitor the pulse shape and frequency throughout the test.}
    \label{fig:degradationsetup}
\end{figure}

High-purity quartz (fused silica) fiber is highly  resistant to UV degradation in most applications. However, prolonged exposure to high-intensity, deep-UV radiation can cause a gradual increase in light absorption (attenuation) within the fiber, a process known as solarization~\cite{Caplinq:HighPuritySilica,FOS:SolarizationSilicaFibers,Avantes:IntroFiberOptics,FirebirdOptics:OpticalWindows}.
In this section we expose multiple fiber types to \qty{275}{nm} or \qty{367}{nm} UV light, while measuring each fiber transparency before and after exposure. A controllable UV-light LED pulsed source, analogous to the pulsed used with the CERN cold-box~\cite{Paudel:2025bfh} was used in the degradation tests, with the test setups presented in Figure~\ref{fig:test_stand_1} and Figure~\ref{fig:degradationsetup}.
This LED pulser circuitry operates at \qty{5}{V} with capability to control the light intensity by allowing to change the LED bias and pulse width. An external trigger was used to configure the timing of the pulses delivered, and a digital oscilloscope connected to the monitoring photodiode, encapsulated with the UV LED, was connected to verify the operational performance.

First simpler test, described with Figure~\ref{fig:test_stand_1} setup, focused on FVP600660710 fibers deployed in cryogenic application with ProtoDUNE SP~\cite{Abi2020} and/or ProtoDUNE HD~\cite{Soto-Oton:2024apm}.
The DUNE-grade 5.7m-long silica fiber (FVP600660710) was tested by transmitting some 90 million of 50 ns long 275 nm and 367 nm light pulses through it, while continuously monitoring the light transmission over the measurement period. With an exposure corresponding to more than 30 years of DUNE calibration UV-light exposure, no fiber attenuation degradation was observed. Upper limits on fiber degradation at the 68\% confidence level were set at 2.8\% for 275 nm light and 1.2\% for 367 nm light, limited by measurement uncertainty.

While the above test focused on FVP600660710 fibers alone, the second test was performed to evaluate the long-term durability of various other optical fibers under prolonged UV light exposure. This later UV degradation test was performed with the test setup presented in Figure~\ref{fig:degradationsetup}, to simulate the cumulative exposure estimated over the DUNE light calibration system over the $20$-year operational lifetime projection. The test used the seven jacketed optical fibers characterized in Section~\ref{sec:OFQC}. In the degradation test setup, the UV light from a \qty{275}{nm} LED was delivered through a permanently installed bare optical fiber (FVP600660710) connected to the \qty{275}{nm} LED pulser. Each jacketed fiber under test was connected to this bare fiber using an SMA-to-SMA mating sleeve, identical to the one used in the transparency measurements described in Section~\ref{sec:OFQC}. This ensured precise optical alignment and mechanical stability during the test.
Figure~\ref{fig:degradationsetup} shows the experimental setup, including the LED source, optical connections, and optical fiber, and illustrates the pulse configuration used for the degradation test, consisting of a \qty{50}{ns} pulse with a \qty{200}{ns} period, yielding a \qty{5}{MHz} repetition rate. Each optical fiber was exposed to roughly $30$ million UV pulses, simulating an approximate total exposure expected over $20$ years of DUNE operation ($150$ light calibration runs per year, $10{,}000$ pulses per run).

\begin{figure}[htpb]
    \centering
    \includegraphics[scale=0.295]{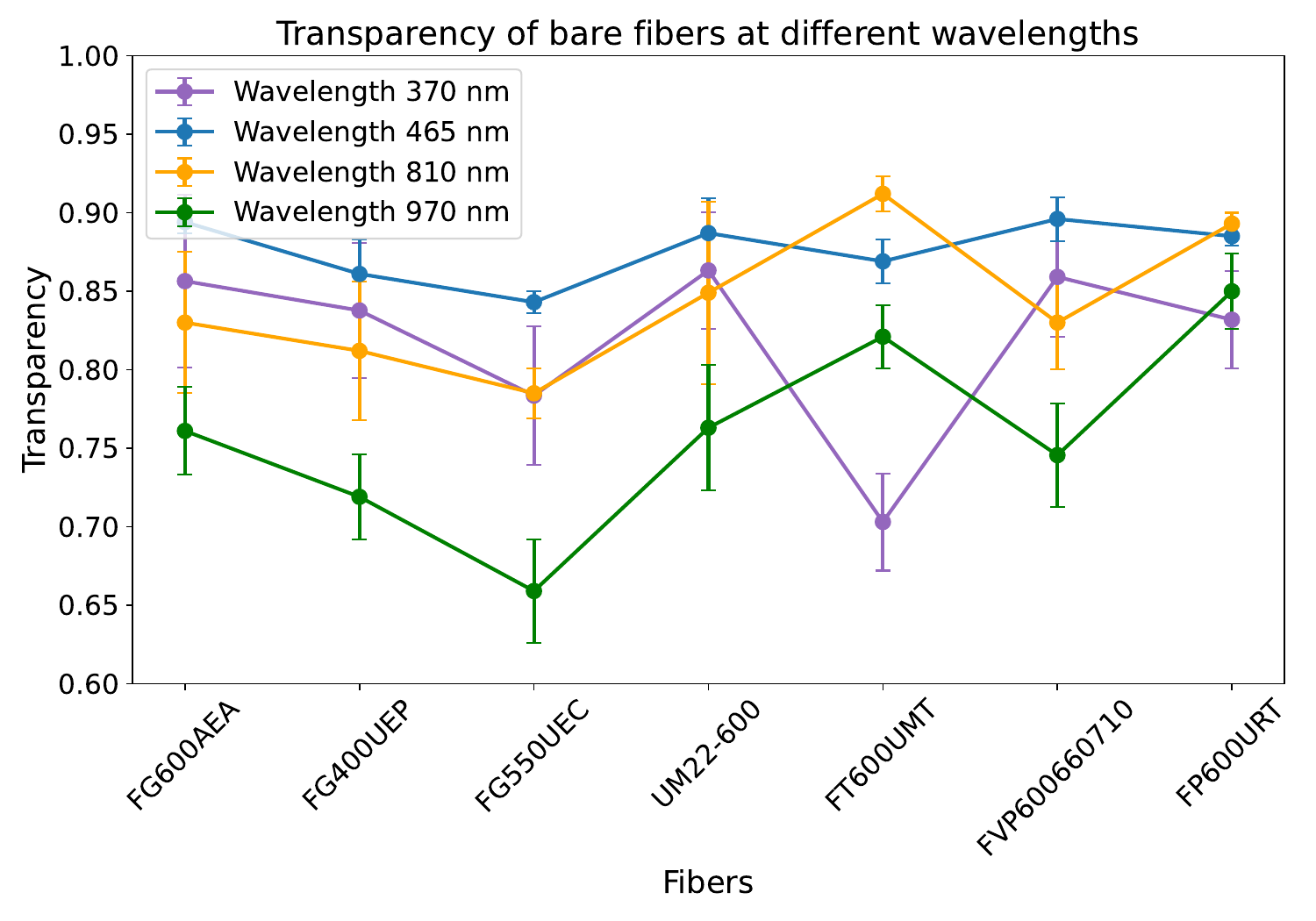}
    \includegraphics[scale=0.295]{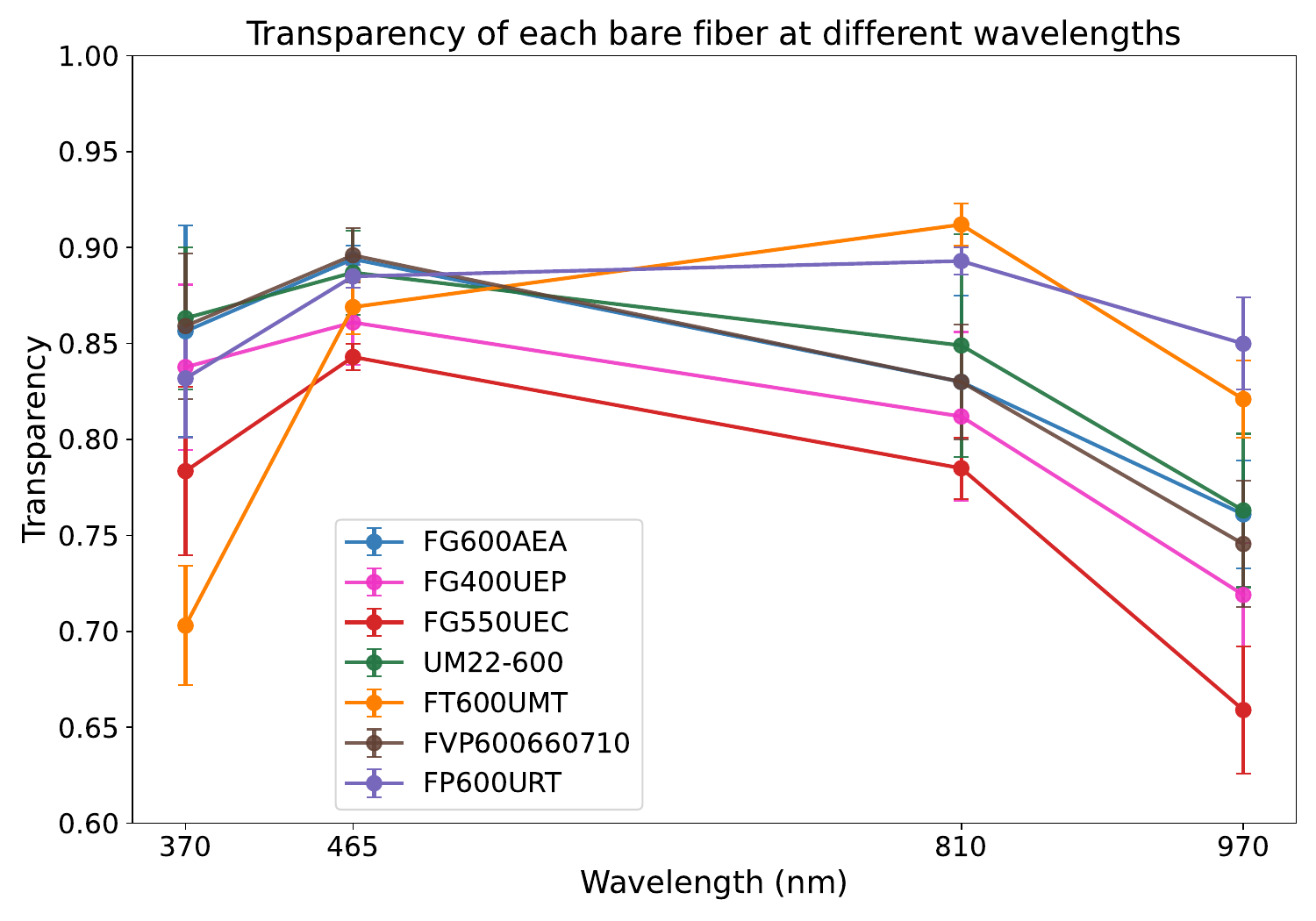}
    \caption{Left: Transparency at four wavelengths (370~nm, 465~nm, 810~nm, and 970~nm) for each jacketed fiber type after degradation test. Right: Transparency of each jacketed fiber as a function of wavelength after degradation test. Error bars represent the statistical uncertainty associated with each measurement.}
    \label{fig:degradation}
\end{figure}

The transparency values measured at four wavelengths \qty{370}{nm}, \qty{465}{nm}, \qty{810}{nm}, and \qty{970}{nm}, for each jacketed fiber after the degradation test are presented in Table~\ref{tab:jacketed_afterdeg}. These measurements provide a direct assessment of the fibers’ optical performance following sustained exposure to high-frequency UV light. These results indicate that all fibers maintained high transparency, with differences across wavelengths and fiber types falling within the expected range of statistical uncertainty. These results are shown in Figure~\ref{fig:degradation}. The consistent performance across the fibers, particularly in the UV and visible regions, demonstrates that the pulsed UV exposure caused no significant degradation, strongly supporting their suitability for long-term operation in harsh environments and indicating that these fibers could reliably perform after the number of UV pulses expected during the $20$-year projected operation of the light calibration system.

\begin{table}[htpb]
\centering
\caption{Transparency measurements for different jacketed fibers after degradation test. Measurements were taken at four wavelengths, and the corresponding statistical errors.}
\label{tab:jacketed_afterdeg}

\begin{adjustbox}{width=\columnwidth,center}
\begin{tabular}{l|c|c|c|c}
\toprule
\multicolumn{5}{c}{\textbf{\shortstack{Transparency results for jacketed fibers plus SMA-to-SMA connector after degradation test}}} \\
\midrule
\textbf{Fiber} & \textbf{370~nm} $(T \pm \Delta T)$ & \textbf{465~nm} $(T \pm \Delta T)$ & \textbf{810~nm} $(T \pm \Delta T)$ & \textbf{970~nm} $(T \pm \Delta T)$ \\
\midrule
FG600AEA     & $0.856 \pm 0.055$ & $0.894 \pm 0.007$ & $0.830 \pm 0.045$ & $0.761 \pm 0.028$ \\
FG400UEP     & $0.838 \pm 0.043$ & $0.861 \pm 0.022$ & $0.812 \pm 0.044$ & $0.719 \pm 0.027$ \\
FG550UEC     & $0.784 \pm 0.044$ & $0.843 \pm 0.007$ & $0.785 \pm 0.016$ & $0.659 \pm 0.033$ \\
UM22-600     & $0.863 \pm 0.037$ & $0.887 \pm 0.022$ & $0.849 \pm 0.058$ & $0.763 \pm 0.040$ \\
FT600UMT     & $0.703 \pm 0.031$ & $0.869 \pm 0.014$ & $0.912 \pm 0.011$ & $0.821 \pm 0.020$ \\
FVP600660710 & $0.859 \pm 0.038$ & $0.896 \pm 0.014$ & $0.830 \pm 0.030$ & $0.746 \pm 0.033$ \\
FP600URT     & $0.832 \pm 0.031$ & $0.885 \pm 0.006$ & $0.893 \pm 0.007$ & $0.850 \pm 0.024$ \\
\bottomrule
\end{tabular}
\end{adjustbox}
\end{table}

\begin{table}[htpb]
\centering
\caption{Ratio of transparency values measured before and after the degradation test for each jacketed fiber. The errors represent the propagated standard deviation from repeated measurements at each wavelength.}
\label{tab:ratiodeg}

\begin{adjustbox}{max width=\columnwidth,center}
\begin{tabular}{l|c|c|c|c}
\toprule
\multicolumn{5}{c}{\textbf{\shortstack{Transparency ratio (after / before degradation test) for\\ jacketed fibers + SMA-to-SMA connector}}} \\
\midrule
\textbf{Fiber} & \textbf{370~nm}  & \textbf{465~nm} & \textbf{810~nm} & \textbf{970~nm} \\
\midrule
FG600AEA     & $1.006 \pm 0.091$ & $1.010 \pm 0.027$ & $1.004 \pm 0.073$ & $0.977 \pm 0.060$ \\
FG400UEP     & $0.988 \pm 0.071$ & $1.003 \pm 0.037$ & $0.997 \pm 0.087$ & $0.991 \pm 0.061$ \\
FG550UEC     & $1.013 \pm 0.077$ & $1.014 \pm 0.022$ & $1.017 \pm 0.037$ & $0.978 \pm 0.075$ \\
UM22-600     & $1.010 \pm 0.063$ & $1.021 \pm 0.039$ & $0.993 \pm 0.096$ & $0.981 \pm 0.082$ \\
FT600UMT     & $0.998 \pm 0.063$ & $1.016 \pm 0.034$ & $0.995 \pm 0.019$ & $0.978 \pm 0.037$ \\
FVP600660710 & $1.009 \pm 0.068$ & $1.014 \pm 0.038$ & $0.976 \pm 0.043$ & $1.022 \pm 0.068$ \\
FP600URT     & $0.981 \pm 0.055$ & $0.997 \pm 0.017$ & $0.990 \pm 0.013$ & $0.981 \pm 0.046$ \\
\bottomrule
\end{tabular}
\end{adjustbox}

\end{table}

\begin{figure}[htpb]
    \centering
    \includegraphics[scale=0.295]{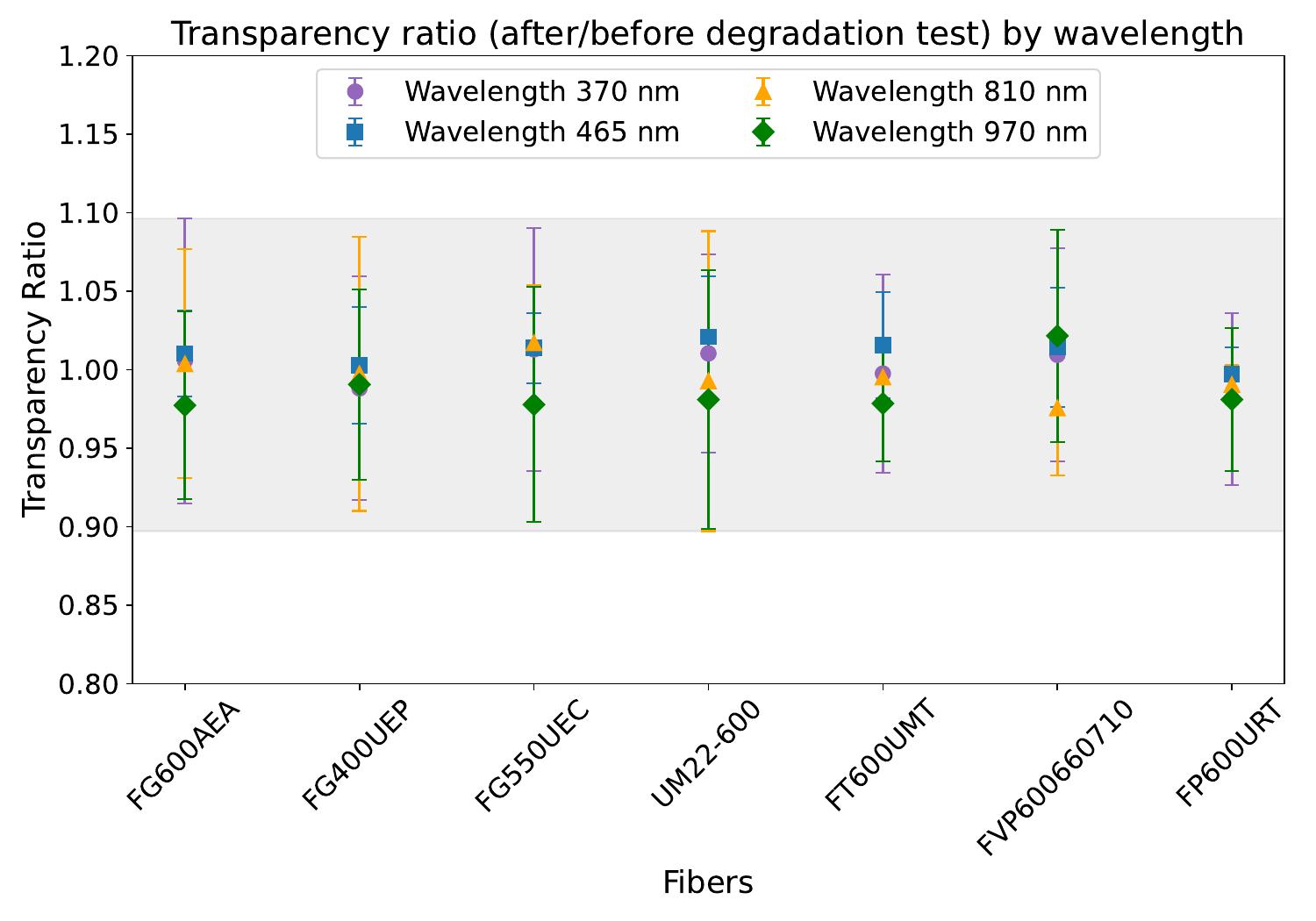}
    \includegraphics[scale=0.295]{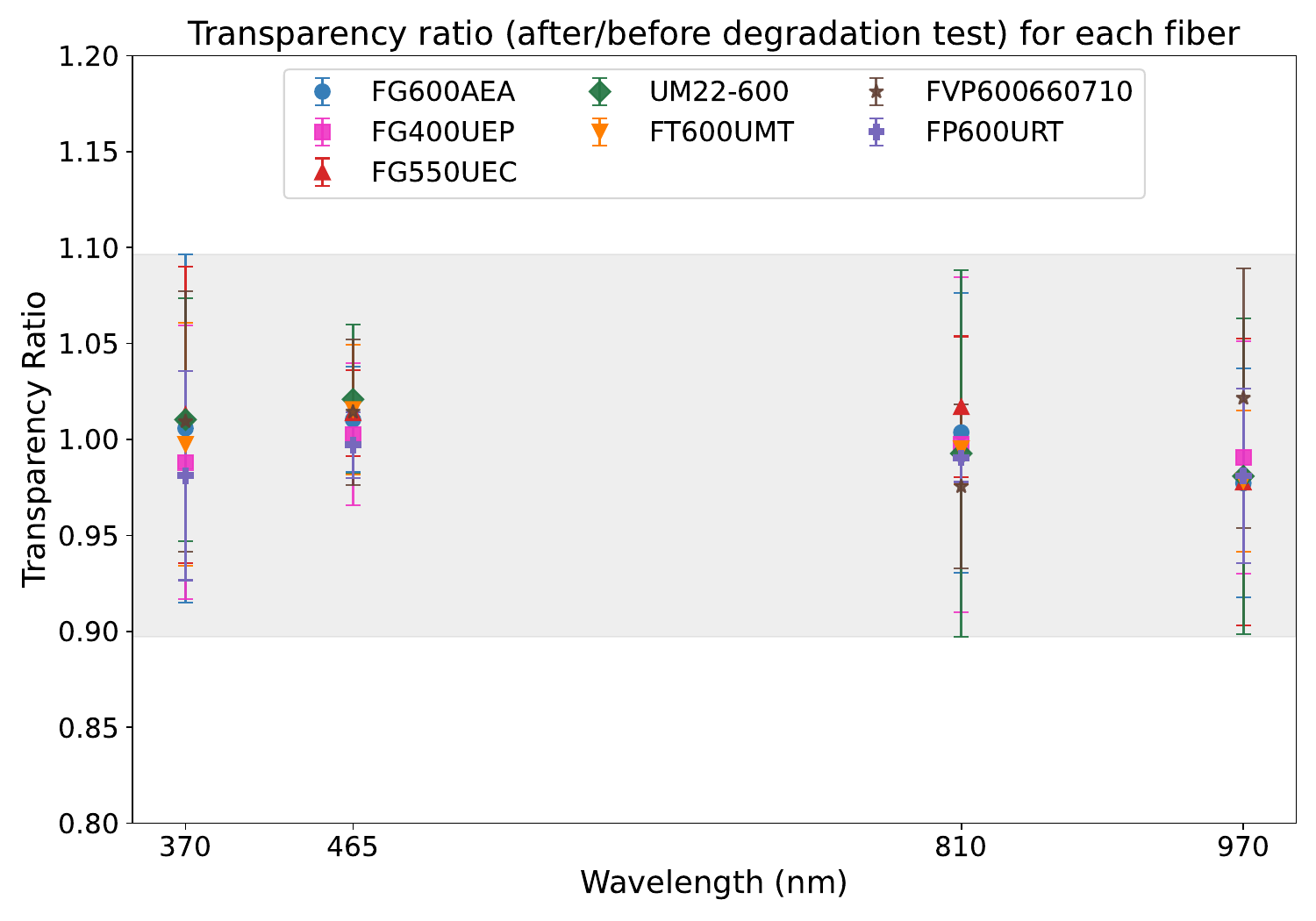}
    \caption{Ratio of transparency values measured before and after degradation test for each jacketed fiber. The shaded region represents an approximate $\pm10\%$ range around unity, within which all measurements fall when uncertainties are included, indicating no significant degradation in optical performance due to this test. Left: Transparency ratios grouped by wavelength, showing the consistency of each fiber’s performance covering the tested spectral range. Right:  Ratios grouped by fiber type, showing the stability of transparency at each wavelength.
}
    \label{fig:Ratiojacket}
\end{figure}

To quantify the stability of the fibers after the degradation test, Table~\ref{tab:ratiodeg} presents the transparency ratios defined as the transmission measured after the degradation test divided by that measured before. This normalized metric captures potential losses in optical performance and is consistently close to unity across all fibers tested, with deviations remaining within $\pm10\%$. Figure~\ref{fig:Ratiojacket} shows on the left side these ratios grouped by wavelength and the right side by fiber type. The shaded band around a ratio of $1$ highlights the region of statistical consistency. All measurements fall within this band, reinforcing the conclusion that none of the fibers experienced significant performance loss following UV pulse exposure. 

Notably, the FVP600660710 fiber demonstrated good stability, further validating their selection for applications involving repeated UV pulsing in cryogenic environments, such as the proposed DUNE UV light calibration system~\cite{fermilab2018design}.

\section{Light Diffuser Characterization}
\label{sec:LDIFF}
Light diffusers are essential components in optical calibration systems used in particle detectors. The diffusers are used to scatter and broaden the emitted light to ensure well-understood illumination across a wider photo-detector sensor area, improving the uniformity of light calibration signals across large particle detector volumes. Materials commonly used for this purpose include ground glass and quartz diffusers, which can be selected with various grit levels (typically ranging from $120$ to $1500$) to control the angular distribution, scattering uniformity, and transmission efficiency. Depending on the grit and thickness, the diffusion pattern can be tailored to  match the light calibration needs of specific particle detector geometries.
Previous designs of the light diffuser used during DUNE prototypes operation were machined in stainless steel, which, while robust, added cost and machining complexity~\cite{Abi2020}.
These diffusers required more specialized manufactured process which could present a challenge when need to be replicated by other experiments. To overcome these challenges the design has been iteratively updated by implementing compact, integrated geometries, self-aligning SMA barrel connectors, and snap-fit mounting features tailored to 3D-printed form factors. The result of these efforts is a single-piece 3D-printed diffuser housing that consolidates all optical alignment and mounting interfaces natively, eliminating post-machining and minimizing part count. This streamlined solution not only could accelerate production but also standardizes light distribution performance across large-scale light calibration systems such as those proposed to be used in DUNE.

\begin{figure}[htpb]
    \centering
    \includegraphics[scale=0.2]{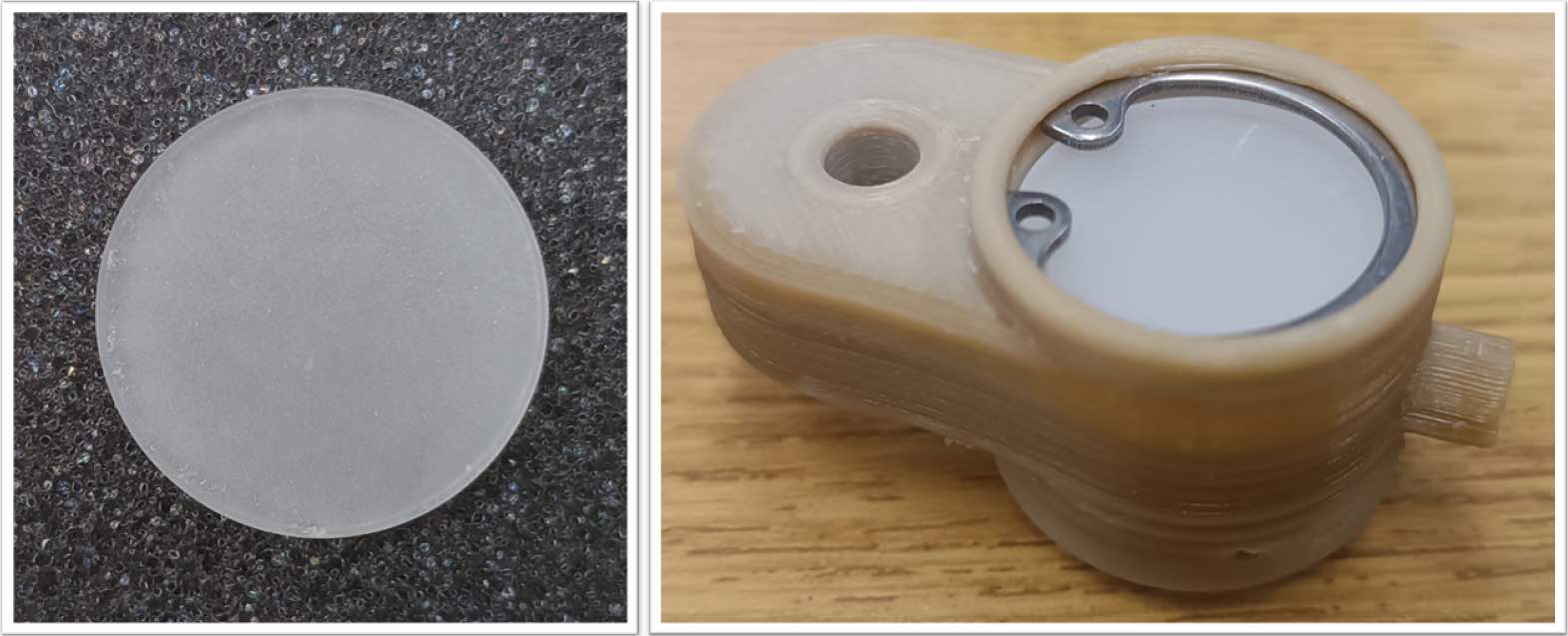}
    \caption{Left: Customized 0.75" diameter ground glass diffuser (Thorlabs DGUV-19MM-220S1S2-SP) with 220-grit finish on both sides. Right: Final diffuser assembly consisting of two stacked diffuser glasses housed within a 3D-printed PEEK holder designed for stable mounting and optical alignment in cryogenic calibration systems.}
    \label{fig:configdiff}
\end{figure}

Measurements to evaluate the spatial light distribution were performed using a customized diffuser tailored to meet the optical and mechanical requirements for cryogenic experiments. The diffuser is made from UV-grade fused silica (UVFS), has a diameter of \qty{19}{mm} (\qty{0.75}{inch}), and features a $220$-grit uniform fine surface finish on both sides, producing a well-characterized diffusion pattern. The optic also provides a clear aperture exceeding $90\%$ of its diameter, ensuring uniform transmission with minimal edge loss. Three different configurations were used to study the spatial light distribution measurements: 1) a single diffuser glass, 2) two diffuser glasses stacked, and 3) two diffuser glasses housed within a 3D-printed PEEK design (see Figure~\ref{fig:configdiff}, right). The housing was designed to securely mount the diffusers and maintain optical alignment, leveraging the benefits of PEEK, a high-performance thermoplastic with a melting point of \qty{\sim 343}{\degreeCelsius} with demonstrated thermal and mechanical stability across a wide range of temperatures, including cryogenic environments. The purpose of testing the 3D-printed PEEK housing is to facilitate application of these diffusers in particle detectors operating at cryogenic environments, as 3D-printed option provides a cost-effective and customizable solution with competitive thermal and mechanical properties.

\begin{figure}[htpb]
    \centering
    \includegraphics[scale = 0.225]{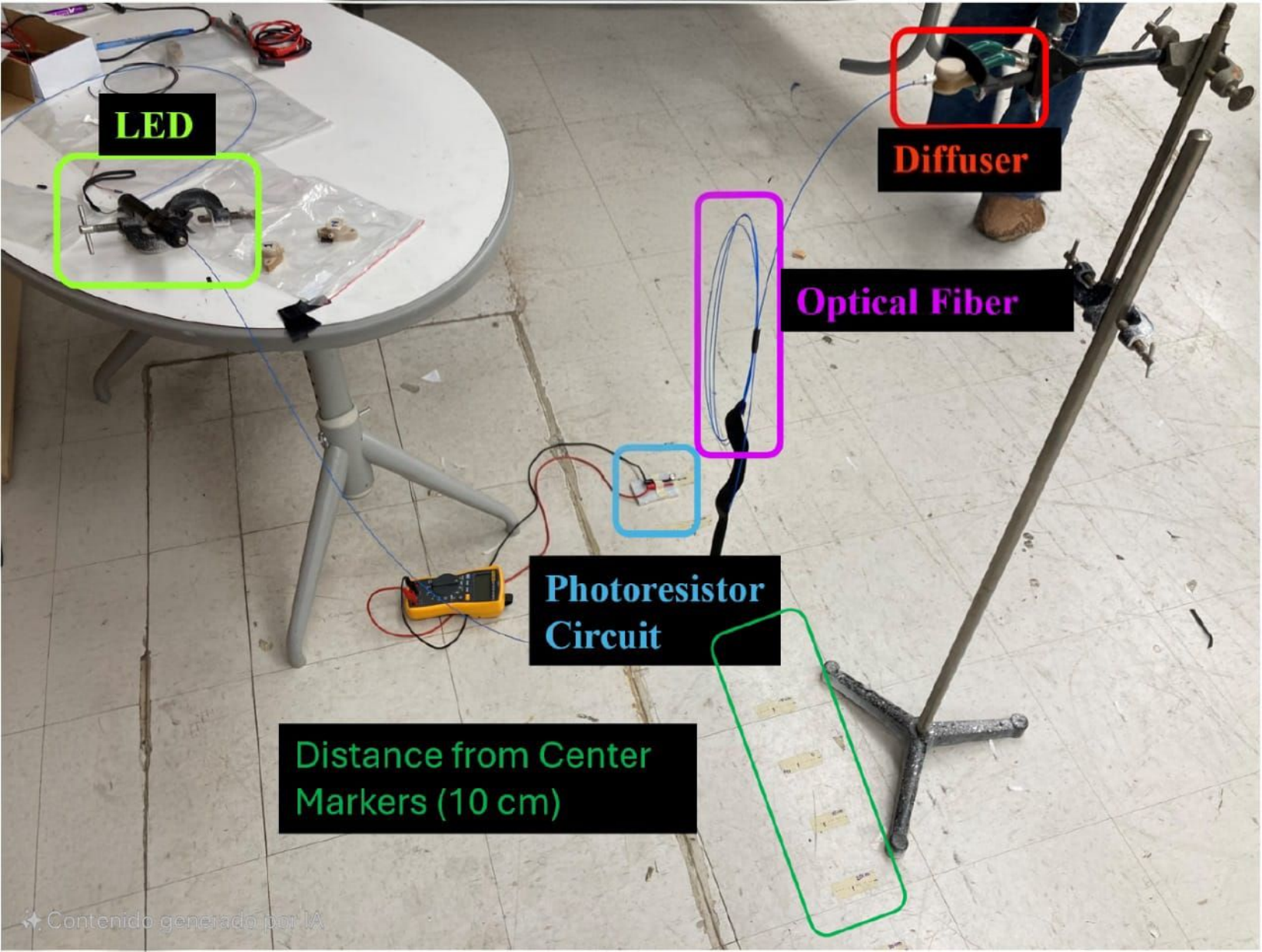}
    \includegraphics[scale=0.255]{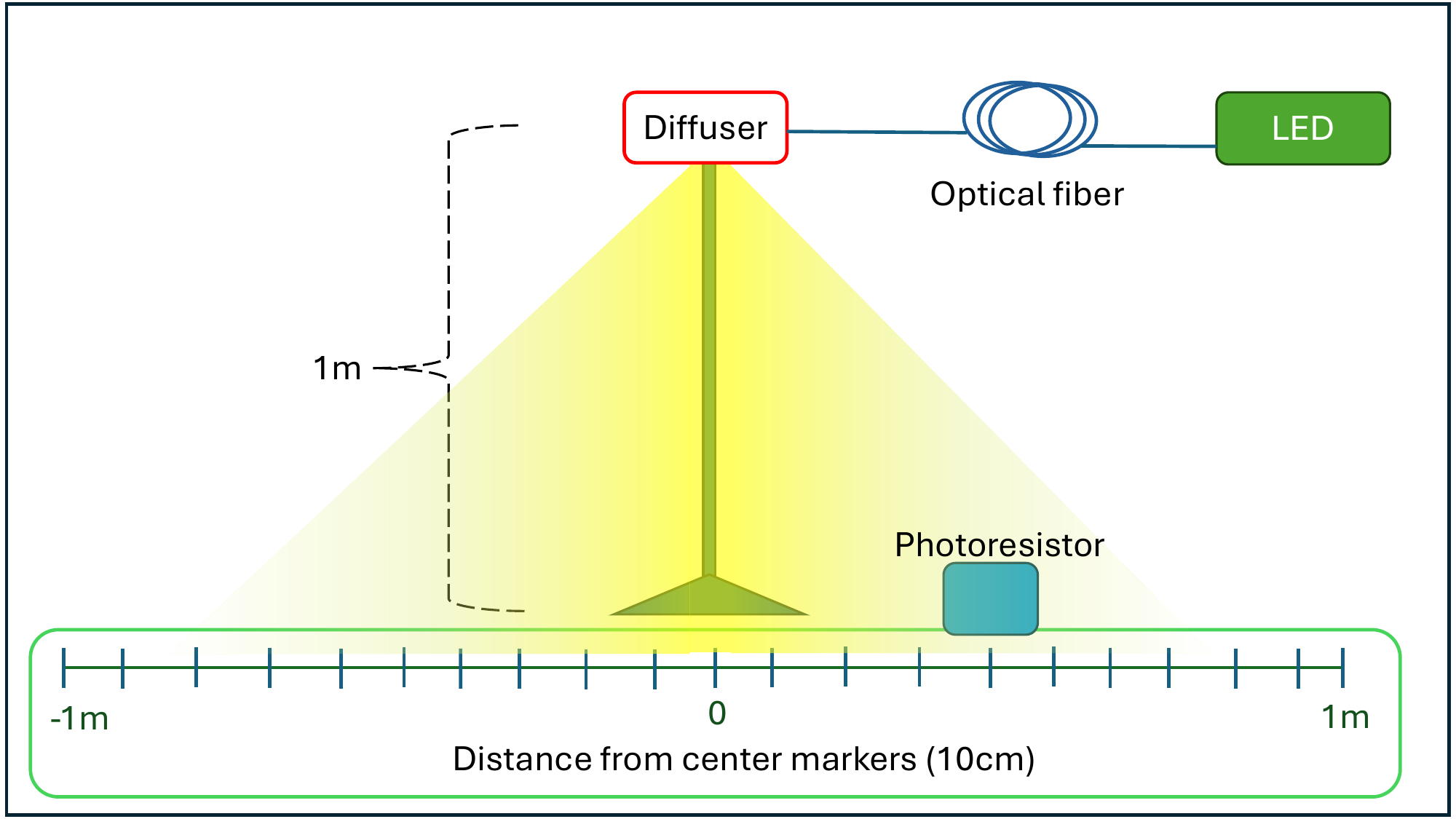}
    \caption{Left: Photograph of the experimental setup showing the LED light source, optical fiber, diffuser mounted 1 m above the floor, and the photoresistor circuit aligned along marked 10 cm intervals. Right: Schematic diagram of the setup showing the vertical diffuser position, the horizontal measurement line along the floor, and the optical path (yellow area) from the light diffused to the photoresistor.}
    \label{fig:DiffuserSetup}
\end{figure}

The light diffuser spatial distribution measurement setup is shown in Figure~\ref{fig:DiffuserSetup}, where light from an LED source was delivered through an optical fiber terminated with an SMA connector. The fiber was inserted into the diffuser housing, where it directed light toward a small internal UV mirror ($10 \times 10$\,\unit{mm}, UV Enhanced Aluminum, $\lambda/4$, Edmund Optics). The reflected beam then passed through the diffuser glass elements, scattering the light outward. The diffuser was mounted vertically in a tripod, with its center positioned at a height of \qty{1}{m} above the floor, which is defined as a "zero"-position in the horizontal coordinate system used for scanning. A photoresistor was used to detect the light intensity as a function of distance from the diffuser center along the floor. The photoresistor located in the floor was moved horizontally in \qty{10}{cm} intervals, and its resistance was recorded at each position. Since the resistance of the photoresistor is inversely proportional to the square of the incident light intensity, this could allow to measure the light attenuation for the $3$ diffuser configurations explained above.

\begin{figure}[htpb]
    \centering
    \includegraphics[scale=0.478]{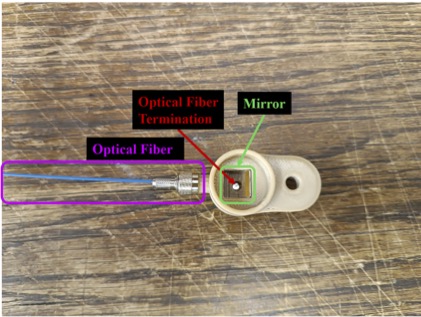}
    \includegraphics[scale=0.42]{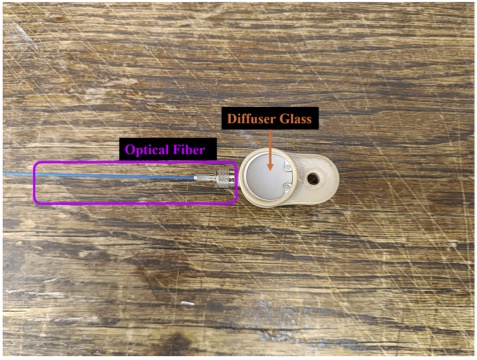}
    \caption{Left: Internal view of the 3D printed PEEK diffuser housing showing the optical fiber termination and internal UV mirror. Right: Fully assembled diffuser showing the optical fiber and installed diffuser glass inside the housing.}
    \label{fig:Diffuser-glass}
\end{figure}

Details of the internal structure of the diffuser configuration are shown in Figure~\ref{fig:Diffuser-glass}, which illustrates the housing and the alignment of the optical fiber, UV mirror, and diffuser glass. The fully assembled version is also shown.

\begin{figure}[htpb]
    \centering
    \includegraphics[scale = 0.5]{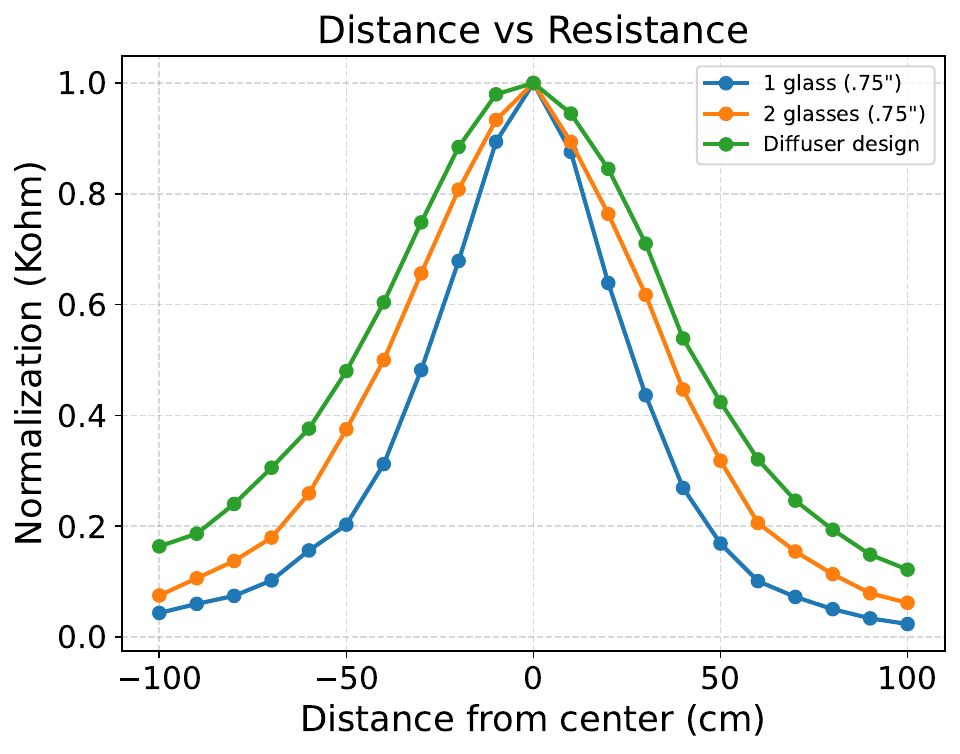}
    \caption{Normalized light intensity profiles spatial distribution measured for three diffuser configurations as a function of distance from the diffuser center. The configurations include: a single 0.75" diffuser glass (blue), two stacked 0.75" diffuser glasses (orange), and the final diffuser design consisting of two diffuser glasses housed within a 3D-printed PEEK holder (green). The data show how the combination of diffuser layering and mechanical housing affects the spatial spread of the emitted light.}
    \label{fig:Diffuserplot}
\end{figure}

Results of the light intensity recorded with the photoresistor in function of the distance with respect to the center of the diffuser are shown in Figure~\ref{fig:Diffuserplot}. The 3D printed diffuser design printed in PEEK shows the widest and most uniform light spread, confirming the effectiveness of combining diffuser glasses and a structured housing. In contrast, the unmounted glass tests ($1$ and $2$ glasses) demonstrate that while additional diffuser layers increase the spread, housing geometry also plays a critical role in directing the light uniformly.
These results support the adoption of the new diffuser design for DUNE, ensuring more even light delivery across a wide area for calibration purposes. 
The PEEK 3D printed diffuser prototype was successfully deployed and used during the operation  of ProtoDUNE HD~\cite{Soto-Oton:2024apm}.

Assuming that the diffuser scatters incident light uniformly in all directions, the observed intensity at a point depends on both the emission angle and the distance from the source. According to Lambert's cosine law, if the diffuser behaves as a Lambertian surface, the intensity at a point $x$ falls off as:
\begin{equation}
I(x) \propto \frac{\cos\theta}{d^2},
\end{equation}
where $d$ is the distance from the diffuser to the measurement point and $\theta$ is the angle of incidence (between the surface normal and the direction to that measurement point). This equation shows how intensity decreases quickly with both increasing angle and distance, combining effects of angular attenuation and geometric spreading. 

\begin{figure}[htpb]
    \centering
    \includegraphics[scale=0.55]{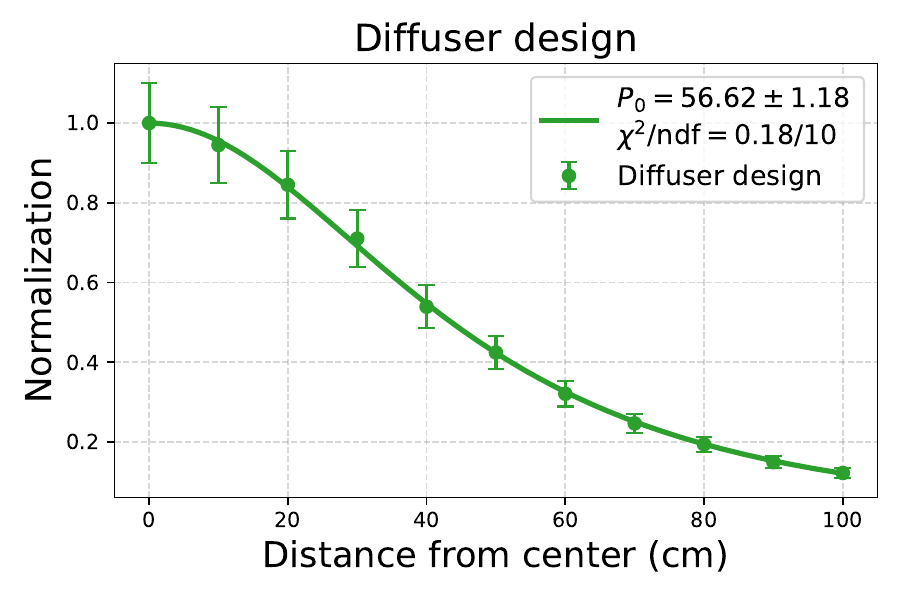}
  \caption{Spatial distribution of normalized light intensity from the diffuser design configuration, compared with Lambert’s cosine law fit. The fitted $P_0$ value and $\chi^2/\text{ndf}$ indicate that the PEEK-housing diffuser two-glass design provides a wider and the most uniform angular emission, compared with the two-glass and single-glass diffusers (orange and blue lines in Fig. \ref{fig:Diffuser-glass}) which present a narrow emission.}
    \label{fig:fitsLambert}
\end{figure}

Figure~\ref{fig:fitsLambert} shows the spatial distribution of normalized light intensity as a function of horizontal distance from the diffuser center, measured at a fixed vertical height of one meter. The data were collected using a photoresistor moved laterally and are fitted using Lambert’s cosine law,
$P(\theta) = P_0 \cos\theta$,
where $P(\theta)$ denotes the measured light intensity (or  photoresistors response) at an angle $\theta = \tan^{-1}(x/(\qty{1}{m}))$, and $P_0$ represents the peak photoresistors response at the center of the illuminated region ($\theta = \ang{0}$). The cosine factor accounts for the angular fall-off of the emitted light intensity, characteristic of an ideal Lambertian source. Since the emission is assumed to be azimuthally symmetric, the cosine dependence effectively models how light intensity spreads across the plane below the diffuser.

The diffuser design yields a light distribution with a fit value of $P_0 = 56.62 \pm 1.18$ and a reduced chi-squared of $\chi^2/\text{ndf} = 0.18/10$, indicating the good agreement with the Lambertian distribution model and a highly uniform angular emission profile. These results validate the effectiveness of layering  diffuser glasses and mechanical housing in shaping the spatial distribution and uniformity of the emitted light.

\section{Summary} 
\label{sec:Summary}
Large liquid argon time projection chambers (LArTPCs) require stable, well-characterized ultraviolet (UV) light delivery to calibrate photon detectors in situ under cryogenic conditions. We have performed a comprehensive, end-to-end characterization of the optical chain components used for such calibration systems—fused-silica fibers, SMA-to-SMA connectors, optical feedthroughs, and diffuser assemblies—covering UV, visible, and near-infrared wavelengths and including bench tests, cryogenic thermal cycling, and accelerated aging studies. The results provide quantitative guidance on component selection and system design for robust, uniform light delivery in large detectors.

We reported measurements of light transmission for different optical fiber types with multiple core, cladding and coating diameters and materials, numerical apertures, and a variety of operating wavelengths ranging from UV to near-infra red.
Connector and feedthrough losses were measured at \qty{275}{nm} and \qty{367}{nm} by inserting five-channel optical feedthroughs into a controlled test stand. The single SMA-to-SMA connector loss, derived from feedthrough measurements, is $(15.2 \pm 5.1)\%$ at \qty{275}{nm} and $(12.4 \pm 5.0)\%$ at \qty{367}{nm}; LED wavelengths were verified with a spectrometer to be centered at \qty{275}{nm} and \qty{367}{nm}. Using these references, we disentangled SMA-to-SMA connector versus bulk-fiber attenuation for representative fiber types. For a \qty{5.7}{m} high-OH fused-silica fiber (FVP600660710) with one SMA-to-SMA connector, the total transmission is $0.637 \pm 0.033$ (\qty{275}{nm}) and $0.748 \pm 0.024$ (\qty{367}{nm}). Correcting for SMA-to-SMA connector loss yields the fiber survival fractions of $0.751 \pm 0.060$ (\qty{275}{nm}) and $0.854 \pm 0.056$ (\qty{367}{nm}), corresponding to attenuation coefficients of $(0.050 \pm 0.014)$\,\unit{\per\metre} and $(0.028 \pm 0.011)$\,\unit{\per\metre}, respectively. A long Tefzel-coated fiber (FP600URT, \qty{9.55}{m}) exhibits strong UV attenuation at \qty{275}{nm}, as expected from specifications, with a fiber survival of $(0.130 \pm 0.012)$ and an attenuation coefficient of $(0.214 \pm 0.016)$\,\unit{\per\metre}; at \qty{367}{nm}, its attenuation coefficient was found to be $(0.013 \pm 0.010)$\,\unit{\per\metre}.
Across shorter test lengths, both FP600URT and FVP600660710 demonstrate strong UV transmission at \qty{370}{nm} and broadly uniform performance into the visible and near-IR.

We also evaluated system-level losses in a DUNE-style mockup consisting of four \qty{4.7}{m} high-OH fibers connected in series through an optical feedthrough (five SMA-to-SMA connectors total). The surviving light fraction through the full chain is $(0.209 \pm 0.026)$ at \qty{275}{nm} and $(0.448 \pm 0.020)$ at \qty{367}{nm}. From these, the single-fiber survival fractions for the \qty{4.7}{m} segments are $(0.831 \pm 0.060)$ (\qty{275}{nm}) and $(0.965 \pm 0.060)$ (\qty{367}{nm}), consistent with the single-fiber attenuation coefficients derived above. These measurements underscore the importance of SMA-to-SMA connector and feedthrough contributions to the light budget, particularly for longer paths and shorter wavelengths.

Cryogenic suitability and long-term stability were assessed through $30$ liquid-nitrogen thermal cycles and high-rate pulsed UV exposure. Thermal cycling produced no statistically significant degradation: the post-/pre-transparency ratios for all bare fibers at \qty{370}{nm}, \qty{465}{nm}, \qty{810}{nm}, and \qty{970}{nm} are consistent with unity within uncertainties, and visual inspection revealed no mechanical damage. A pulsed-aging test delivering $\sim 30$ million \qty{275}{nm} LED pulses to jacketed fibers likewise showed no measurable transparency loss across the same spectral bands; after/before ratios remained within $\sim 10\%$ with all central values near unity. The FVP600660710 fiber was characterized with upper limits on UV-light caused degradation to be 2.8\% for 275 nm light and 1.2\% for 367 nm light over 30 year calibration exposure.
These results indicate good mechanical resilience and optical stability under repeated cryogenic cycling and sustained UV pulsing.

Finally, we characterized diffuser spatial profiles and housing designs for uniform illumination. A compact, 3D-printed PEEK housing with stacked UV-grade fused-silica diffusers (two \qty{0.75}{inch} UVFS, $220$-grit) produces the most wider and uniform angular distribution and Lambertian emission among configurations tested. The integrated light diffuser design simplifies production, preserves alignment, and has been successfully deployed in ProtoDUNE HD operation.

In summary, we have quantified SMA-to-SMA connector, feedthrough, and fiber losses at UV and visible wavelengths; established attenuation coefficients for representative fibers; and demonstrated cryogenic robustness and pulsed-aging stability. High-OH fused-silica fibers (e.g., FVP600660710) and Tefzel-coated fibers (e.g., FP600URT) both perform well for \qtyrange{367}{370}{nm} delivery, with high-OH fibers favored for deeper UV ($\approx \qty{275}{nm}$). The compact PEEK diffuser assembly provides uniform illumination suitable for large-area calibration. These results inform component choices and calibration procedures for uniform, reliable UV light delivery in large cryogenic detectors, and have already guided successful deployments in ProtoDUNE SP and ProtoDUNE HD. Future work will extend these studies to long-duration cryogenic soaks, radiation tolerance, in-argon operation, and full-channel mapping in integrated calibration modules for forthcoming detectors.

\acknowledgments
Work at Argonne National Laboratory was supported by the U.S. Department of Energy, Office of High Energy Physics. Argonne National Laboratory, a U.S. Department of Energy Office of Science Laboratory, is operated by University of Chicago Argonne LLC under contract no. DE-AC02-06CH11357.
Work at South Dakota School of Mines and Technology was produced by Fermi Forward Discovery Group, LLC under Contract No. 892430-\hspace{0pt}24CSC00\hspace{0pt}0002 with the U.S. Department of Energy, Office of Science, Office of High Energy Physics. Publisher acknowledges the U.S. Government license to provide public access under the DOE Public Access Plan. Work at South Dakota School of Mines and Technology has also been supported by funding under the U.S. Department of Energy, Office of Science, Office of High Energy Physics under Award Number DE-SC0014223.

\bibliographystyle{JHEP}
\bibliography{bibliography}

@article{fermilab2018design,
    author        = "Abi, B. and others",
    collaboration = "DUNE",
    title         = "{Volume IV. The DUNE Far Detector Single-Phase Technology}",
    journal       = "JINST",
    volume        = "15",
    number        = "08",
    pages         = "T08010",
    year          = "2020",
    doi           = {10.1088/1748-0221/15/08/T08010},
    url           = {https://doi.org/10.1088/1748-0221/15/08/T08010}
}

@article{Abi2020,
    author         = "Abi, B. and others",
    collaboration  = "DUNE",
    title          = "{First Results on ProtoDUNE-SP Liquid Argon Time Projection Chamber Performance from a Beam Test at the CERN Neutrino Platform}",
    journal        = "JINST",
    volume         = "15",
    number         = "12",
    pages          = "P12004",
    year           = "2020",
    doi            = "10.1088/1748-0221/15/12/P12004",
    url            = "https://dx.doi.org/10.1088/1748-0221/15/12/P12004"
}

@inproceedings{Soto-Oton:2024apm,
    author = "Soto-Oton, J.",
    collaboration = "DUNE",
    title = "{ProtoDUNE Photon Detection System}",
    booktitle = "{25th International Workshop on Neutrinos from Accelerators}",
    eprint = "2412.15154",
    archivePrefix = "arXiv",
    primaryClass = "hep-ex",
    doi = "10.48550/arXiv.2412.15154",
    month = "12",
    year = "2024"
}

@misc{CERN:ProtoDUNE,
    title         = "{ProtoDUNE: Paving the Way for the Deep Underground Neutrino Experiment}",
    howpublished  = {\url{https://ep-news.web.cern.ch/content/protodune-paving-way-deep-underground-neutrino-experiment}}
}

@article{fenta2021fibre,
    author         = "Fenta, M. C. and Potter, D. K. and Szanyi, J.",
    title          = "{Fibre Optic Methods of Prospecting: A Comprehensive and Modern Branch of Geophysics}",
    journal        = "Surveys in Geophysics",
    volume         = "42",
    number         = "3",
    pages          = "551--584",
    year           = "2021",
    doi            = "10.1007/s10712-021-09634-8",
    url            = "https://doi.org/10.1007/s10712-021-09634-8"
}

@book{okoshi2012optical,
    author         = "Okoshi, T.",
    title          = "{Optical Fibers}",
    publisher      = "Elsevier",
    year           = "2012"
}

@article{keiser2014review,
    author         = "Keiser, G. and Xiong, F. and Cui, Y. and Shum, P. P.",
    title          = "{Review of Diverse Optical Fibers Used in Biomedical Research and Clinical Practice}",
    journal        = "Journal of Biomedical Optics",
    volume         = "19",
    number         = "8",
    pages          = "080902",
    year           = "2014",
    doi            = "10.1117/1.JBO.19.8.080902",
    url            = "https://doi.org/10.1117/1.JBO.19.8.080902"
}

@article{addanki2018review,
    author         = "Addanki, S. and Amiri, I. S. and Yupapin, P.",
    title          = "{Review of Optical Fibers: Introduction and Applications in Fiber Lasers}",
    journal        = "Results in Physics",
    volume         = "10",
    pages          = "743--750",
    year           = "2018",
    doi            = "10.1016/j.rinp.2018.07.028",
    url            = "https://doi.org/10.1016/j.rinp.2018.07.028"
}

@misc{Thorlabs:LEDsSpecific,
    title         = "{Light-Emitting Diodes (LEDs): 370 nm, 465 nm, 810 nm, and 970 nm}",
    howpublished  = {\url{https://www.thorlabs.com/newgrouppage9.cfm?objectgroup_id=2814}}
}

@misc{Caplinq:HighPuritySilica,
    title         = "{High Purity Silica: Sand, Quartz, Crystalline and Spherical Micro Quartz Powder}",
    howpublished  = {\url{https://www.caplinq.com/high-purity-silica.html}}
}

@misc{FOS:SolarizationSilicaFibers,
    title         = "{Solarization Effects in Silica-Silica Fibers with UV-C Wavelengths: An In-Depth Exploration of Atomic-Level Changes and Electron Behavior}",
    howpublished  = {\url{https://fosoptics.de/solarization-effects-in-silica-silica-fibers-with-uvc-wavelengths/}}
}

@misc{Avantes:IntroFiberOptics,
    title         = "{Introduction to Fiber Optics}",
    howpublished  = {\url{https://www.avantes.com/support/theoretical-background/introduction-to-fiber-optics/}}
}

@misc{FirebirdOptics:OpticalWindows,
    title         = "{Optical Windows}",
    howpublished  = {\url{https://www.firebirdoptics.com/optical-windows/fused-silica-uv-windows}}
}

@article{Paudel:2025bfh,
    author = "Paudel, A. and others",
    title = "{Modeling Light Signals Using Data from the First Pulsed Neutron Source Program at the DUNE Vertical Drift ColdBox Test Facility at CERN Neutrino Platform}",
    eprint = "2512.10790",
    archivePrefix = "arXiv",
    primaryClass = "hep-ex",
    reportNumber = "FERMILAB-PUB-25-0856-LBNF",
    month = "12",
    year = "2025"
}

@techreport{NP02:2025AnnualReport,
    author         = "{The NP02 Collaboration}",
    title          = "{NP02 2025 Annual Report to the CERN-SPSC}",
    institution    = "CERN",
    year           = "2025",
    url            = "https://cds.cern.ch/record/2932899/files/SPSC-SR-366.pdf"
}
\end{document}